\documentclass[preprint,3p,12pt]{elsarticle}

\usepackage[utf8]{inputenc}
\usepackage[T1]{fontenc}
\usepackage{calc}
\usepackage{subfigure}
\usepackage{titlesec}
\usepackage{graphicx}
\usepackage{mathtools}
\usepackage{xcolor}
\usepackage{multirow}
\usepackage{comment}
\usepackage{ulem}
\usepackage{url}

\usepackage[linesnumbered,ruled,vlined]{algorithm2e}
\usepackage{amssymb}

\usepackage{lineno}

\journal{Elsevier}
\SetKwInput{KwInput}{Input}                
\SetKwInput{KwOutput}{Output}              

\graphicspath{{img/}}

\begin{document}

\begin{frontmatter}


  \title{Convolution finite element-based digital image correlation for  displacement and strain measurements}


  \author[label1]{Ye Lu\corref{cor1}}
  \cortext[cor1]{Corresponding author}
  \ead{yelu@umbc.edu}
  \author[label1]{Weidong Zhu}

  \address[label1]{Department of Mechanical Engineering, University of Maryland Baltimore County, Baltimore, USA}


  \begin{abstract}
   This work presents a novel global digital image correlation (DIC) method, based on a newly developed convolution finite element (C-FE) approximation. The convolution approximation can rely on the mesh of linear finite elements and enables arbitrarily high order approximations without adding more degrees of freedom.  Therefore, the C-FE based DIC can be more accurate than {the} usual FE based DIC by providing highly smooth and accurate displacement and strain results with the same element size. The detailed formulation and implementation of the method have been discussed in this work. The controlling parameters in the method include the polynomial order, patch size, and dilation. A general choice of the parameters and their potential adaptivity have been discussed. The proposed DIC method has been tested by several representative examples, including the DIC challenge 2.0 benchmark problems, with comparison to the usual FE based DIC. C-FE outperformed FE in all the DIC results for the tested examples. This work {demonstrates the potential of C-FE and} opens a new avenue to enable highly smooth, accurate, and robust DIC analysis for full-field displacement and strain measurements.
  
  \end{abstract}


\end{frontmatter}


\newcommand{\revision}[2]{\sout{#1} \textcolor{red}{(#2)}}

\section{Introduction}
Data-driven or hybrid data-driven and physics-based modeling and computations have attracted great interests in the community of computational science and engineering in recent years (see e.g., \cite{liu2022eighty,kirchdoerfer2016data,chinesta2020virtual,xie2021mechanistic,fang2022data,prume2023model}), due to the increasing popularity of  machine learning, artificial intelligence, and big data techniques. Ideally, experimental data can be used directly in  numerical models to replace any unknown or even uncertain physical relationship, such as the so-called data-driven computational mechanics \cite{kirchdoerfer2016data}. Therefore, the accuracy of the experimental data becomes vital with their integration into numerical models. 

Digital image correlation (DIC), a non-contact full-field measurement method, has been widely adopted for  displacement and strain measurements in experiments \cite{burt1982local,peters1982digital,sutton2009image,hild2006digital,pan2018digital}. The general idea of DIC is to use a camera or any other digital imaging device to acquire images used to measure in-plane displacement by correlating the gray levels between the reference and the deformed image. Strain can be then calculated from the displacement field in a subsequent step. DIC is often used in material testing for material characterization, model calibration, and validation. In the new paradigm of computational mechanics, it is {natural} to expect DIC to provide accurate displacement and strain data to be integrated with numerical simulations for model updating or surrogate.

DIC techniques can be classified into two main families: local DIC and global DIC. In general, local DIC \cite{burt1982local,sutton2009image} splits the image into some subsets and then performs the gray level correlation analysis in a subset-wise fashion. Therefore, local DIC does not guarantee the continuity {(compatibility)} of the measured displacement at intersections of subsets. {The comparison with simulated displacement fields might need additional smoothing techniques and a proper interpolation on a simulation mesh.} Nevertheless, local DIC is widely adopted for commercial and research oriented DIC packages because of its simplicity of implementation and efficiency. There is an intensive literature on the development and improvement of local DIC techniques, including the studies on subset shape functions \cite{schreier2002systematic}, grayscale interpolation \cite{schreier2000systematic}, correlation criteria \cite{tong2005evaluation}, subset size \cite{pan2008study}, which can strongly affect the accuracy of the local DIC results. {It is worthwhile mentioning that the displacement compatibility can be imposed in local DIC through a constrained optimization problem, though some convergence concerns may arise \cite{yang2019augmented}. In this case, it involves the solution of some global problems, leading to a method with a mixed nature of local and global methods.}  More recently, machine learning based DIC has also been developed for improved accuracy of local DIC \cite{duan2023digital,yang2022deep}. {In terms of 3D stereo-DIC \cite{balcaen2017stereo,dufour2015cad}, local DIC seems more advanced than global methods, especially when the out-of-plane displacement and strain become significant, which makes local DIC much more popular in this scenario.} However, it should be noted that the local DIC method must use a relatively large subset size to have enough {measurement} accuracy, which usually restricts its application to measure relatively homogeneous materials with slightly varying strain gradients. In addition, local DIC does not provide results directly {linked} to finite element (FE) based simulations. The integration of local DIC data with numerical models might not be straightforward.

Global DIC, also known as FE based DIC \cite{besnard2006finite}, {as an alternative to local DIC}, is thus developed to directly link the experimentally measured displacement to the simulated one.  Global DIC can use the same FE shape function and FE mesh as simulations for analyzing the displacement and makes the {connection} to FE simulations very straightforward. This offers great advantages for its integration with numerical models.  In the meantime, the FE-DIC measured displacement maintains {automatically} the global continuity across the subsets (or elements). As the element size is usually smaller than local DIC,  global DIC is preferable for heterogeneous materials with complex displacement fields. To improve the accuracy and robustness of  global DIC, works have been done through different aspects, such as using local DIC results as initial {guesses} in  global DIC \cite{wang2015some}, multi-grid formulation with regularization \cite{fedele2013global}, {and} multiscale FE-DIC approach \cite{passieux2015multiscale}. However, FE-DIC usually employs low order shape functions for the displacement field, which limits the accuracy and smoothness of the DIC results especially for the strain measurement, although higher order (quadratic) elements based DIC \cite{besnard2006finite, ma2012mesh} and Non-Uniform Rational B-Spline (NURBS) based IsoGeometric-DIC (IG-DIC) \cite{elguedj2011isogeometric} have been developed.

This work proposes a novel global DIC method, which is expected to enable a unified correlation analysis framework with arbitrarily high order approximations. The proposed global DIC method is based on the newly developed convolution FE approximation \cite{lu2023convolution}. The convolution FE method was originally developed for solving partial differential equations (PDEs). It has been demonstrated that the convolution approximation can enable superior accuracy with arbitrarily high order convergence rates for the solutions of PDEs, without adding more degrees of freedom into the FE mesh. {This is based on the fact that the solutions of many problems are smooth and regular, and therefore  high order approximations can help improve the accuracy and convergence. Moreover, the method} can also improve the solution stability in nonlinear analysis \cite{lu2023convolution}, resolve locking problems with low order elements \cite{park2023convolution}, and provide a built-in filter in topology optimization \cite{li2023convolution}. By incorporating the convolution approximation into the DIC, we can obtain similar advantages, such as improved accuracy for both displacement and strain without modifying the original FE mesh, and improved robustness with respect to noise due to the convolution "filter", as demonstrated by the numerical examples.  {Different from the conventional} IG-DIC, the C-FE based DIC {has the following features}: 1) high order approximation without modifying the mesh, 2) flexible local/global adaptivity, 3) {direct} use of linear FE mesh, {and 4) Kronecker delta property that facilitates the implementation of Dirichlet-type  conditions. Some special treatments are usually required to have similar features for IG based methods \cite{de2018role,schillinger2016lagrange,fromm2023interpolation,chapelier2021free}. These methods can also be combined with C-FE as they make direct links between FE and IG methods.} {In case of fracture or bi-material interface, the C-FE based DIC can be extended to account for the displacement discontinuities, based on the concept of eXtended FE \cite{moes1999finite,rethore2008extended}. This will be investigated in our future work.} The proposed global DIC framework {has great potential} to provide accurate FE analysis-suitable data for the next-generation computational algorithms that rely on intensive interactions between numerical models and experimental data. 

{The objective of this paper is to demonstrate the benefits of using C-FE approximation in DIC applications, especially in terms of offering high order approximations, a built-in length scale filter, and flexible potential adaptivity.} The paper is organized as follows. We start with {a general concept of convolution FE approximation in Section \ref{section:concept}}. The formulation and implementation of C-FE based DIC is discussed in Section \ref{section:CFEDIC}. Section \ref{sec:example} presents the numerical results of the proposed DIC with comparison to the usual FE based DIC. The paper will be closed with some concluding remarks. 

\section{Convolution finite element approximation}
\label{section:concept}
\begin{figure}[htbp]
\centering 
{\includegraphics[scale=0.35]{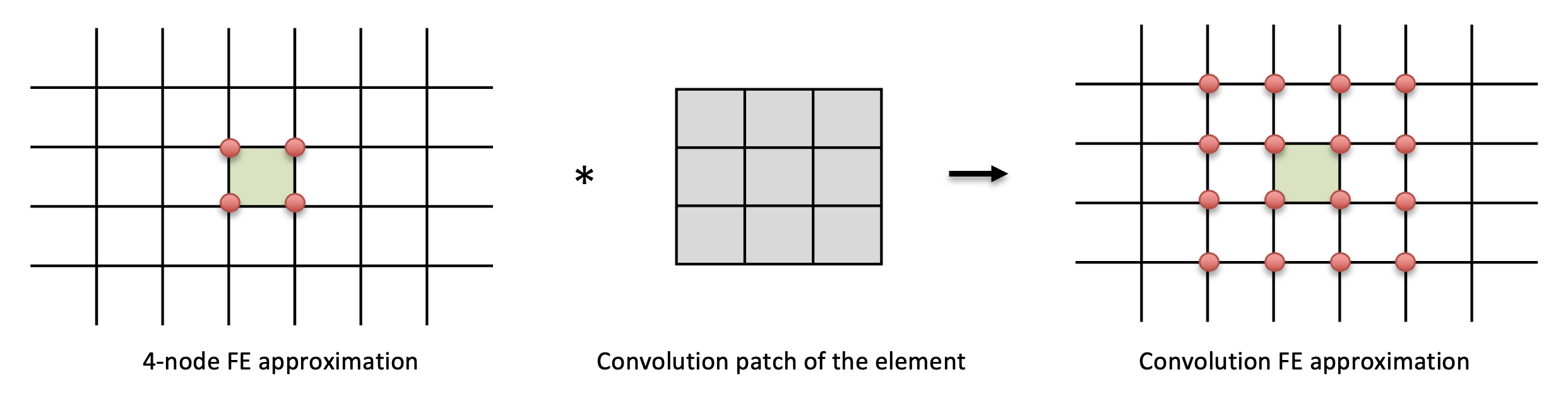}}
\caption{Supporting nodes for  finite element and convolution finite element approximations}
\label{fig:convapprox}
\end{figure}

The convolution finite element (C-FE) method is based on a newly developed convolution approximation method \cite{lu2023convolution} that combines the finite element and meshfree radial basis interpolations. To illustrate the idea of convolution approximation, let us consider a 2D finite element grid shown in \figurename~\ref{fig:convapprox}. The usual FE approximation uses a 4-node support $A^e$ and defines a linear type interpolation: $u(\boldsymbol{\xi})=\sum_{i\in A^e} N_i(\boldsymbol{\xi}) u_i$, whereas the C-FE uses, in addition, the neighboring nodes to construct a high order approximation with the help of a smooth convolution patch function $W$. As a result, the convolution approximation can be written as
{
\begin{equation}
\displaystyle
\label{eq:convolution}
     u^c (\boldsymbol{\xi})=(u*W)(\boldsymbol{\xi})=\int_{\Omega_{\text{patch}}} u(\boldsymbol{t})W(\boldsymbol{\xi-\boldsymbol{t}})d\boldsymbol{t}
\end{equation}}
where $\Omega_{\text{patch}}$ is the convolution patch domain, which can span outside an element $\Omega_e$ with a predefined patch size. Since $W$ can be an arbitrary high order smooth function, the resultant approximation $u^c$ can also be {of} arbitrarily high order.  \figurename~\ref{fig:convapprox} is a discrete version of the above convolution operation.  

{The above convolution formulation can be seen as a general and effective way to construct high order approximations without modifying the total number of nodes. C-FE provides a specific way to construct an equivalent approximation to the discrete version of convolution. In addition, C-FE approximation satisfies the partition of unity and the Kronecker delta property, and therefore fits perfectly the FE framework.}

{Among various choices to define the convolution patch function $W$  \cite{lu2023convolution}, we adopt here the radial basis interpolation (see \ref{apdx:radialbasis}) as they can ensure the desired properties for the final approximation functions. Furthermore, we consider that the elemental patch domain $\Omega_{\text{patch}}$ is formed by several nodal patches. The corresponding convolution approximation then reads } 

\begin{equation}
\label{eq:C-FEM}
    u^{c}(\boldsymbol{\xi})=\sum_{i\in A^e}{N}_{{i}}(\boldsymbol{\xi})\sum_{j\in A^i_s} {W}^{\boldsymbol{\xi}_i}_{a,j} (\boldsymbol{\xi}) u_j=\sum_{k\in A^e_s} \tilde{{N}}_k(\boldsymbol{\xi}) u_k
\end{equation}
where $A_e$ is the original FE support, $A^i_s$ is the nodal patch defined in the radial basis interpolation, $A^e_s=\underset{i\in A^e}{\bigcup}  A^i_s$ contains all the supporting nodes in the convolution element patch $\Omega_{\text{patch}}$, {and} $\tilde{{N}}_k$ and $u_k$ denote the final C-FE shape function and the nodal solution, respectively. 
In  this convolution approximation, three controllable parameters can be defined: polynomial order $p$ of $W$, dilation parameter $a \in \mathbb{R}^+$ , and patch size $s\in \mathbb{Z}^+$. The dilation $a$ controls the window size {(influencing or non-zero domain)} of kernel functions in $W$. The patch size $s$ defines the number of layers of elements outside the original (4-node) element.

\begin{figure}[htbp]
\centering 
{\includegraphics[scale=0.3]{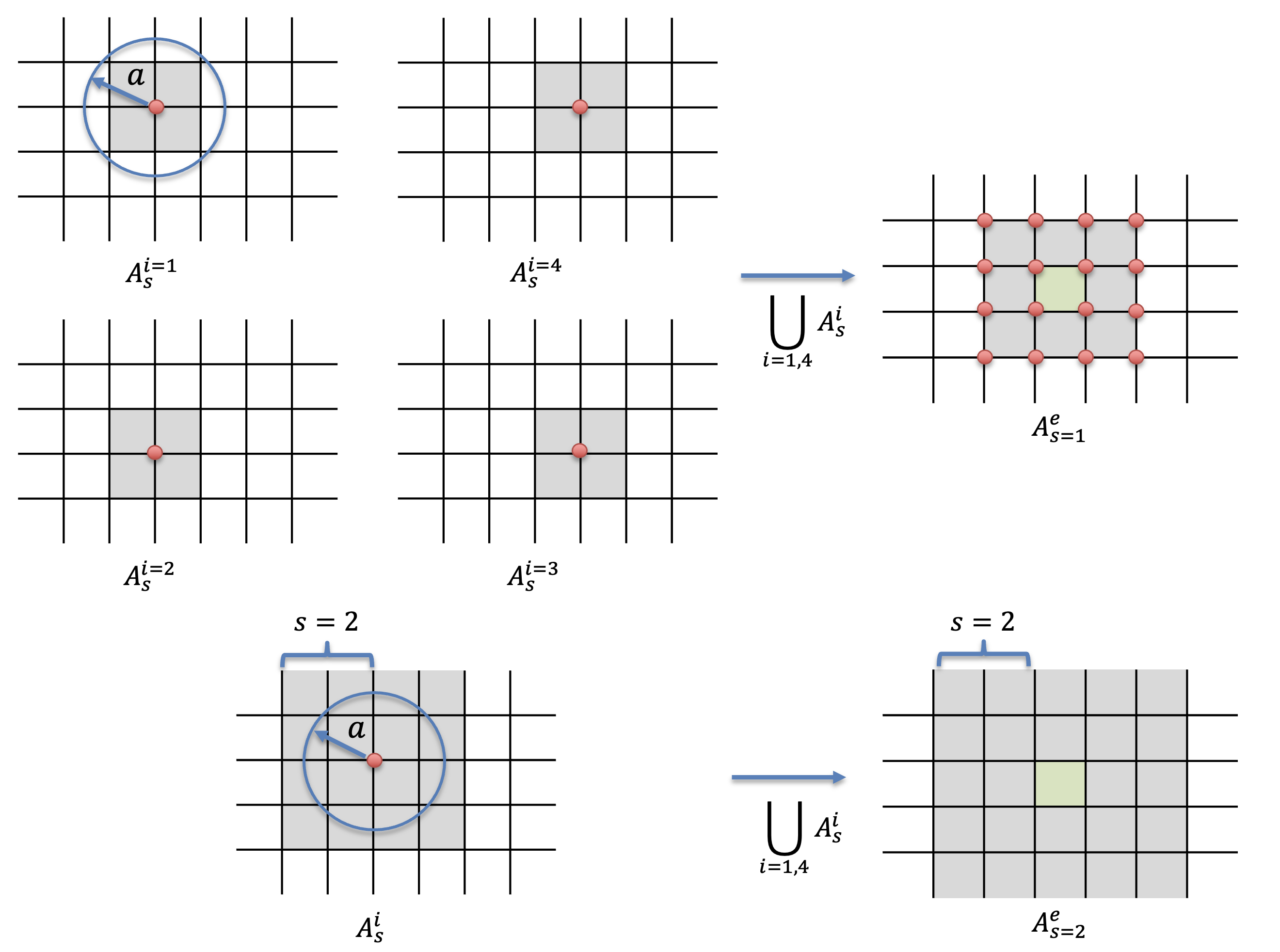}}
\caption{{Illustration of different patch sizes and the influencing domain of kernel function defined by the dilation}}
\label{fig:CFEsupport}
\end{figure}

{\figurename~\ref{fig:CFEsupport} illustrates examples of two different patch sizes of the convolution approximation resulting from Eq. \eqref{eq:C-FEM}. The final convolution  patch is formed by 4 nodal patches, and each nodal patch is associated with a dilation parameter $a$ and a patch size $s$. In particular, we can see that the dilation  can be smaller than the patch size in general cases. Since the dilation defines the actual influencing domain of the kernel functions used in $W$, it can be expected that some of the patch nodes might have zero contribution to the convolution interpolation. In another word, a large patch size does not necessarily mean a large set of influencing nodes. The  patch size and dilation have to be increased together to have a large influencing domain for an convolution element.  This also offers the flexibility in the choice of parameters to control the final approximation. It has been observed that the approximation accuracy depends on both parameters and the convolution approximation can potentially degenerate to a FE approximation using a very small dilation \cite{lu2023convolution}. A discussion about the degeneration and parametric studies of different combination of the parameters can be found in \cite{lu2023convolution}. An example of constructing specific 1D convolution shape functions can be found in \ref{apdx:1DCFEshapefunction}.
}

\begin{figure}[htbp]
\centering
\subfigure[$p=1, a=1$]{\includegraphics[scale=0.35]{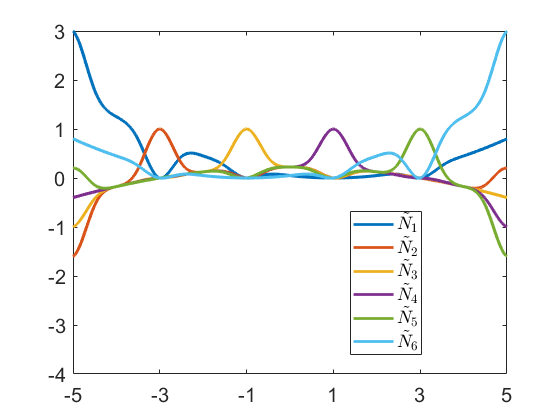}}
\subfigure[$p=1, a=2$]{\includegraphics[scale=0.35]{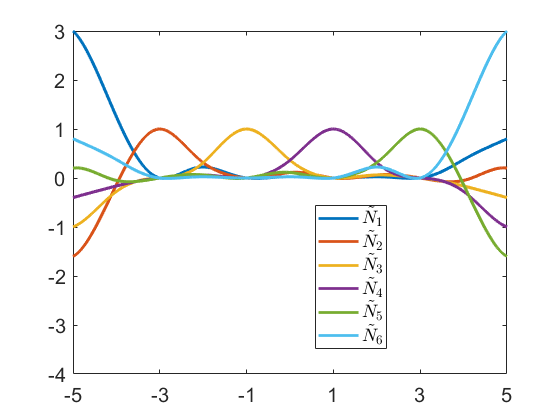}}\\
\subfigure[$p=2, a=1$ ]{\includegraphics[scale=0.35]{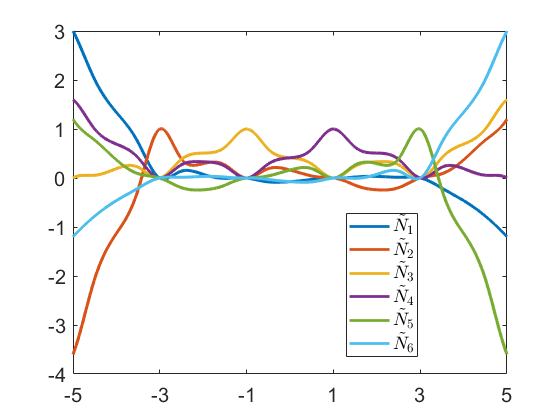}}
\subfigure[$p=2, a=2$]{\includegraphics[scale=0.35]{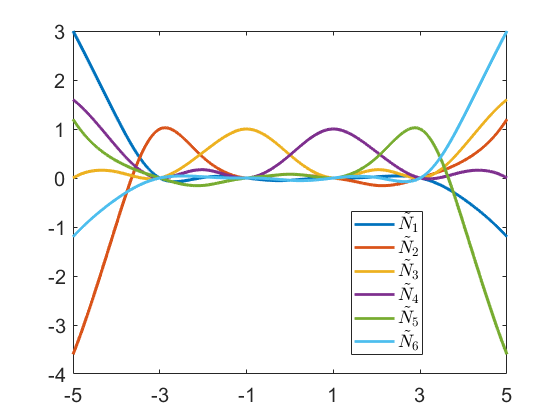}}
\caption{Illustration of the convolution finite element shape function, where the patch size $s=2$, the support nodes locate at $\{-5,-3,-1, 1, 3, 5\}$, $p$ denotes the polynomial order of $W$. The original FE shape function remains linear for all the cases, whereas the convolution shape function is nonlinear with higher-order smoothness.}
\label{fig:convshapefunctions}
\end{figure}

\begin{figure}[htbp]
\centering
{\includegraphics[scale=0.35]{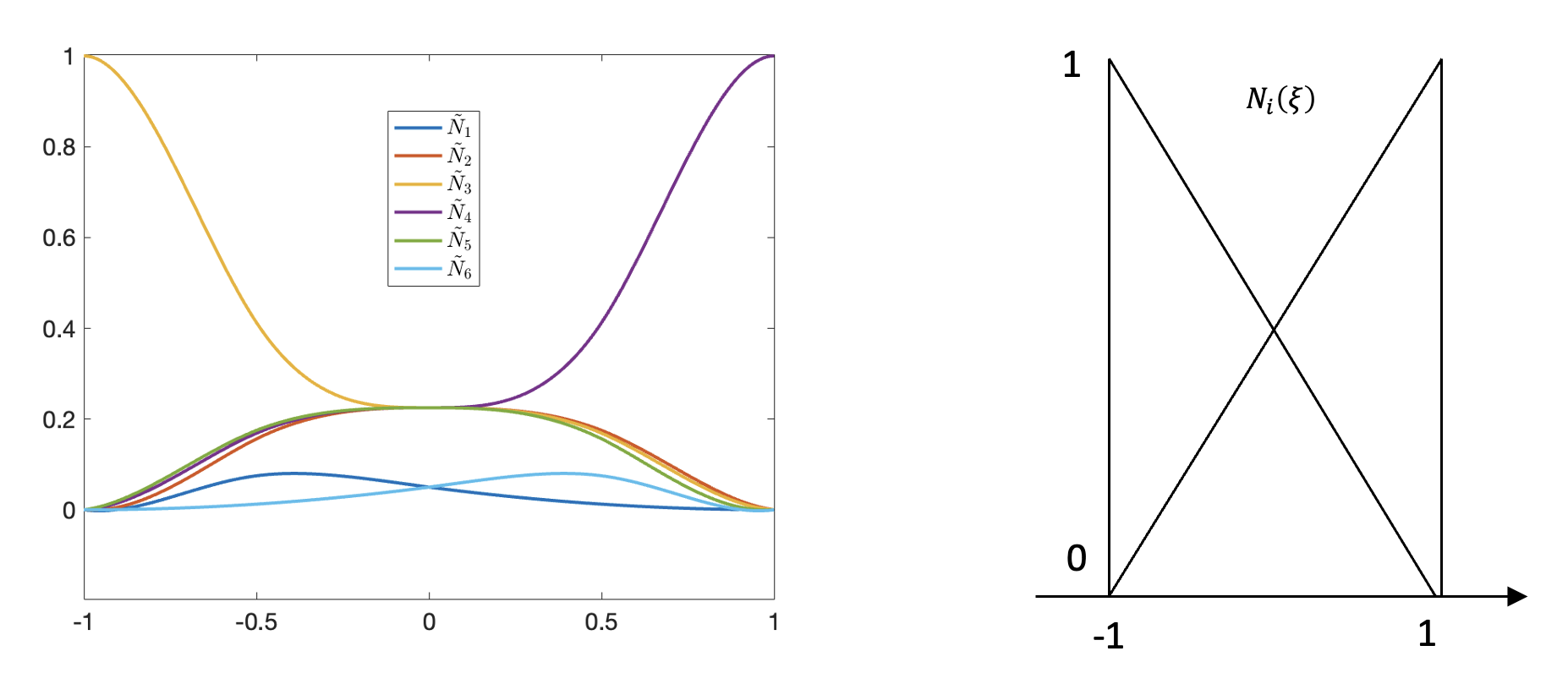}}
\caption{{C-FE shape functions over the element region $[-1,1]$. Left: C-FE with $s=2, p=1, a=1$, Right: FE.}}
\label{fig:convshapefunctionscut}
\end{figure}

To illustrate the effect of $p$ and $a$, we can look at the 1D C-FE shape functions shown in \figurename~\ref{fig:convshapefunctions}. The patch size $s$ is fixed at 2 in these examples. The original element is defined by two nodes $A^e=\{-1, 1\}$. With additional two layers ($s=2$) outside the element , the support of the convolution patch has in total six nodes: $A^e_s=\{-5,-3,-1, 1, 3, 5\}$ and therefore is associated with six C-FE shape functions. \figurename~\ref{fig:convshapefunctions} shows four combinations of $p$ and $a$ for the 1D C-FE shape functions. It can be seen that the C-FE shape functions are smooth functions with a higher continuity with respect to linear FE functions. Moreover, it shows that a large $a$ with a high order $p$ can lead better behaviors of the shape functions. {Note that although the C-FE functions can span outside the original element, the area of interest for the final functions is still $[-1,1]$. The difference with respect to FE is that there will be 6 functions contributing to the interpolation. The Kronecker delta property is ensured, as shown in \figurename~\ref{fig:convshapefunctionscut}.  }

\textbf{Remark}: the C-FE approximation can use the same mesh as linear FE without introducing more global DoFs. Increasing the order of approximation only needs a higher order $p$ and a larger dilation $a$ and potentially a larger patch size $s$. Although the DoFs of the global discrete system is not modified, the  increasing connectivity of each element can lead to a larger bandwidth in the final sparse discrete matrix of the system. A relatively smaller patch size $s$ is preferred when the accuracy is satisfactory.

Similarly to {FE method (FEM)}, we can define the isoparametric mapping between the natural (parametric) coordinate $\boldsymbol{\xi}$ and the physical coordinate $\boldsymbol{x}$ using the  C-FE shape functions
\begin{equation}
\label{eq:isoparametricmap}
    \boldsymbol{x}(\boldsymbol{\xi})=\sum_{i\in A^e}{N}_{{i}}(\boldsymbol{\xi})\sum_{j\in A^i_s} {W}^{\boldsymbol{\xi}_i}_{a,j} (\boldsymbol{\xi}) \boldsymbol{x}_j=\sum_{k\in A^e_s} \tilde{{N}}_k(\boldsymbol{\xi}) \boldsymbol{x}_k
\end{equation}

The convolution approximation has proven to be a promising way to improve the solution accuracy, smoothness, and stability. Various examples of using {C-FE method (C-FEM)} for solving PDEs have been shown in \cite{lu2023convolution}. An implementation of C-FEM  using GPU (graphics processing unit) is also discussed in \cite{park2023convolution}. This work presents a novel C-FE approximation based DIC technique. We expect highly accurate, smooth, and robust displacement and strain measurements with the proposed DIC analysis.

\section{Convolution finite element for digital image correlation}
\label{section:CFEDIC}
This section presents in detail the C-FE based DIC formulation and the numerical implementation. The potential adaptivity enabled by the convolution approximation is also discussed.
\subsection{Problem formulation}
The principle of the proposed DIC remains similar to that of original FE-DIC \cite{besnard2006finite}. Considering two speckle images with one as the reference (undeformed) image and another as the deformed image, the grayscale of the reference and deformed images is denoted, respectively, by $f(\boldsymbol{x})$ and $g(\boldsymbol{x})$. The relationship of the two images can be defined by the conservation of the optical flow
\begin{equation}
\displaystyle
\label{eq:opticalflow}
     g(\boldsymbol{x})=f(\boldsymbol{x}{-}\boldsymbol{u})
\end{equation}
where $\boldsymbol{u}$ is the displacement vector associated with the material point originally occupied the position $\boldsymbol{x}$ in the reference image. The goal of DIC is to find out the displacement $\boldsymbol{u}$ such that the difference between the two grayscales is minimized
\begin{equation}
\displaystyle
\label{eq:lossfunction}
     \min\ \mathcal{L}=\int_\Omega (g(\boldsymbol{x})-f(\boldsymbol{x}{-}\boldsymbol{u}))^2 d\boldsymbol{x}
\end{equation}
where $\Omega$ denotes the zone of interest (ZoI) in the image. Now, considering the Taylor's expansion of $f$ and ignoring the high order terms
\begin{equation}
\displaystyle
\label{eq:linearization}
     f(\boldsymbol{x}{-}\boldsymbol{u})=f(\boldsymbol{x}){-} \boldsymbol{u}\cdot\nabla f 
\end{equation}
Substituting  \eqref{eq:linearization} into \eqref{eq:lossfunction} leads to 
\begin{equation}
\displaystyle
\label{eq:lossfunction_linear}
     \min \ \mathcal{L}=\int_\Omega (g(\boldsymbol{x})-f(\boldsymbol{x}){+}\boldsymbol{u}\cdot\nabla f)^2 d\boldsymbol{x}
\end{equation}
Taking the variation of $\mathcal{L}$ and let $\delta\mathcal{L}=0$, we have
\begin{equation}
\displaystyle
\label{eq:lossfunction_linear_variation}
     \delta \mathcal{L}=\int_\Omega 2(g-f{+}\boldsymbol{u}\cdot\nabla f) \nabla f\cdot \delta \boldsymbol{u}\ d\boldsymbol{x}=0
\end{equation}

Solving the above  problem needs an approximation of the displacement $\boldsymbol{u}$. Traditional ways to solve this problem uses FE approximation, including both Q4 linear element and Q8 quadratic element. Their performance against the proposed C-FE approximation will be discussed later in the numerical experiments.

\subsection{Convolution FE discretization}
Let us consider a discretized ZoI, denoted by $\Omega^h$ with the element size $h$.  The element-wise displacement is approximated by the C-FE shape function
\begin{equation}
\label{eq:C-FEM-DIC}
    \boldsymbol{u}^e(\boldsymbol{\xi})=\sum_{i\in A^e}{N}_{{i}}(\boldsymbol{\xi})\sum_{j\in A^i_s} {W}^{\boldsymbol{\xi}_i}_{a,j} (\boldsymbol{\xi}) \boldsymbol{u}_j=\sum_{k\in A^e_s} \tilde{{N}}_k(\boldsymbol{\xi}) \boldsymbol{u}_k
\end{equation}
where $\boldsymbol{\xi} \in \Omega^e\vcentcolon=\Omega_\xi \times \Omega_\eta = [-1,1]\times[-1,1]$. The displacement vector $\boldsymbol{u}=[u, v]^T$ with $u$ as the horizontal component and $v$ as the vertical component. The element-wise interpolation can be written in a vector form by introducing a convolution shape function vector, like the usual FEM
\begin{equation}
\label{eq:C-FEM-shape-vector}
    \tilde{\boldsymbol{N}}(\boldsymbol{\xi}) =\begin{pmatrix}
     \tilde{{N}}_1 &\   \tilde{{N}}_2 &\ \cdots  &\ \tilde{{N}}_K  \\
       &\   &\  &\ &\  \tilde{{N}}_1 &\   \tilde{{N}}_2 &\ \cdots  &\ \tilde{{N}}_K  
    \end{pmatrix}\\
\end{equation}
where we assume the convolution element patch has $K$ nodes and $A^e_s=[1,\dots,K]$. The element-wise interpolation \eqref{eq:C-FEM-DIC} is then
\begin{equation}
\label{eq:C-FEM-DIC-vector}
    \boldsymbol{u}^e(\boldsymbol{\xi})= \tilde{\boldsymbol{N}}(\boldsymbol{\xi}) \boldsymbol{U}^e
\end{equation}
with 
\begin{equation}
    \boldsymbol{U}^e=[u_1, u_2, \dots, u_K, v_1, v_2, \dots, v_K]^T
\end{equation}

The discretized form of \eqref{eq:lossfunction_linear_variation} reads then
\begin{equation}
\displaystyle
\label{eq:lossfunction_linear_discrete}
     \sum_e \int_{\Omega^e} (g-f{+}(\tilde{\boldsymbol{N}}\boldsymbol{U}^e)^T\nabla f) \nabla^T f \tilde{\boldsymbol{N}}\delta\boldsymbol{U}^e J d\boldsymbol{\xi}=0
\end{equation}
where the determinant of Jacobian matrix $J=|\boldsymbol{J}|=|\frac{\partial\boldsymbol{x}}{\partial \boldsymbol{\xi}}|$ can be derived from \eqref{eq:isoparametricmap}. Rearranging the above equation leads to
\begin{equation}
\displaystyle
     (\sum_e \int_{\Omega^e} (g-f)\nabla^T f \tilde{\boldsymbol{N}}J d\boldsymbol{\xi} {+} \sum_e \int_{\Omega^e}\boldsymbol{U}^{eT}\tilde{\boldsymbol{N}}^T\nabla f\  \nabla^T f \tilde{\boldsymbol{N}} J d\boldsymbol{\xi})\ \delta\boldsymbol{U}=0
\end{equation}
or
\begin{equation}
\displaystyle
     (\boldsymbol{Q}^T {+} \boldsymbol{U}^T \boldsymbol{K}^T )\ \delta\boldsymbol{U}=0
\end{equation}
with 
\begin{equation}
\displaystyle
\begin{aligned}
    \begin{cases}
       \boldsymbol{Q} =\sum_e \int_{\Omega^e} \tilde{\boldsymbol{N}}^T(g-f)\nabla f J d\boldsymbol{\xi}  \\
       \boldsymbol{K} = \sum_e \int_{\Omega^e}\tilde{\boldsymbol{N}}^T\nabla f\  \nabla^T f \tilde{\boldsymbol{N}} J d\boldsymbol{\xi}
    \end{cases}
\end{aligned}
\end{equation}
Therefore, the nodal displacement vector $\boldsymbol{U}$ can be found by 
\begin{equation}
\displaystyle
\label{eq:DIC-discrete}
     \boldsymbol{K}\boldsymbol{U} ={-}\boldsymbol{Q} 
\end{equation}
For problems with large deformation, the above equation might need to be solved repetitively together with a multi-grid strategy \cite{elguedj2011isogeometric,passieux2019classic}. We will investigate this point in our future work. The implementation of the C-FEM can be done in a {analogous} way to FEM. {The matrix size of each element is larger than linear FE and is increased with the patch size. They can be integrated by usual matrix/vector operations. The assembly can also be done in the usual way by offering appropriate nodal index information.}

However, some aspects need special treatments as shown below.
\begin{itemize}
    \item \textit{Connectivity for the element patch}: Although the convolution element has the same size as a 4-node element with $\Omega^e\vcentcolon=\Omega_\xi \times \Omega_\eta = [-1,1]\times[-1,1]$, the support of the convolution shape functions expands beyond the element domain. This requires a systematic way to define the connectivity of nodes and their coordinates  for the patch information of each convolution element.  As an example, \figurename~\ref{fig:connectivity} shows the support for a convolution element with the patch size $s=1$. The connectivity vector of this element should contain in total $4\times 4 = 16$ nodes, each of them is associated with a shape function $\tilde{{N}}_k$ in the parametric space. The present parametric coordinates of the nodes are suitable for a regular mesh. They can be adapted if irregular meshes are used, as discussed in \cite{lu2023convolution}. Since we define the shape functions in the reference parametric space, the same shape functions can be used repetitively for different elements.  
\begin{figure}[htbp]
\centering 
{\includegraphics[scale=0.45]{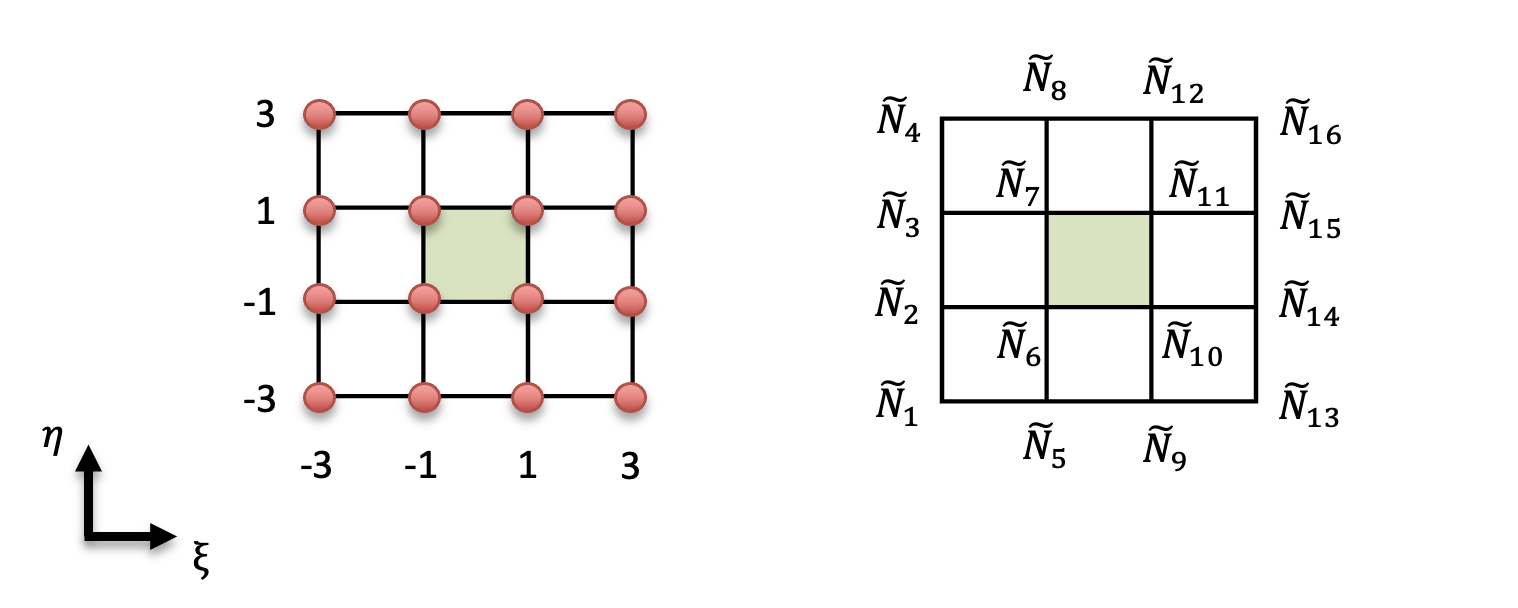}}
\caption{Supporting nodes for the convolution shape functions with the patch size $s=1$. Green color indicates the actual element domain.}
\label{fig:connectivity}
\end{figure}
    \item  \textit{Boundary elements}: The elements on (or close to) the boundary might not have enough nodes for the predefined patch size. In this case, a reduced convolution patch can be used instead by ignoring the patch nodes which expand outside the computational domain. \figurename~\ref{fig:reducedpatch} shows some examples of reduced patches with comparison to the full patch for interior elements.
\begin{figure}[htbp]
\centering 
{\includegraphics[scale=0.35]{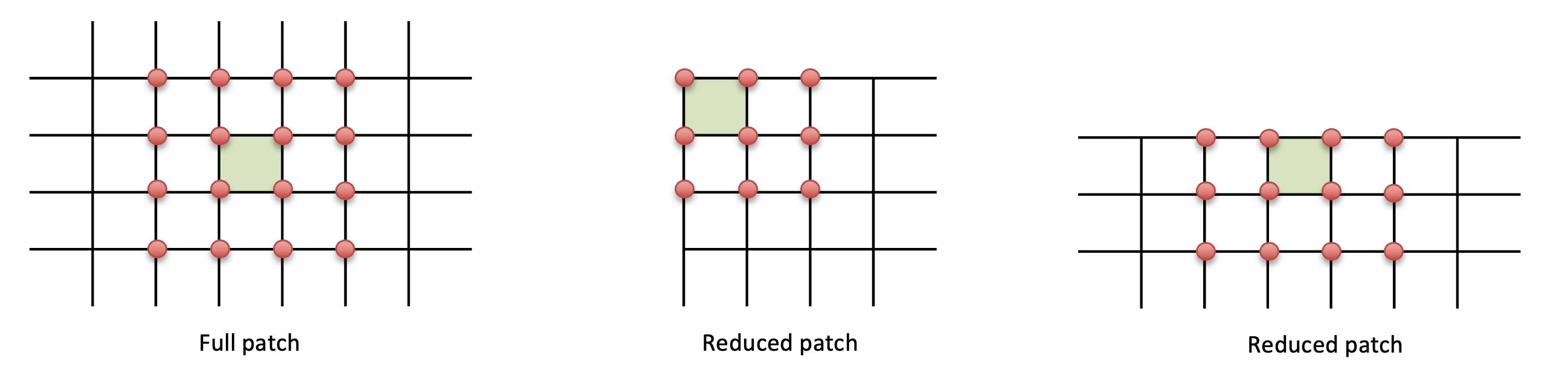}}
\caption{Reduced patches for elements on the boundary with comparison to the full patch for interior elements (left) with $s=1$.}
\label{fig:reducedpatch}
\end{figure}
    \item \textit{Numerical integration}: {The numerical integration can be done using a simple discrete summation with uniformly distributed quadrature points and equal weights \cite{pierre2016unstructured}.} It should be noted that the integration domain is still the domain of element $\Omega^e= [-1,1]\times[-1,1]$, regardless of the patch size. {{In the numerical examples, the number of quadrature points in each direction is the same as the element size, unless otherwise stated.} It seems good enough for an accurate integration in our examples.}
\end{itemize}

{It should be noted that the objective function or correlation criteria \eqref{eq:lossfunction} used in DIC can be modified to include a regularization (a weighted higher order penalty) for improving the conditioning of the final discrete matrix. This is not considered in our work as the choice of penalty requires further experience and can lead to potential biases, as explained in \cite{yang2019augmented}.}

\subsection{Smoothed grayscale interpolation}
The discrete formula \eqref{eq:lossfunction_linear_discrete} of C-FE-DIC needs the gradient information of the grayscale images, i.e., $\nabla f$. The original grayscale images only contain the discrete  grayscale value of each pixel. Therefore, a smoothing technique is needed to compute the gradient of the grayscale. We present two techniques in this work.

\subsubsection{Convolution shape function based interpolation}
As mentioned earlier, the C-FE shape functions offer an arbitrarily high order interpolation for the displacement field. It is convenient to use the same shape function for the grayscale interpolation. Assuming the grayscale value of each grid point in the discretized ZoI is known, the following interpolation can be used to construct a smooth grayscale function
\begin{equation}
\label{eq:C-FEM-grayscale}
    f(\boldsymbol{\xi})=\sum_{i\in A^e}{N}_{{i}}(\boldsymbol{\xi})\sum_{j\in A^i_s} {W}^{\boldsymbol{\xi}_i}_{a,j} (\boldsymbol{\xi}) f_j=\sum_{k\in A^e_s} \tilde{{N}}_k(\boldsymbol{\xi}) f_k
\end{equation}
where $f_k$ denotes the grayscale values on the grid points. Similarly for the deformed image
\begin{equation}
\label{eq:C-FEM-grayscale_g}
    g(\boldsymbol{\xi})=\sum_{k\in A^e_s} \tilde{{N}}_k(\boldsymbol{\xi}) g_k
\end{equation}
The gradient of $f$ reads then 
\begin{equation}
\displaystyle
\label{eq:C-FEM-grayscale-gradient}
    \nabla f  =\begin{pmatrix}
    \frac{\partial f}{\partial x} \\
       \frac{\partial f}{\partial y} 
    \end{pmatrix}=\begin{pmatrix}
    \frac{\partial f}{\partial \xi} \frac{\partial \xi}{\partial x} +\frac{\partial f}{\partial \eta} \frac{\partial \eta}{\partial x} \\
      \frac{\partial f}{\partial \xi} \frac{\partial \xi}{\partial y} +\frac{\partial f}{\partial \eta} \frac{\partial \eta}{\partial y} 
    \end{pmatrix}= \frac{\partial f}{\partial \boldsymbol{\xi}}\frac{\partial \boldsymbol{\xi}}{\partial \boldsymbol{x}}\\=\sum_{k\in A^e_s} \frac{\partial \tilde{{N}}_k}{\partial \boldsymbol{\xi}} \boldsymbol{J}^{-1} f_k 
\end{equation}
where $\boldsymbol{J}$ is the Jacobian matrix. 

{Eq.} \eqref{eq:C-FEM-grayscale}-\eqref{eq:C-FEM-grayscale-gradient} can be used to evaluate the sub-element grayscale and the gradient for computing $\boldsymbol{K}$ and $\boldsymbol{Q}$ in the discretized DIC formulation \eqref{eq:DIC-discrete}. Since the convolution shape function considers the contributions of surrounding nodes with the controlled patch size $s$ and dilation $a$, the approximated grayscale and gradient can be seen as the results of a built-in "filter" related to $s$ and $a$. This feature could be advantageous for noisy images.  We will investigate more on this point in our future work.

\subsubsection{Spline interpolation with separation of variables}
\label{sect:spline}
An alternative way to compute the sub-element grayscale and the gradient is the spline interpolation. In order to simplify the interpolation and reduce the polynomial order, we propose to use the separation of variables so that the 2D interpolation can be done using the products of 1D interpolations. The separation of variables for the grayscale images can be written as 
\begin{equation}
\label{eq:Sep-grayscale}
    f(\boldsymbol{x})=\sum_{m=1}^{M_f} f_x^{(m)}(x)f_y^{(m)}(y)
\end{equation}
and 
\begin{equation}
\label{eq:Sep-grayscale_g}
    g(\boldsymbol{x})=\sum_{m=1}^{M_g} g_x^{(m)}(x)g_y^{(m)}(y)
\end{equation}
where $M_f, M_g$ are the number of modes for $f$ and $g$. $f_x$ and $f_y$ are the separated 1D functions and can be approximated by spline functions of given order. This separation of variables corresponds to a matrix decomposition for the discrete grayscale matrix of an image, and it can be done by the singular value decomposition (SVD) or the high order proper generalized decomposition (HOPGD) \cite{lu2018multi,lu2018adaptive}. {Our} experience showed that the separation of variables can greatly reduce the complexity of the polynomials in the approximation. In theory, the grayscale images can be accurately reproduced if the number of modes is large enough \cite{eckart1936approximation}.

With the separation of variables and the spline interpolation, the gradient of $f$ can be obtained by 
\begin{equation}
\displaystyle
\label{eq:Sep-grayscale-gradient}
    \nabla f=[\sum_{m=1}^{M_f} \frac{\partial f_x^{(m)}}{\partial x} f_y^{(m)} \quad \sum_{m=1}^{M_f} f_x^{(m)}\frac{\partial f_y^{(m)}}{\partial y} ]^T
\end{equation}

We remark that the convolution shape function can also be combined with the separation of variables for the grayscale interpolation. In order to make a fair comparison between proposed C-FE based DIC and the usual FE based DIC, we chose the spline interpolation with the separation of variables for the numerical examples in section \ref{sec:example}. 

\subsection{Discussion on the potential  adaptivity}
Like FEM, the C-FE based DIC can use adaptive approximations for different regions of the ZoI. The adaptive approximation is expected to improve the accuracy of solutions while saving the computational cost. In addition to the potential $r$- and $h$- adaptivity, the convolution approximation offers following unique features.

\begin{itemize}
    \item \textit{$p$-adaptivity}: One of the big advantages of the convolution approximation is the arbitrary order of approximation without modifying the mesh. This can lead to a flexible local/global $p$-adaptivity, i.e., using different orders of approximations for different elements or regions. As shown in \figurename~\ref{fig:padaptivity}, C-FEM allows a flexible choice of approximation order $p$ for an element without modifying the mesh, whereas more nodes have to be added in the FEM to keep a conformal mesh for the $p$-adaptivity.
\begin{figure}[htbp]
\centering 
{\includegraphics[scale=0.35]{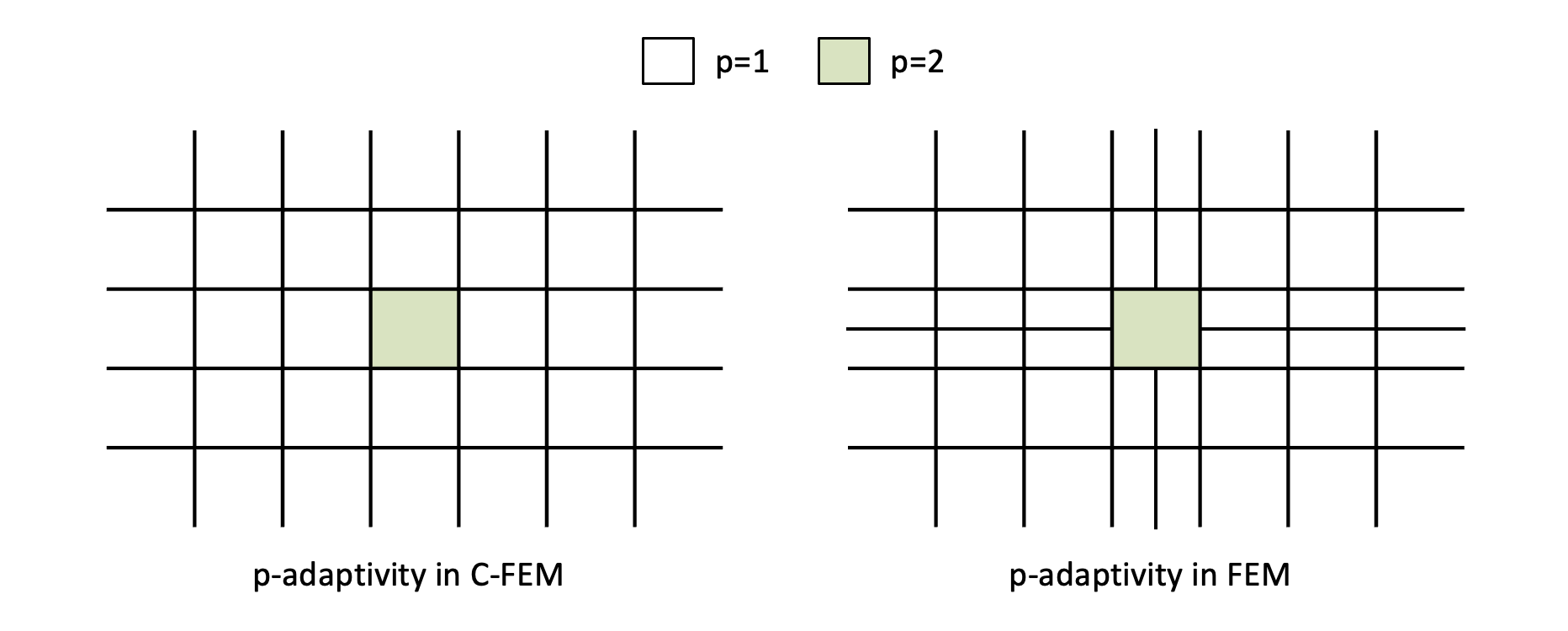}}
\caption{$p$-adaptivity and corresponding mesh in C-FEM and standard FEM.}
\label{fig:padaptivity}
\end{figure}

    \item \textit{$s$-adaptivity}: The patch size $s$ of each element can also be adaptively chosen. As a uniform large $s$ might lead to a large bandwidth in the final discrete matrix $\boldsymbol{K}$, the adaptive selection of $s$ can help minimize the bandwidth and the resulting computational cost. In general, small $s$ is preferred and it should provide a large enough support to construct the desired order of approximation.
    \item \textit{$a$-adaptivity}: The dilation $a$ is controlling the actual window size of the approximation and therefore the approximation length scale. In general, the dilation $a$ should be large enough to get smooth results. But it can be adaptively chosen, together with adaptive $s$, to provide an adaptive length scale filtering in the DIC. 
    
\end{itemize}
The  adaptivity can be done through different ways. For example, it can be done by predefining the adaptively chosen parameters ($p$, $s$, and $a$) based on  prior knowledge and experience. Another way is to use an optimization scheme and automatically choose the parameters while minimizing the loss function $\mathcal{L}$ in Eq. \eqref{eq:lossfunction}. Since the adaptivity is not the focus of this work, we do not consider the adaptive approximation in the numerical examples. The  parameters are chosen as $p=2$, $s=2$, $a=8$, {unless otherwise stated}, for all convolution elements in the next section, and it seems these parameters can provide overall good performance for the C-FE based DIC.

\section{Numerical experiments}
\label{sec:example}
The performance of the C-FE based DIC is tested using four examples. The first two examples are using synthetic speckle images generated by predefined displacement fields. Quantitative errors using the standard root mean square error (RMSE) and the $L^2$-norm error will be computed and compared with FE based DIC results. The last two examples are from the benchmark problems of DIC Challenge 2.0: star1 and star2 \cite{reu2022dic}. {As suggested in \cite{reu2022dic}, the so-called spatial resolution and  metrological efficiency indicator will be used to evaluate the performance of the proposed DIC method on the star examples.} 

\subsection{Example 1: synthetic image with large spots}\label{sec:example1}
\begin{figure}[htbp]
\centering
\subfigure[Reference image]{\includegraphics[scale=0.27]{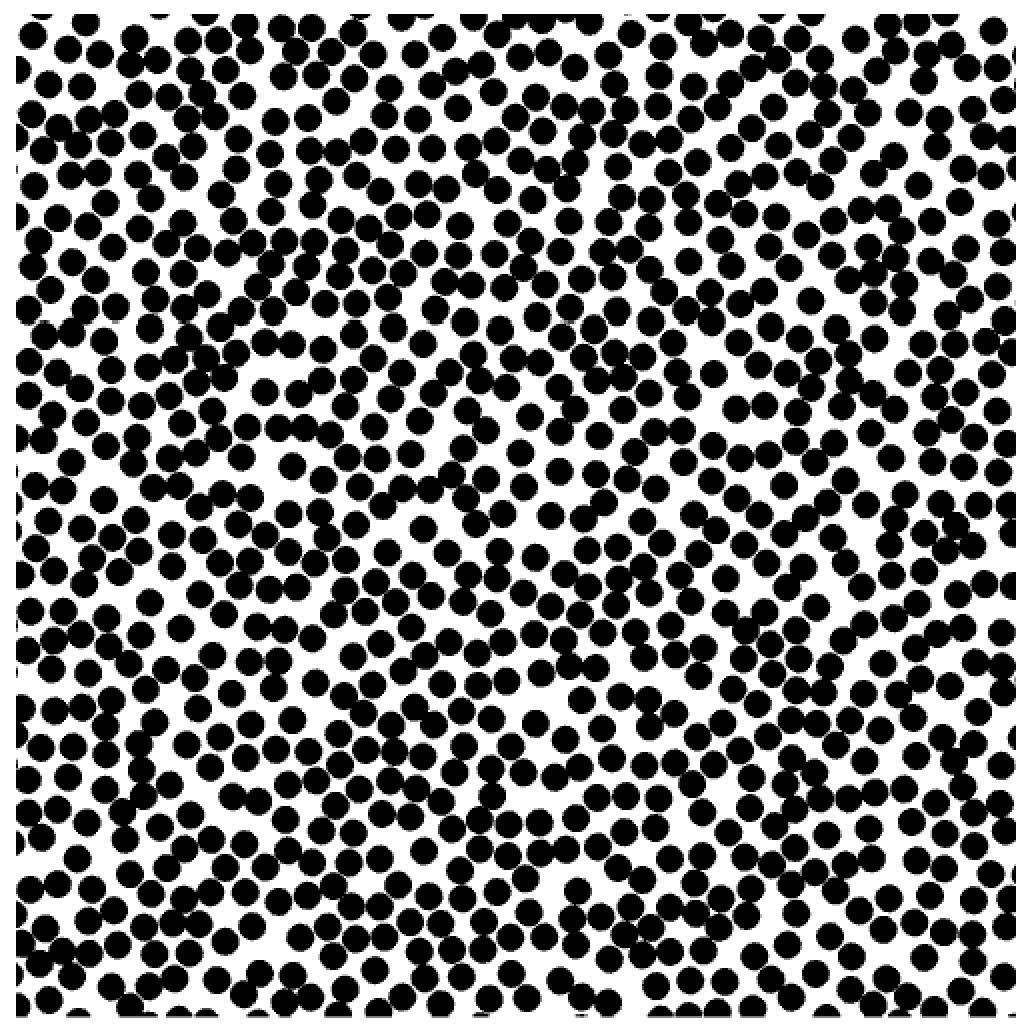}}\quad\quad
\subfigure[Deformed image ]{\includegraphics[scale=0.27]{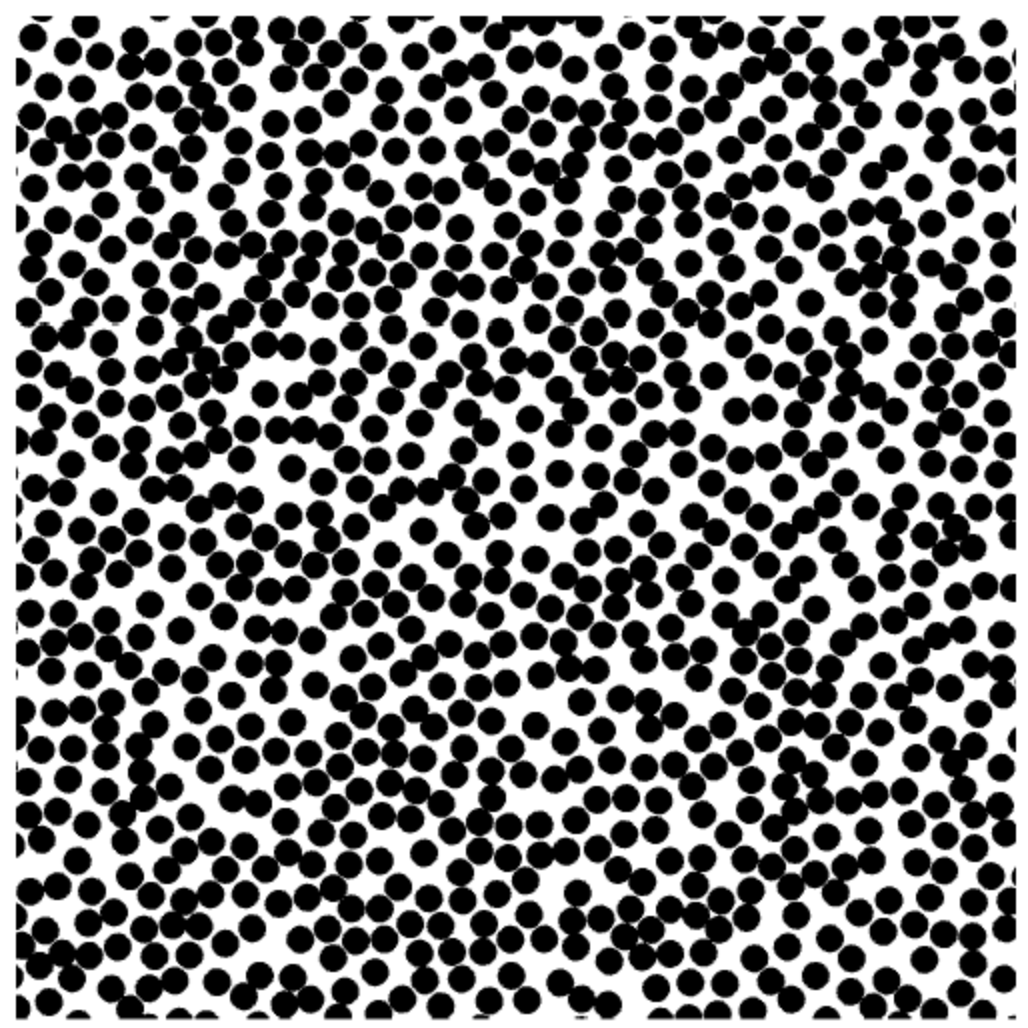}}\quad\\
\subfigure[Ground truth: $u$ ]{\includegraphics[scale=0.15]{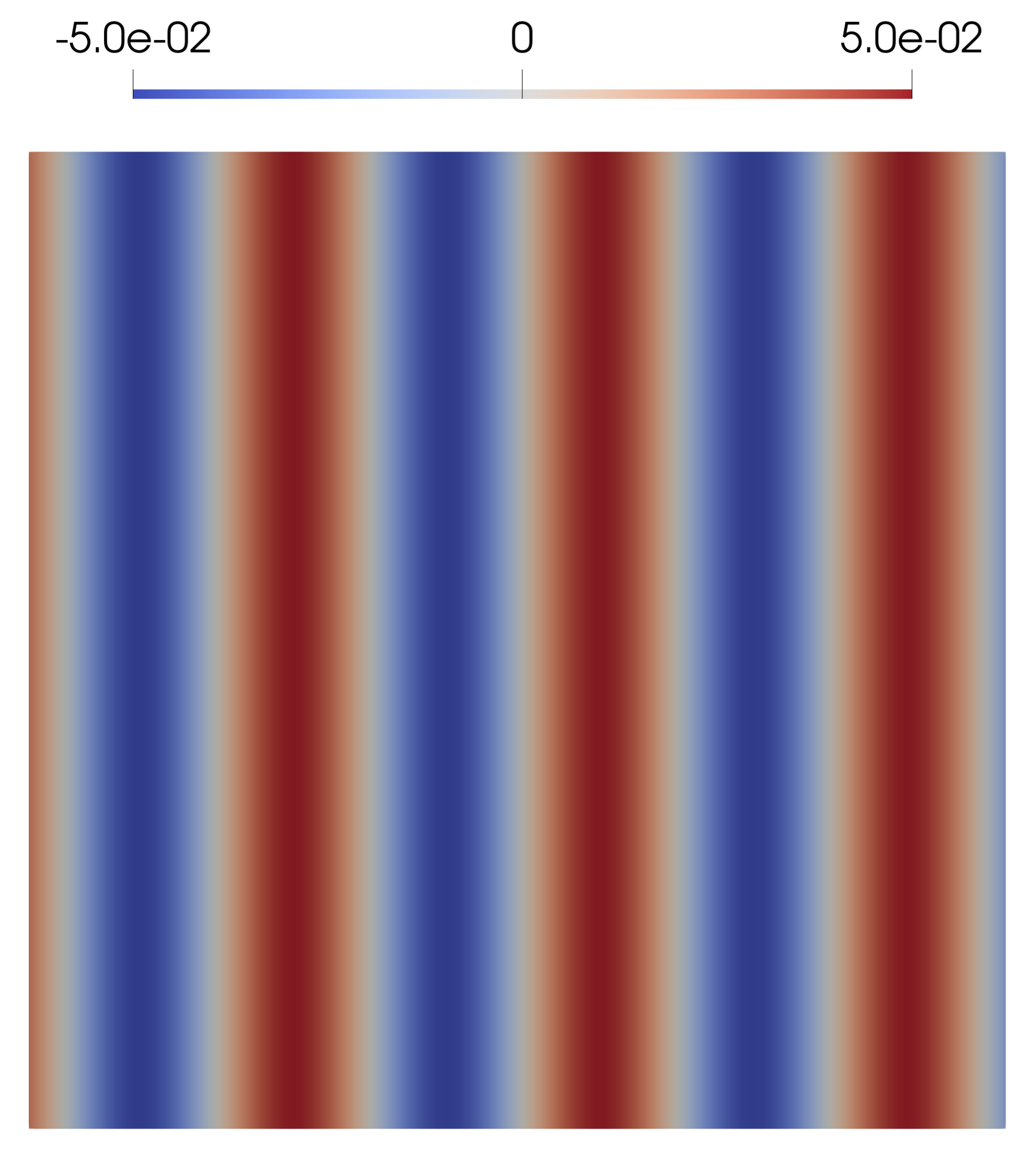}}\quad\quad
\subfigure[C-FE: $u$]{\includegraphics[scale=0.15]{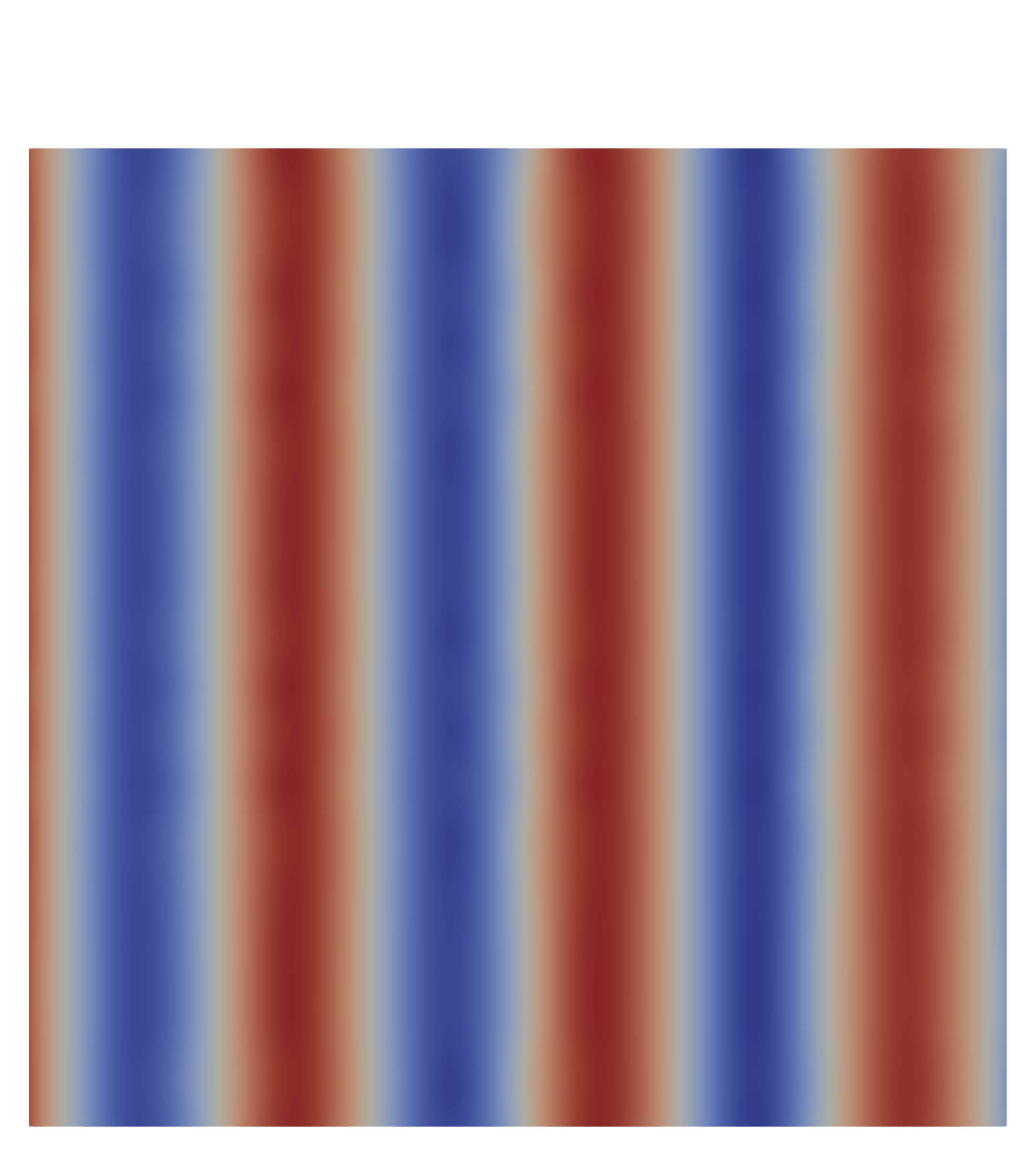}}\quad\quad\\
\subfigure[FE-Q4: $u$]{\includegraphics[scale=0.15]{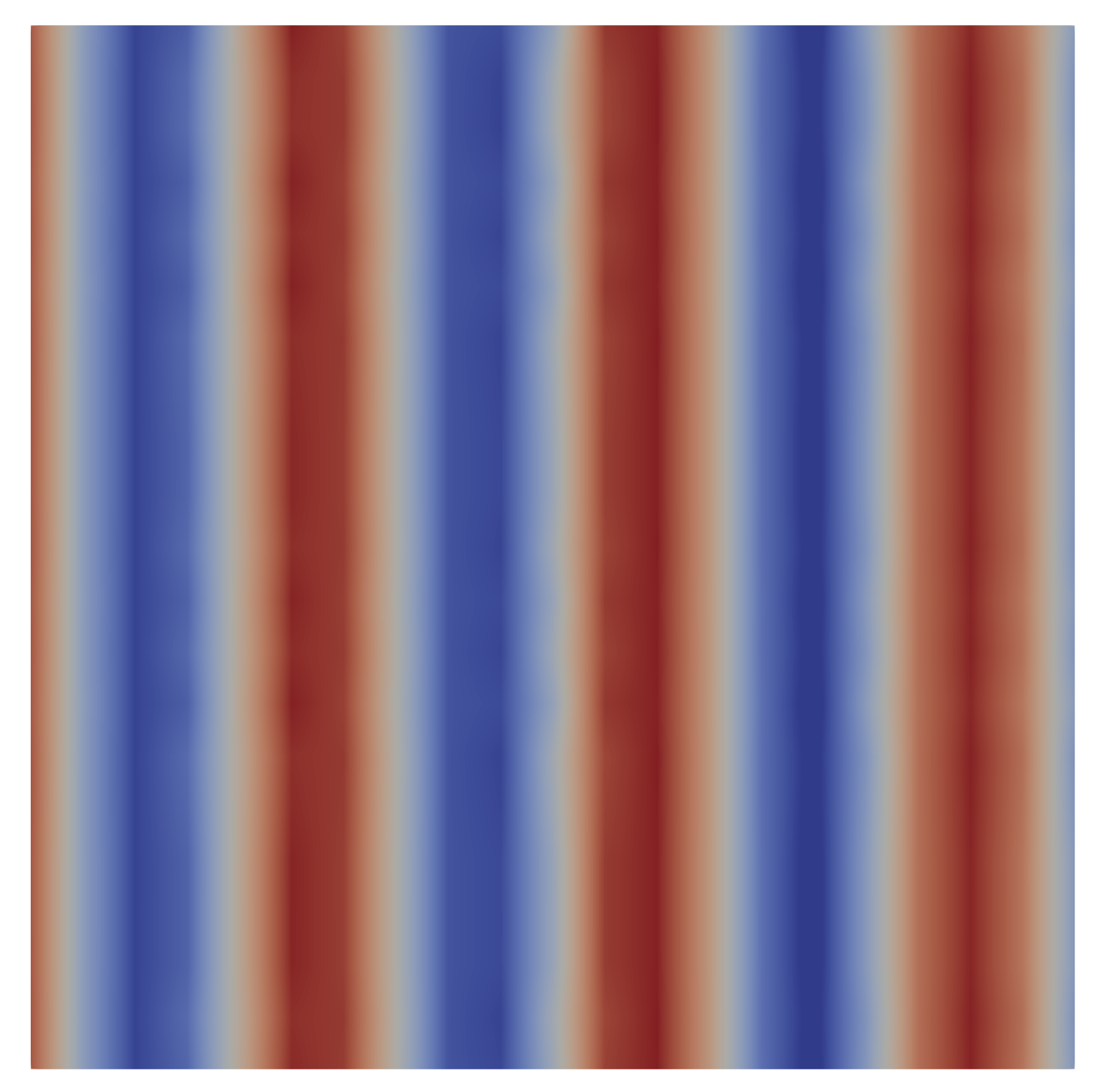}}\quad\quad
\subfigure[FE-Q8: $u$ ]{\includegraphics[scale=0.075]{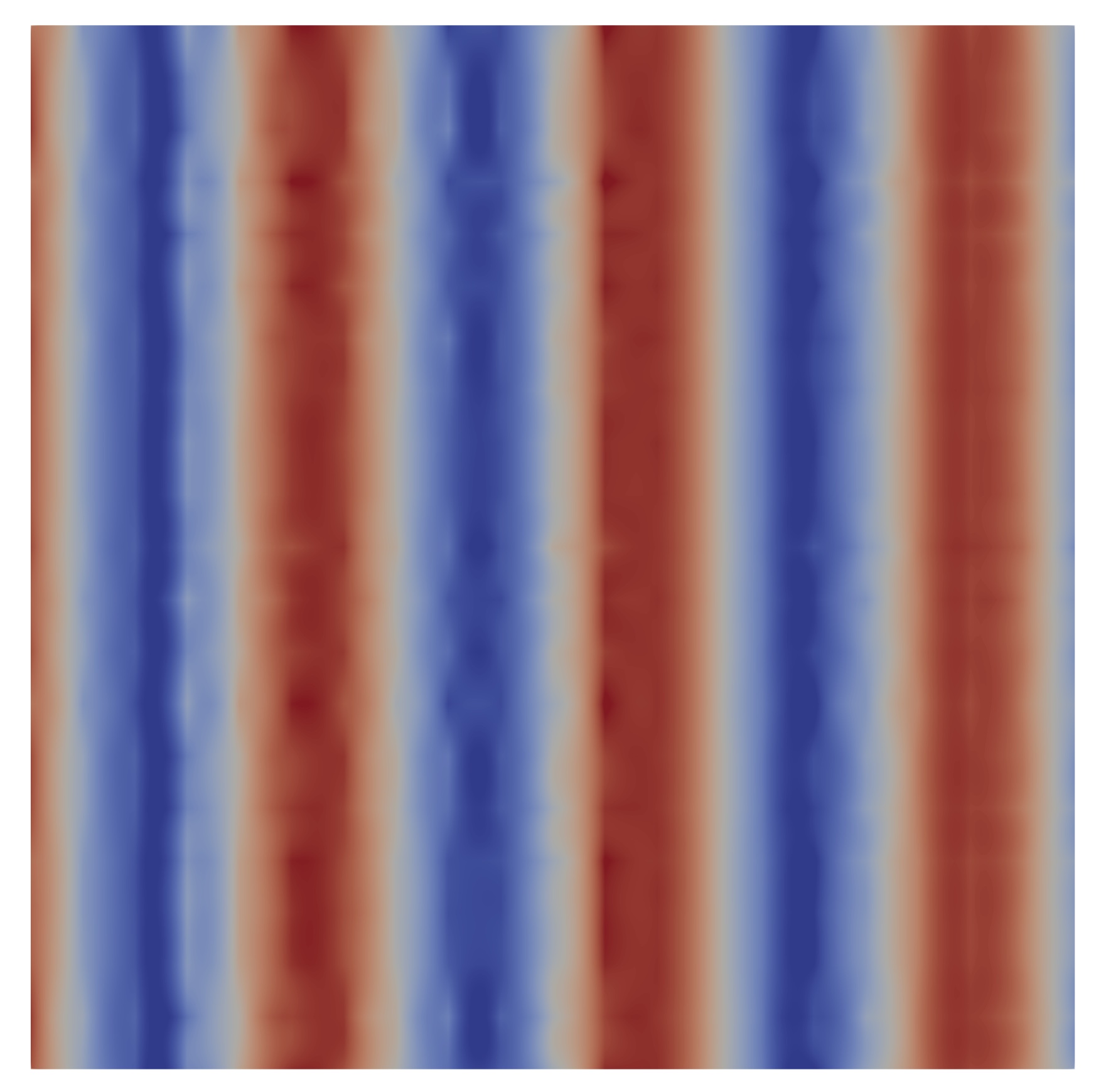}}
\caption{DIC results of the first example. The element size $h=20$ (pixels) in all solutions. FE-Q4 corresponds to the linear 4-node element, FE-Q8 corresponds to the quadratic 8-node element. }
\label{fig:exam1ux}
\end{figure}

This example evaluates the performance of the proposed DIC method on images with relatively large speckle patterns. The initial reference image with randomly distributed spots is generated by the code available at \cite{speckleimage} with the image size $500\times500$ (pixels). The following analytical displacement field is used to generate the deformed image. 

\begin{equation}
\displaystyle
\label{eq:ex1_u_ref}
\begin{cases}
     u(\boldsymbol{x})=A\sin wx\\
     v(\boldsymbol{x})=0
\end{cases}
\end{equation}
where $u$ and $v$ are the horizontal and vertical displacement respectively. The parameters of the function are $x\in [0, 500]$ (pixels), $w=0.05$ (/pixels), $A=0.05$ (pixels). \figurename~\ref{fig:exam1ux}(a) and (b) show the generated speckle patterns.  \figurename~\ref{fig:exam1ux}(c) shows the ground truth horizontal displacement for a ZoI of size $400\times 400$ (pixels) located at the center of the speckle images.

For the DIC analysis, a 4-node FE mesh with  the element size  $h=20$ pixels is created. The spline interpolation with separation of variables presented in section \ref{sect:spline} is used to interpolate the sub-element grayscale and compute the gradient. In this example, the 4th order splines is adopted and seems good enough to represent the graycale variations. This treatment is set to be the same for all following DIC results to ensure a fair comparison.

\begin{figure}[htbp]
\centering
{\includegraphics[scale=0.6]{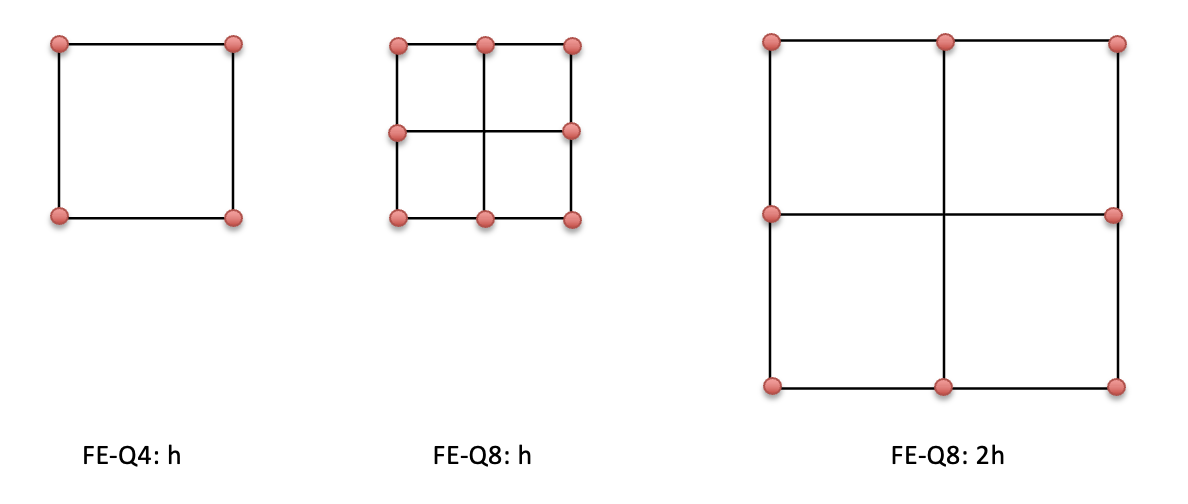}}
\caption{{Support nodes for FE-Q4 and FE-Q8 with different $h$} }
\label{fig:FEQ8}
\end{figure}

\figurename~\ref{fig:exam1ux}(d) shows the displacement result from the proposed C-FE based DIC analysis with the parameters $p=2$, $s=2$, $a=8$ (Same for the other examples).  Compared to the usual 4-node (Q4) FE based DIC result (\figurename~\ref{fig:exam1ux}(e)), the displacement from C-FE is much smoother and agrees well with the ground truth solution. As expected, the C-FE offers a higher order approximation. To further compare with FE based DIC, the quadratic  8-node (Q8) element is also used for the DIC analysis. With the same element size, we obtained the displacement result shown in  \figurename~\ref{fig:exam1ux}(f). The result looks smoother than Q4 FE. However, due to the additional nodes {(i.e., DoFs)} into the FE mesh (see \figurename~\ref{fig:FEQ8}), Q8 elements {tend to capture more small-scale noise or biases, which can} worsen the DIC solution in overall accuracy. {Similar results can be seen in \cite{ma2012mesh}. Hence, it is suggested to use Q8 elements of size $2h$ when comparing with Q4 elements of size $h$ \cite{ma2012mesh}, as they have the same characteristic length scale for filtering the noise or other biases. We have computed the results for coarser Q8 elements, their quantitative errors will be reported later in this subsection.}

\begin{figure}[htbp]
\centering
\subfigure[Ground truth: $\varepsilon_{xx}$ ]{\includegraphics[scale=0.15]{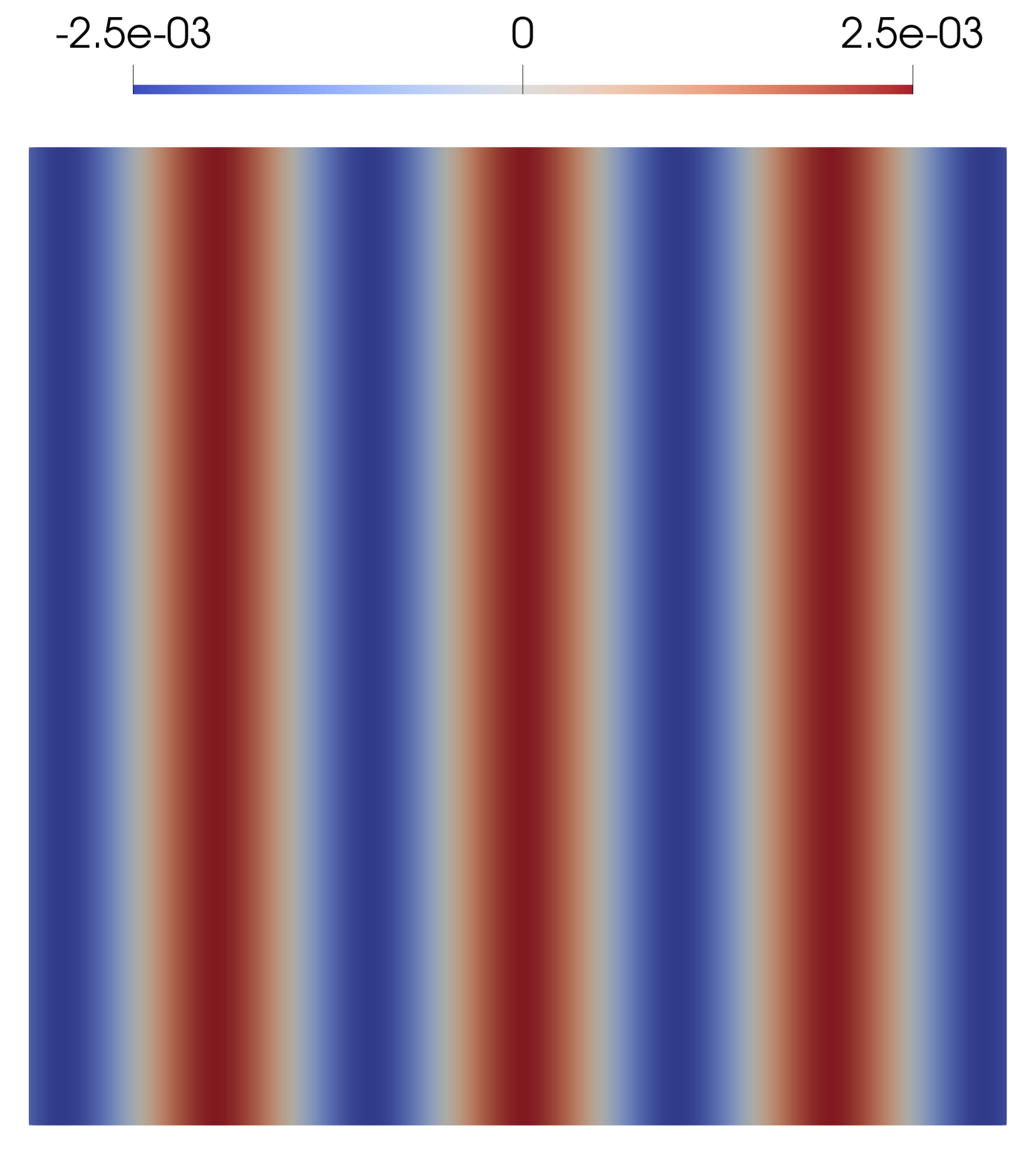}}\quad\quad
\subfigure[C-FE: $\varepsilon_{xx}$]{\includegraphics[scale=0.15]{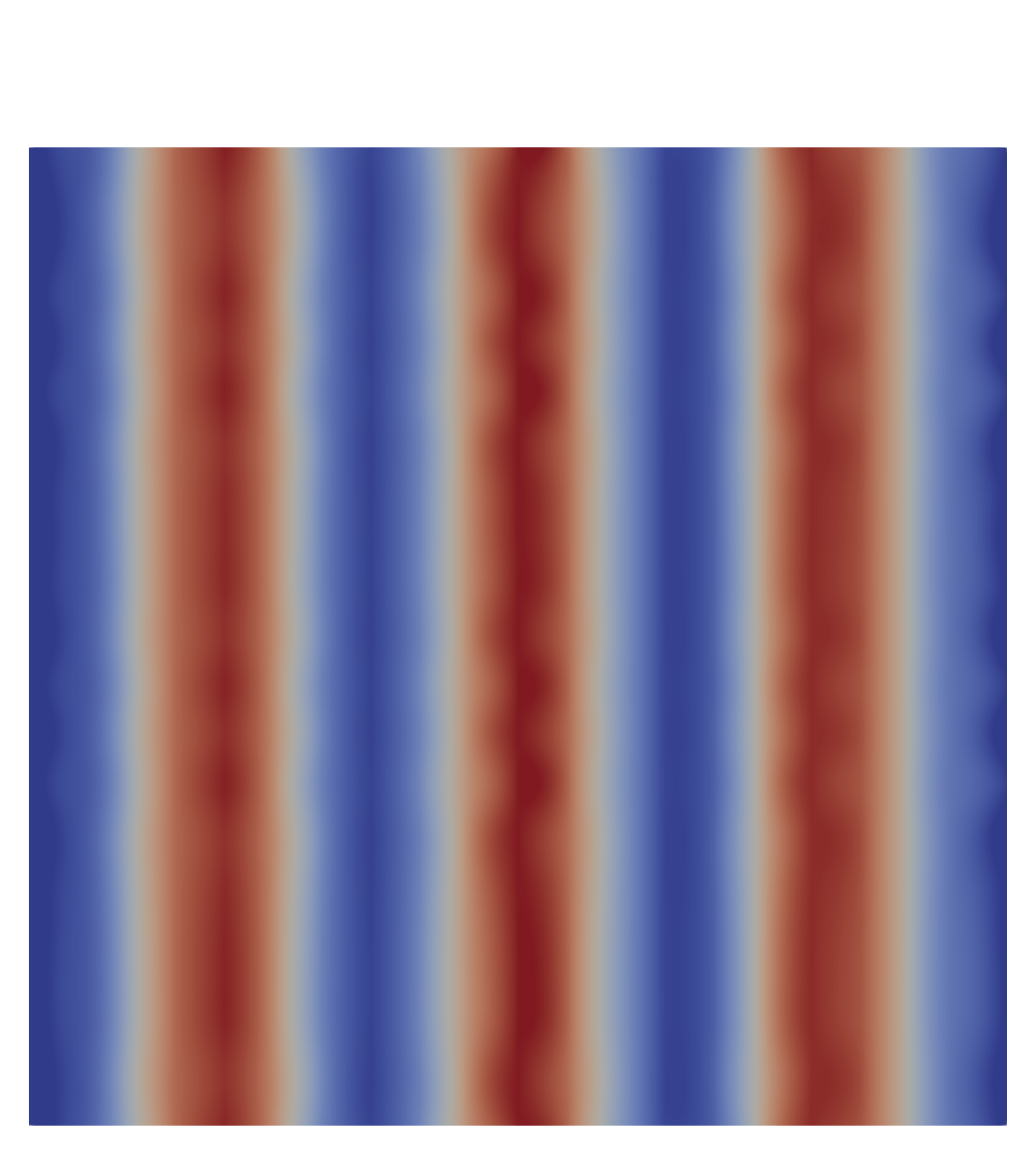}}\quad\quad\\
\subfigure[FE-Q4: $\varepsilon_{xx}$]{\includegraphics[scale=0.15]{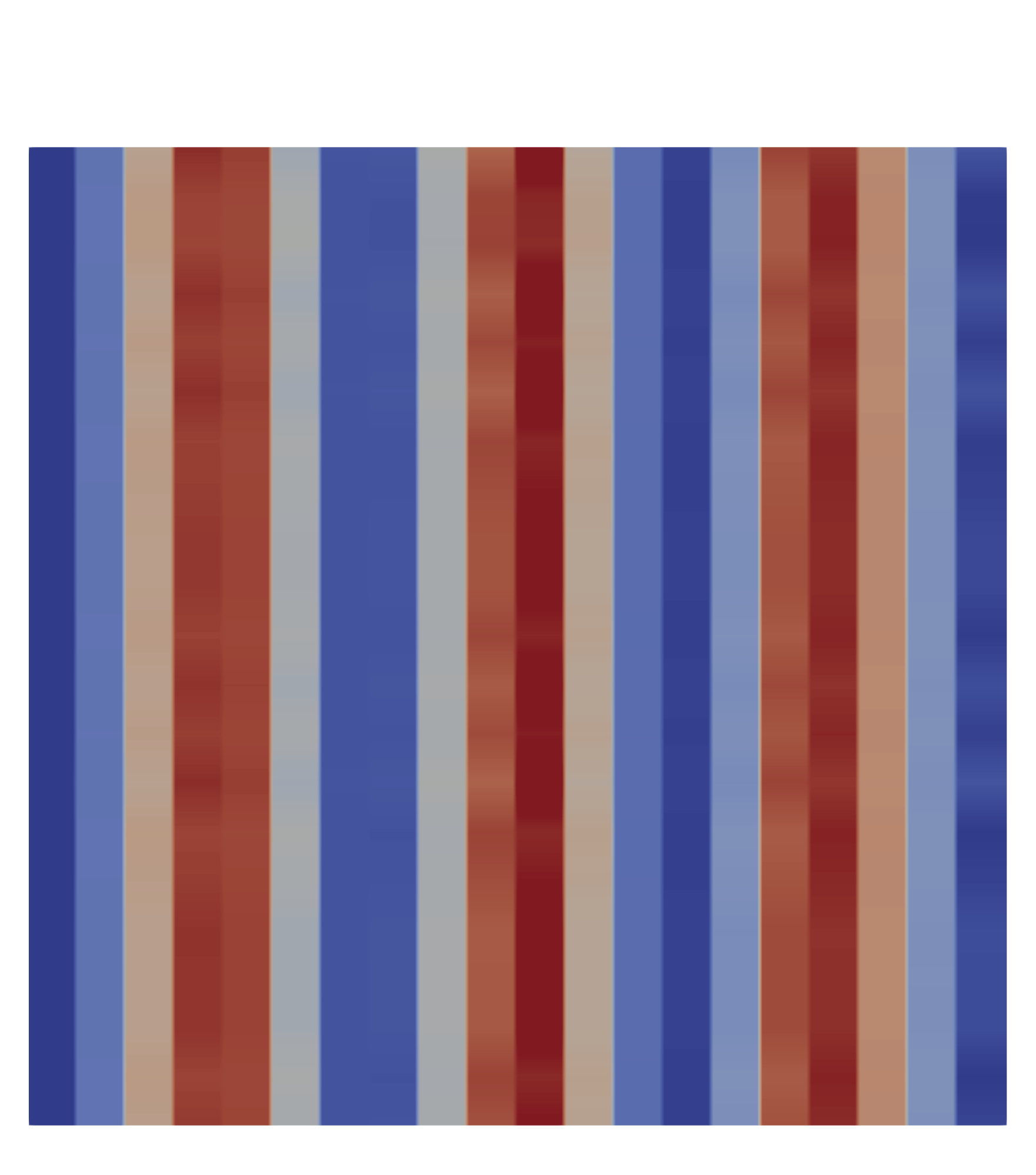}}\quad\quad
\subfigure[FE-Q8: $\varepsilon_{xx}$ ]{\includegraphics[scale=0.075]{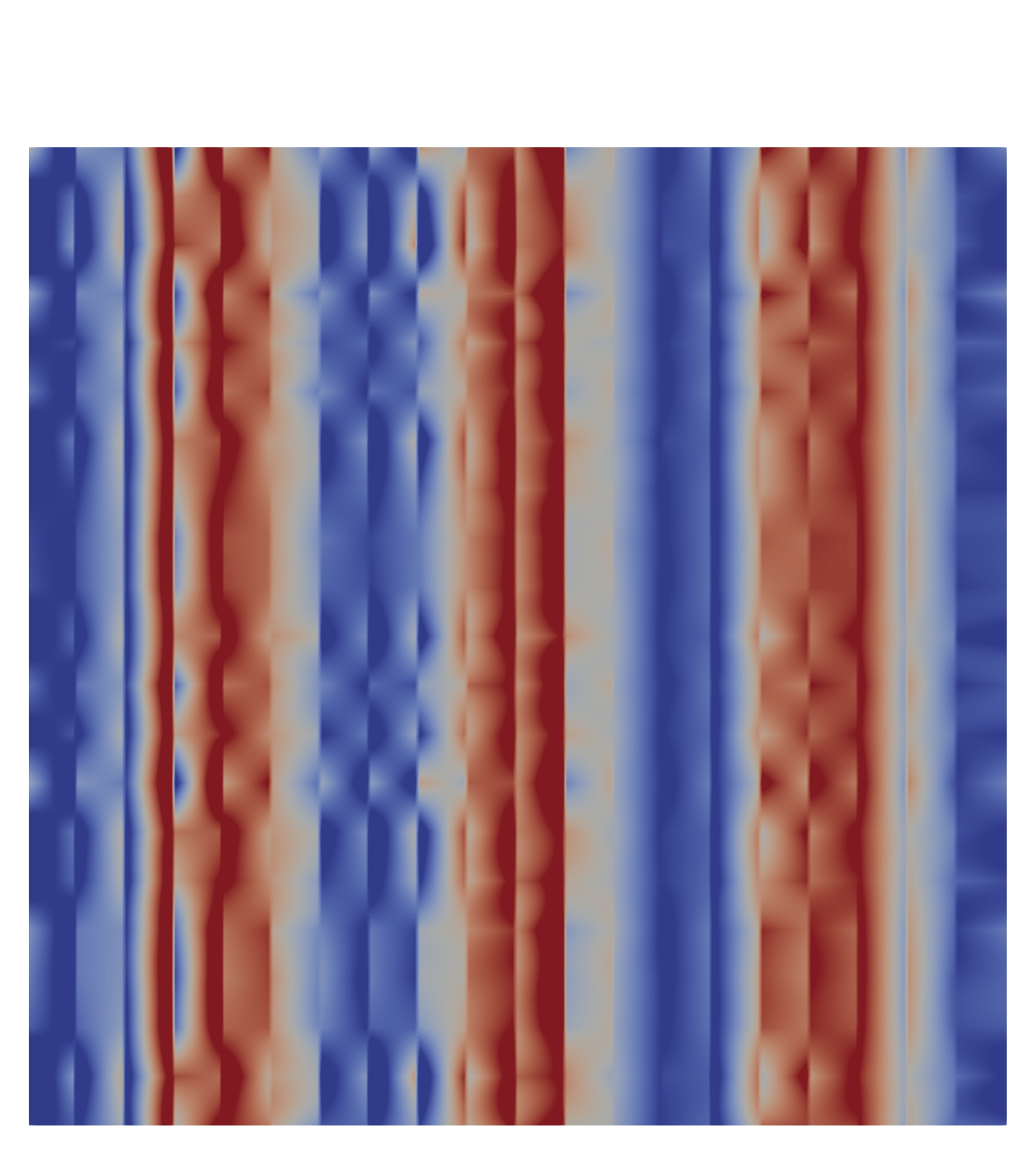}}
\caption{DIC results of the first example for the strain. The element size $h=20$ (pixels) in all solutions. FE-Q4 corresponds to the linear 4-node element, FE-Q8 corresponds to the quadratic 8-node element. }
\label{fig:exam1dux}
\end{figure}

In terms of strain, the C-FE still outperforms the FE including the linear Q4 and quadratic Q8 elements, as shown in \figurename~\ref{fig:exam1dux}. The strain $\varepsilon_{xx}$ derived from the C-FE shows an excellent agreement with the ground truth, whereas FE fails to provide an accurate strain results. 

In order to quantify the accuracy of different methods, we computed the RMSE and $L^2$-norm error as follows
\begin{equation}
\displaystyle
\label{eq:rmse}
     \epsilon_{RMSE}(u)=\sqrt{\sum_{i=1}^{N}(u_i-u^{\text{Ext}}_{i})^2/N}
\end{equation}
where $u^{\text{Ext}}$ denotes the ground truth solution, $N$ is the number of sample points.
\begin{equation}
\displaystyle
\label{eq:L2}
     \epsilon_{L}(u)=\frac{\|u-u^{\text{Ext}}\|_{L^2(\Omega)}}{\|u^{\text{Ext}}\|_{L^2(\Omega)}}=\sqrt{\frac{\int_\Omega (u-u^{\text{Ext}})^2 d\boldsymbol{x}}{{\int_\Omega (u^{\text{Ext}})^2 d\boldsymbol{x}}}}
\end{equation}
where the integration can be done by the Gauss quadrature. Similarly, the error for the strain results can be computed.

The computed errors for both displacement and strain using the two measures are summarized in \tablename~\ref{table:ex1error}. {As mentioned earlier}, the error for Q8 FE results with a coarser element size $h=40$ (pixels) is also computed. The increased element size improves the solution for the quadratic Q8 element. However, it is still less accurate than C-FE. Compared to all other FE results, the C-FE shows a significantly better performance in terms of accuracy, especially for the strain, due to the high order approximation. This confirms the advantages of the proposed C-FE-DIC technique. In addition, we can observe that $L^2$-error seems a better measure for quantifying the error as it takes the relative value to the ground truth solution.

\begin{table}[htbp]
\caption{Error assessment for the DIC results of the first example}
\centering
\begin{tabular}{|c|c|c|c|c|c|c|}
\hline
 Method & Element size  & $\epsilon_{RMSE}(u)$ & $\epsilon_{L}(u)$& $\epsilon_{RMSE}(\varepsilon_{xx})$ & $\epsilon_{L}(\varepsilon_{xx})$\\ \hline
FE-Q4 & 20 &  {0.0032}  & {0.0925} & {$5.07\times 10^{-4}$} &{0.2806} \\ \hline
FE-Q8 & 20  & {0.0038} & {0.1085} & {$6.50\times 10^{-4}$} &{0.3599}  \\ \hline
FE-Q8 & 40 & {0.0032} & {0.0932} & {$3.17\times 10^{-4}$} &{0.1760}\\ \hline
C-FE & 20 & {0.0031} & {0.0886}  & {$2.01\times 10^{-4}$}  &{0.1114}   \\ \hline
\end{tabular}\\
\label{table:ex1error}
\end{table}

\subsection{Example 2: synthetic image with random spots}
\begin{figure}[htbp]
\centering
\subfigure[Reference image]{\includegraphics[scale=0.15]{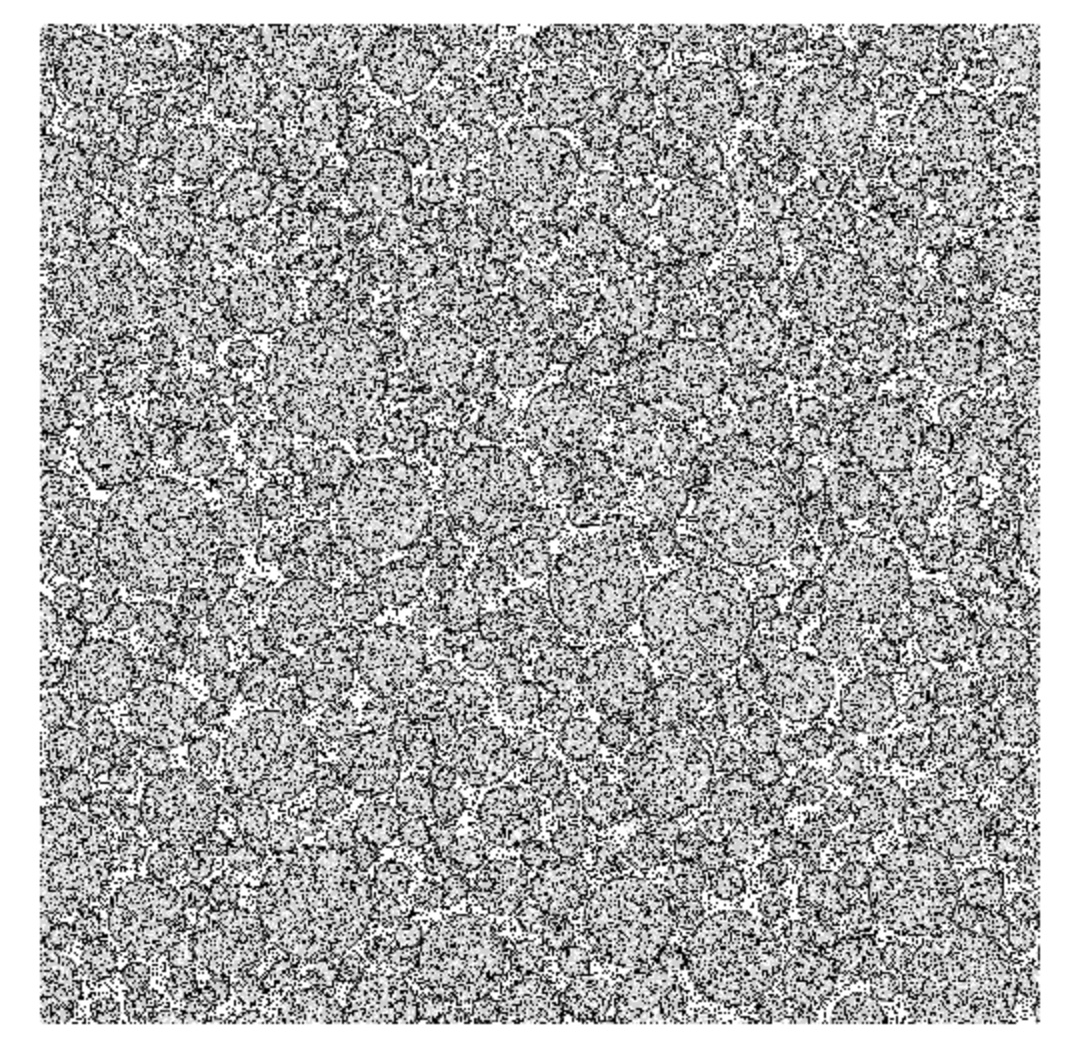}}\quad
\subfigure[Deformed image ]{\includegraphics[scale=0.15]{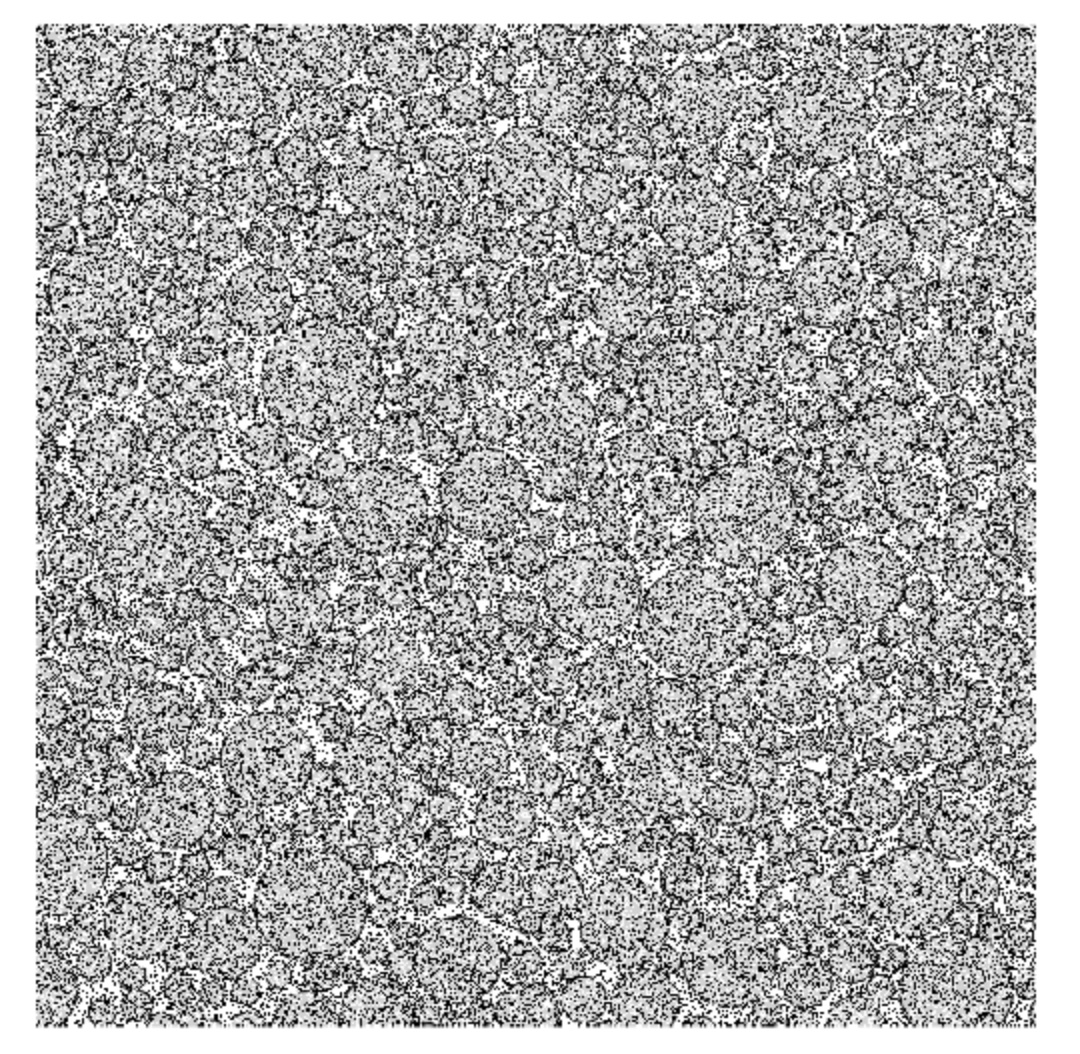}}\quad\\
\subfigure[Ground truth: $u$ ]{\includegraphics[scale=0.15]{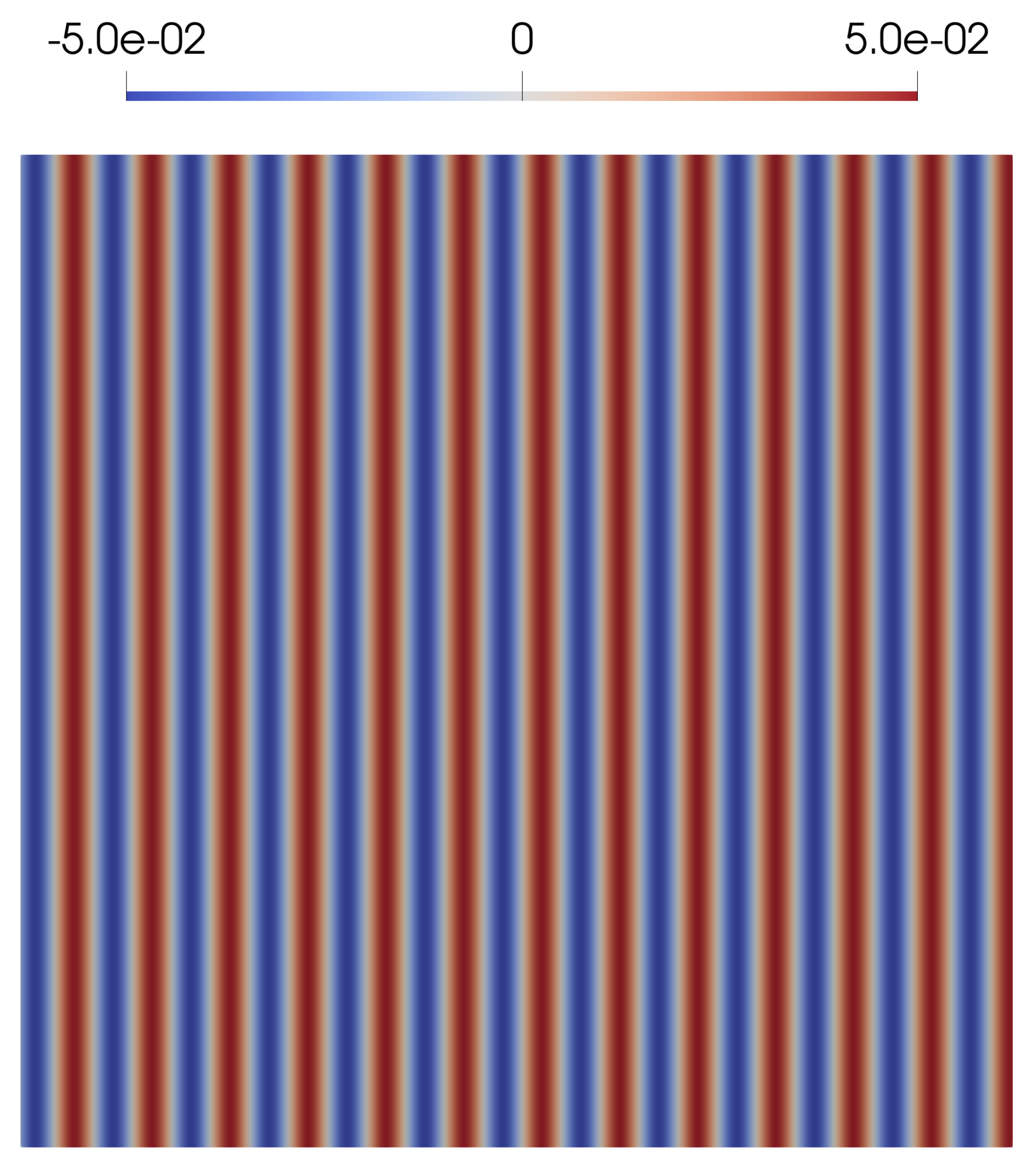}}\quad\quad
\subfigure[C-FE: $u$]{\includegraphics[scale=0.0734]{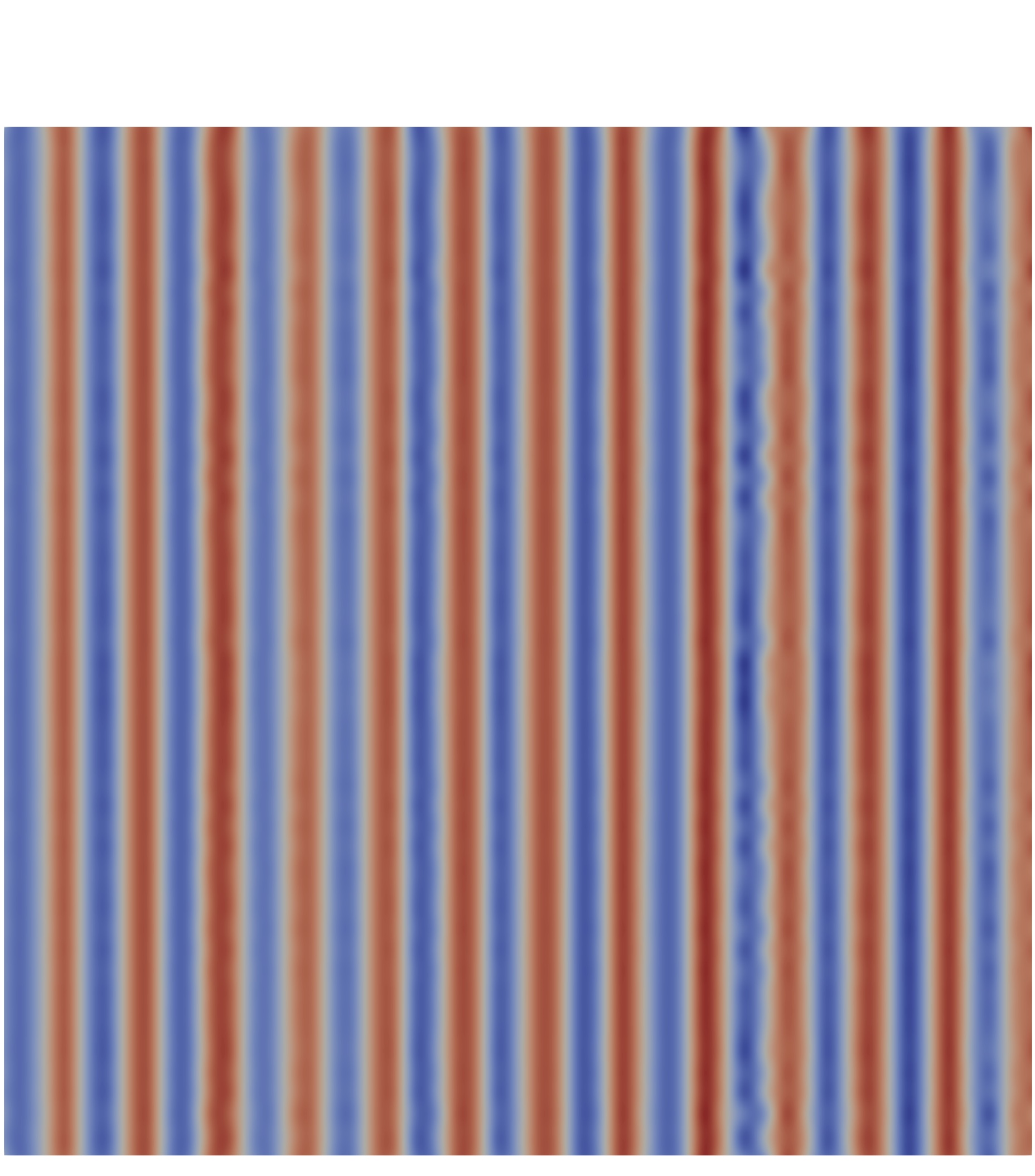}}\quad\quad\\
\subfigure[FE-Q4: $u$]{\includegraphics[scale=0.0734]{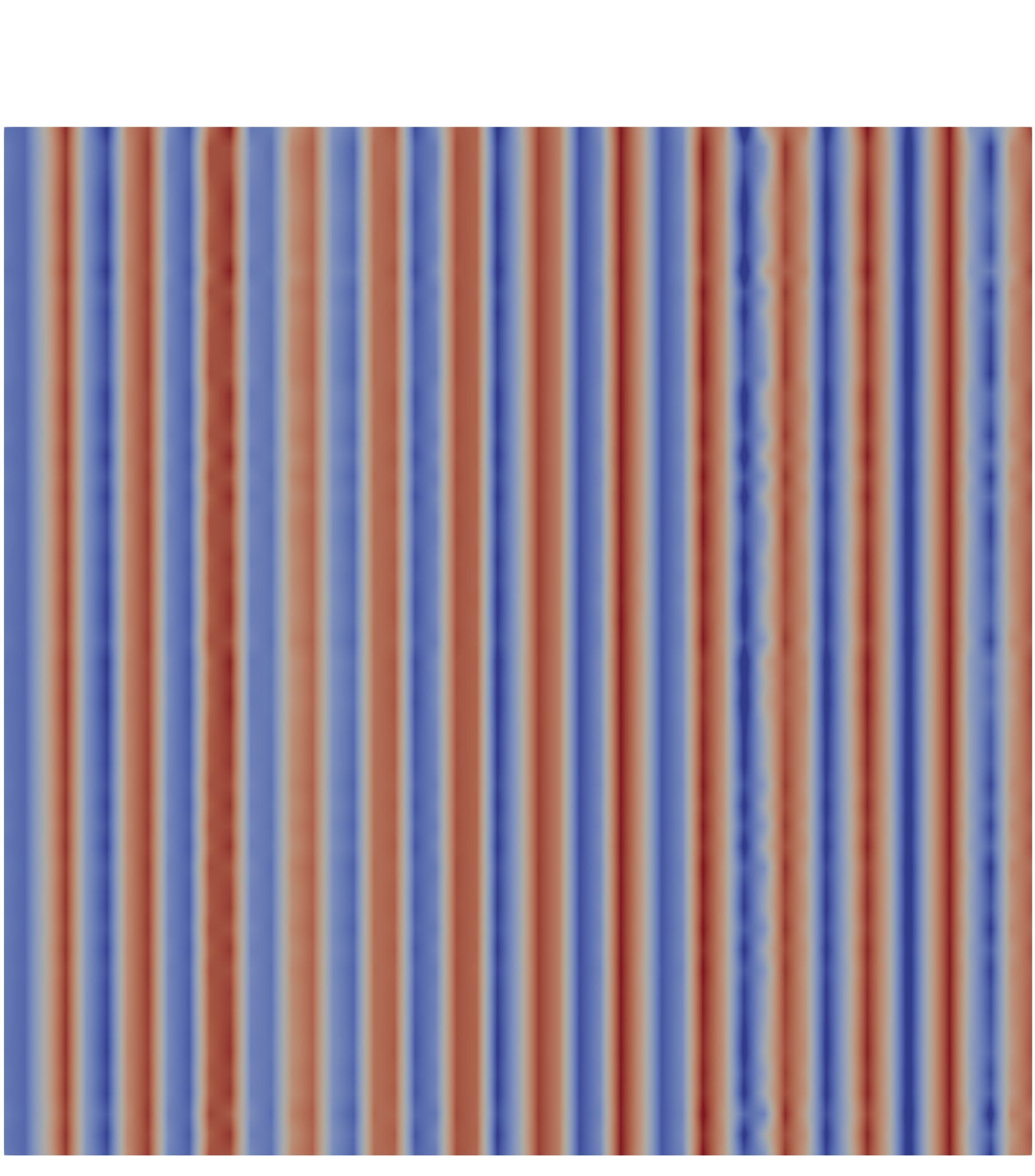}}\quad\quad
\subfigure[FE-Q8: $u$ ]{\includegraphics[scale=0.0745]{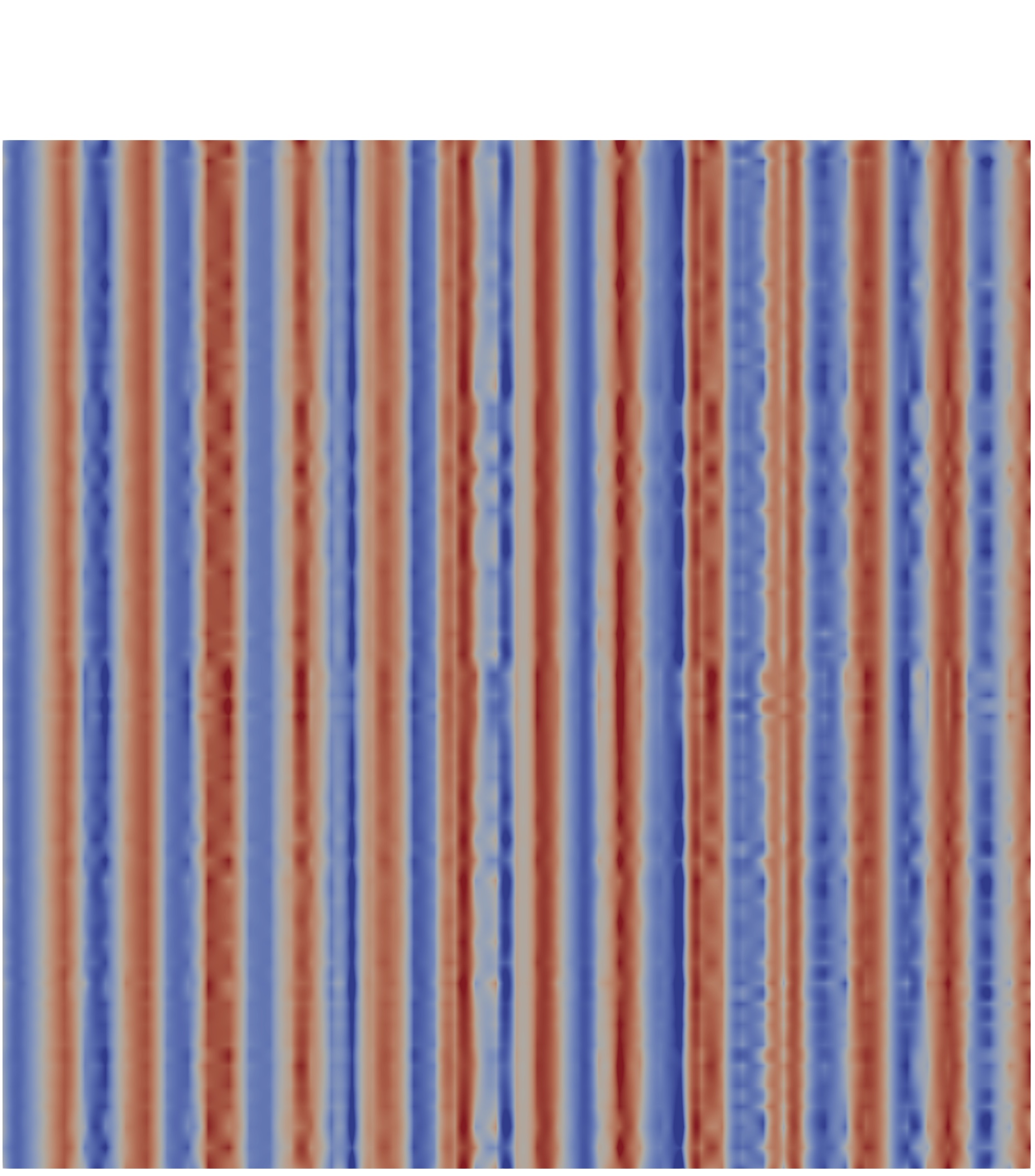}}
\caption{DIC results of the second example. The element size $h=8$ (pixels) in all solutions. FE-Q4 corresponds to the linear 4-node element, FE-Q8 corresponds to the quadratic 8-node element. }
\label{fig:exam2ux}
\end{figure}

The second example exams the DIC method in a more challenging situation on speckle images with random spots. Again, the initial reference image is generated by the code \cite{speckleimage} with random spot size and noise. The image size remains as $500\times500$ (pixels) with the ZoI $400\times400$ (pixels) at the center. The deformed image is generated by the analytical displacement field \eqref{eq:ex1_u_ref} with a modified $w=0.2$ (/pixels). The generated speckle images and the reference ground truth solution for the ZoI are illustrated in \figurename~\ref{fig:exam2ux}(a-c). 

The element size is reduced to $h=8$ (pixels) to capture the variation of solution. Similarly as before, we use spline interpolation for the sub-element grayscale treatment. The DIC results of the displacement field are illustrated in \figurename~\ref{fig:exam2ux}(d-f). Again, the C-FE provides results closer to the ground truth, although the accuracy on the amplitude of the displacement seems degraded due to the quality of the speckle patterns. 

\begin{figure}[htbp]
\centering
\subfigure[Ground truth: $\varepsilon_{xx}$ ]{\includegraphics[scale=0.15]{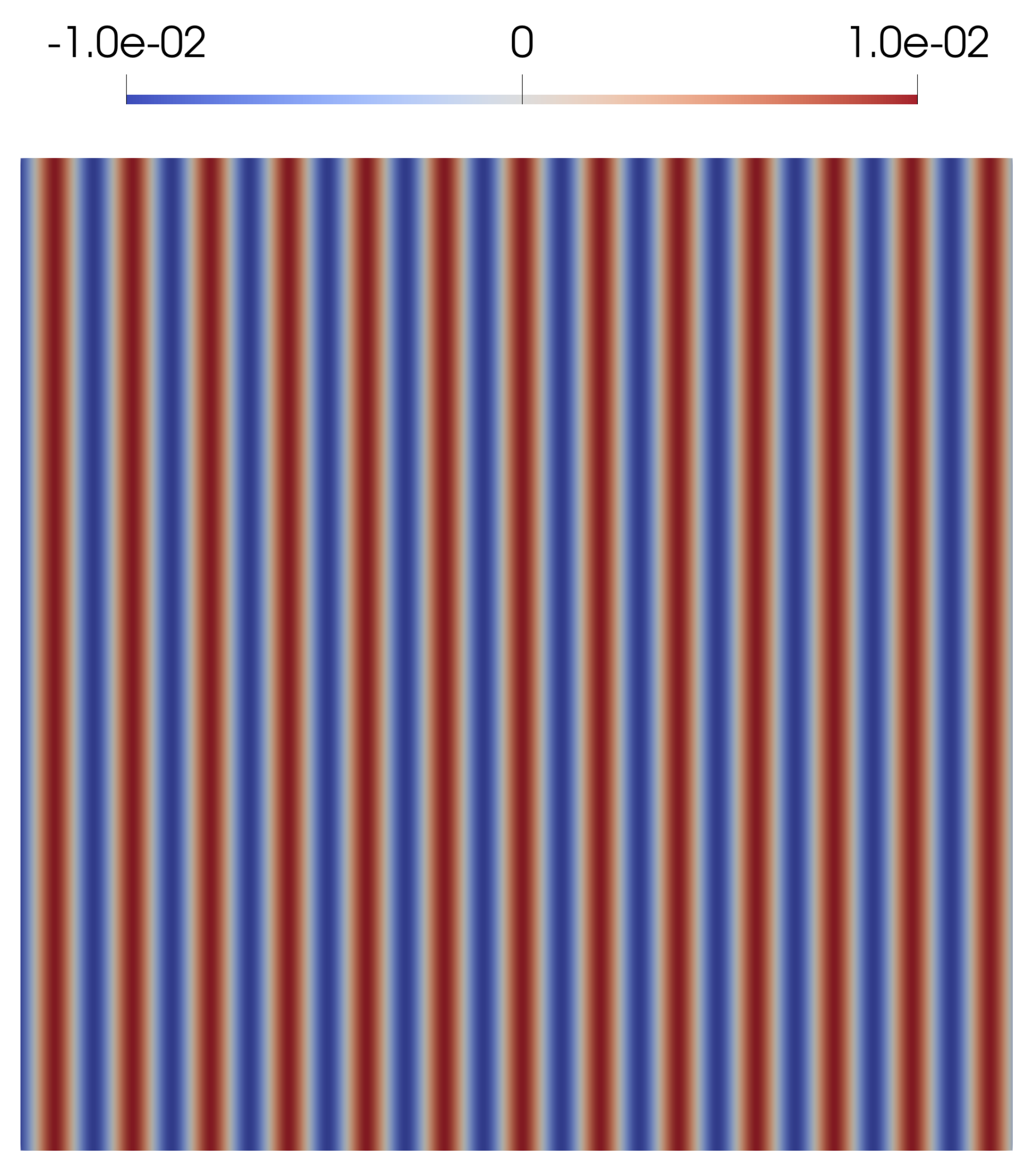}}\quad\quad
\subfigure[C-FE: $\varepsilon_{xx}$]{\includegraphics[scale=0.0734]{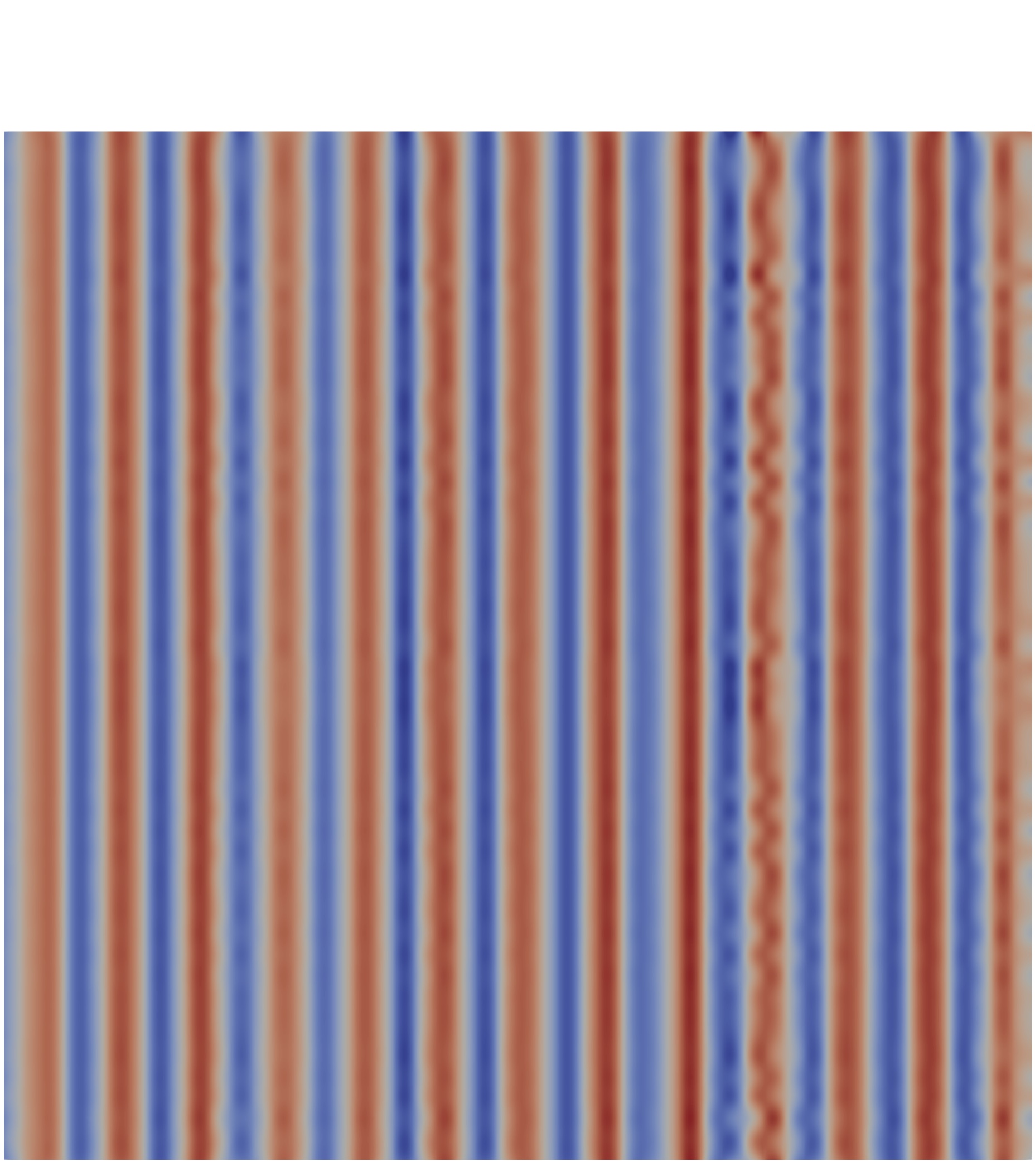}}\quad\quad\\
\subfigure[FE-Q4: $\varepsilon_{xx}$]{\includegraphics[scale=0.146]{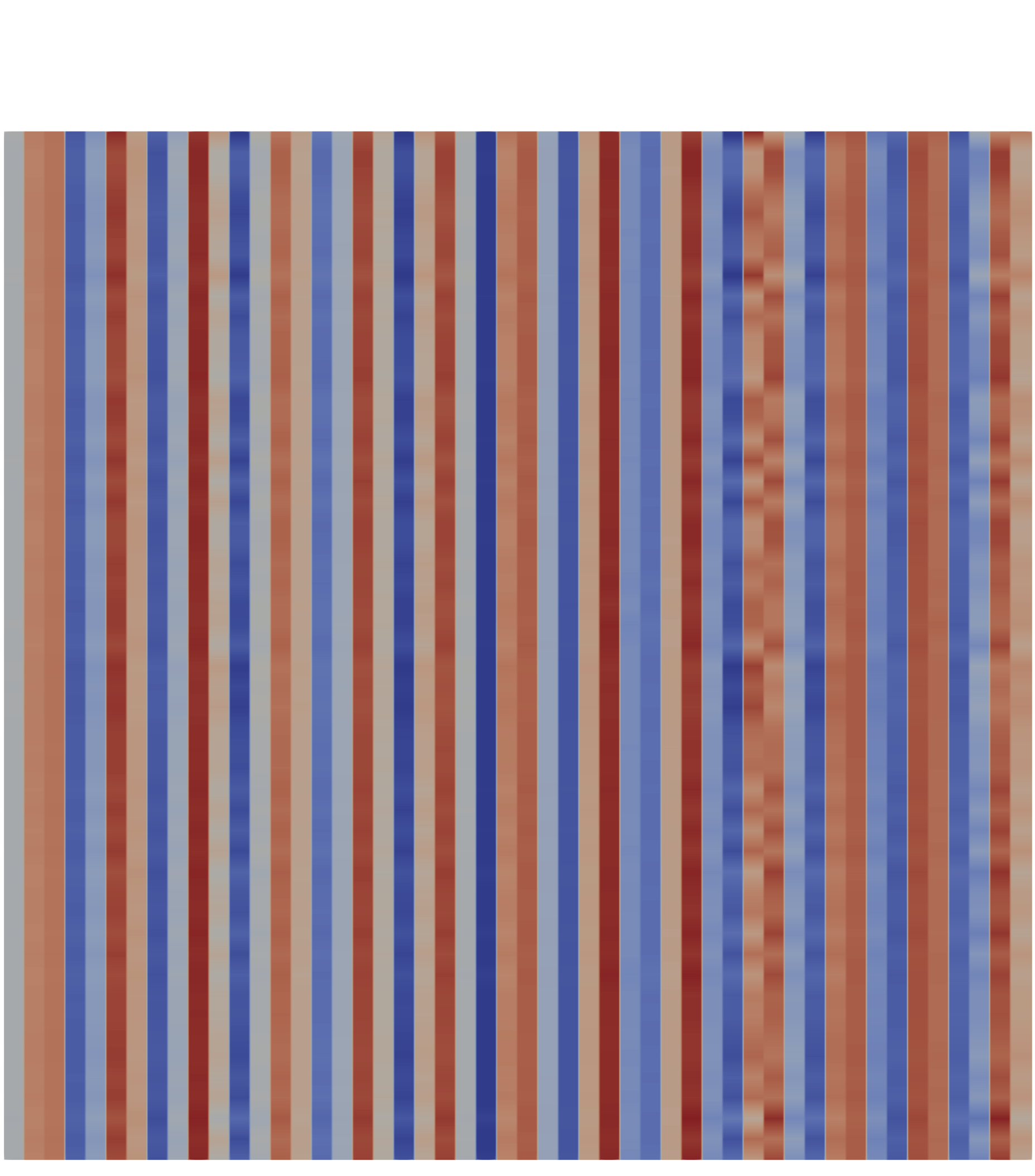}}\quad\quad
\subfigure[FE-Q8: $\varepsilon_{xx}$ ]{\includegraphics[scale=0.0734]{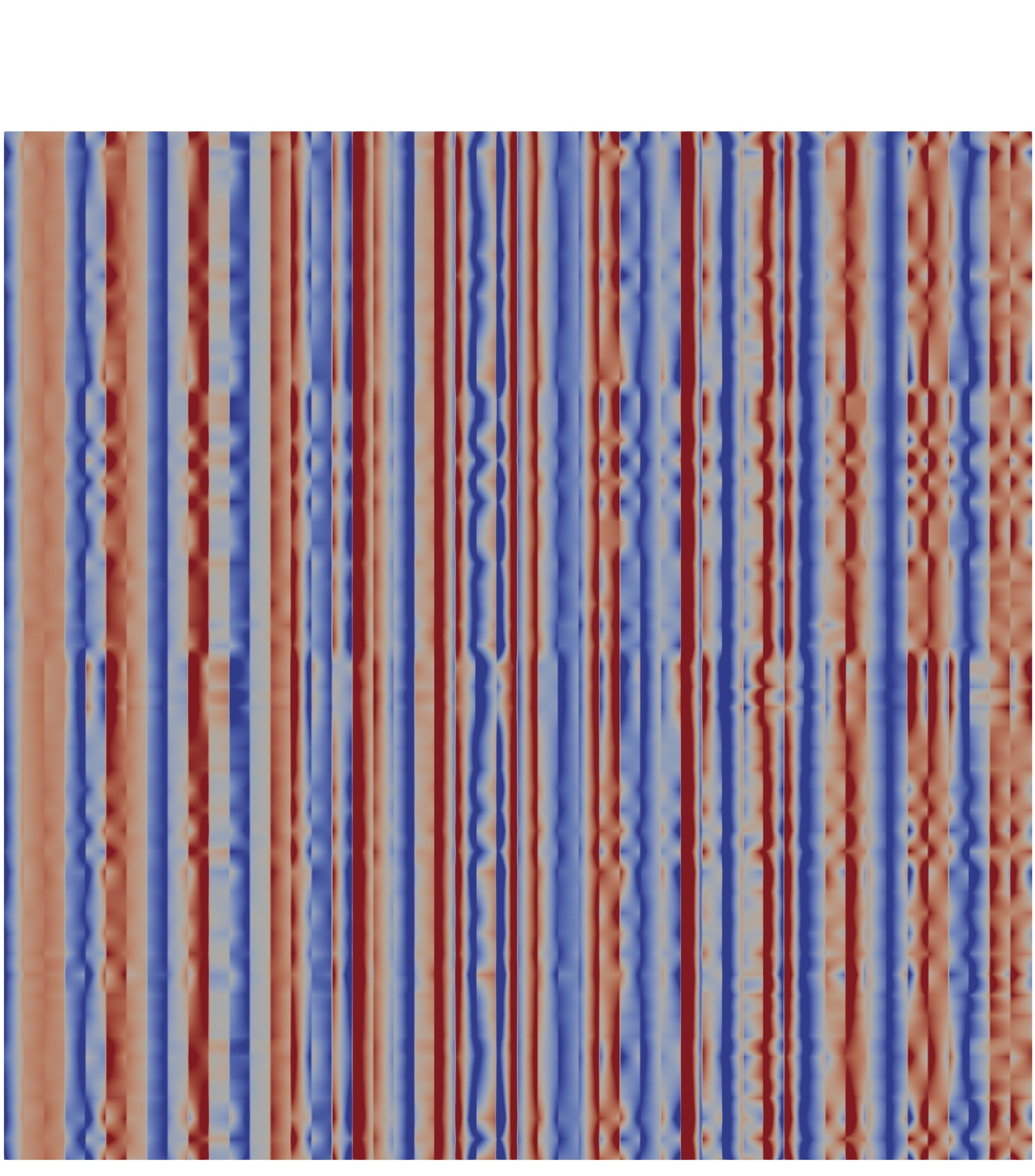}}
\caption{DIC results of the second example for the strain. The element size $h=8$ (pixels) in all solutions. FE-Q4 corresponds to the linear 4-node element, FE-Q8 corresponds to the quadratic 8-node element. }
\label{fig:exam2dux}
\end{figure}

\figurename~\ref{fig:exam2dux} shows the results for the strain measurement. The C-FE still gives relatively satisfactory results while the FE fails to capture the overall strain variation. Since the $L^2$-error seems a better measure for the accuracy, we only compute the $L^2$-error of the DIC results in this example. As shown in \tablename~\ref{table:ex2error}, the C-FE outperforms the other FE based methods. Under the same condition of the speckle patterns, the C-FE is promising to provide relatively more accurate results for both displacement and strain. In this example, the error due to the quality of speckle patterns seems significant. This error should be minimized in practice.

\begin{table}[htbp]
\caption{Error assessment for the DIC results of the second example}
\centering
\begin{tabular}{|c|c|c|c|c|c|c|}
\hline
 Method & Element size   & $\epsilon_{L}(u)$ & $\epsilon_{L}(\varepsilon_{xx})$\\ \hline
FE-Q4 & 8   & {0.278} & {0.497} \\ \hline
FE-Q8 & 8  & {0.362} &{0.875} \\ \hline
FE-Q8 & 16 & {0.269}  & {0.428}\\ \hline
C-FE & 8  & {0.258}  &{0.279}  \\ \hline
\end{tabular}\\
\label{table:ex2error}
\end{table}

\subsection{Example 3: DIC Challenge 2.0 star1}
\begin{figure}[htbp]
\centering
\subfigure[Reference image]{\includegraphics[scale=0.075]{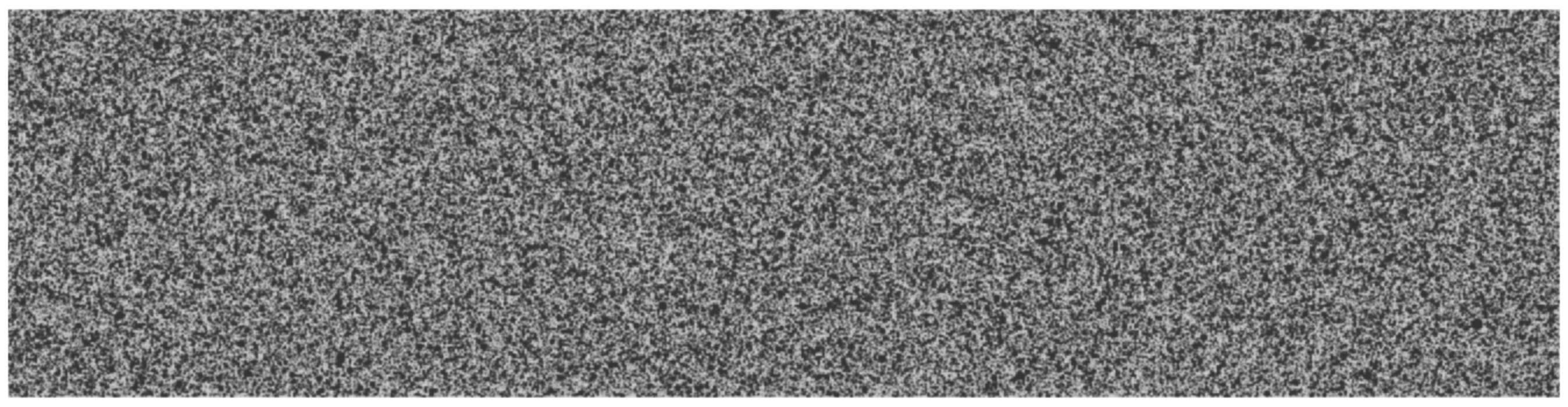}}\quad\quad
\subfigure[Deformed image ]{\includegraphics[scale=0.075]{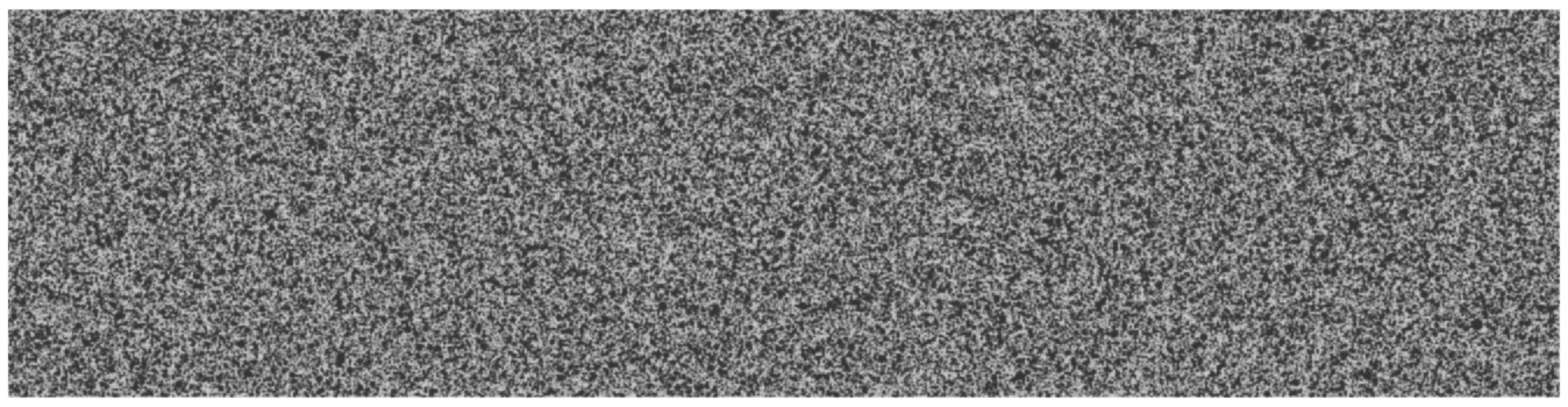}}\quad\\
\subfigure[C-FE: $h=5$ (pixels)  ]{\includegraphics[scale=0.075]{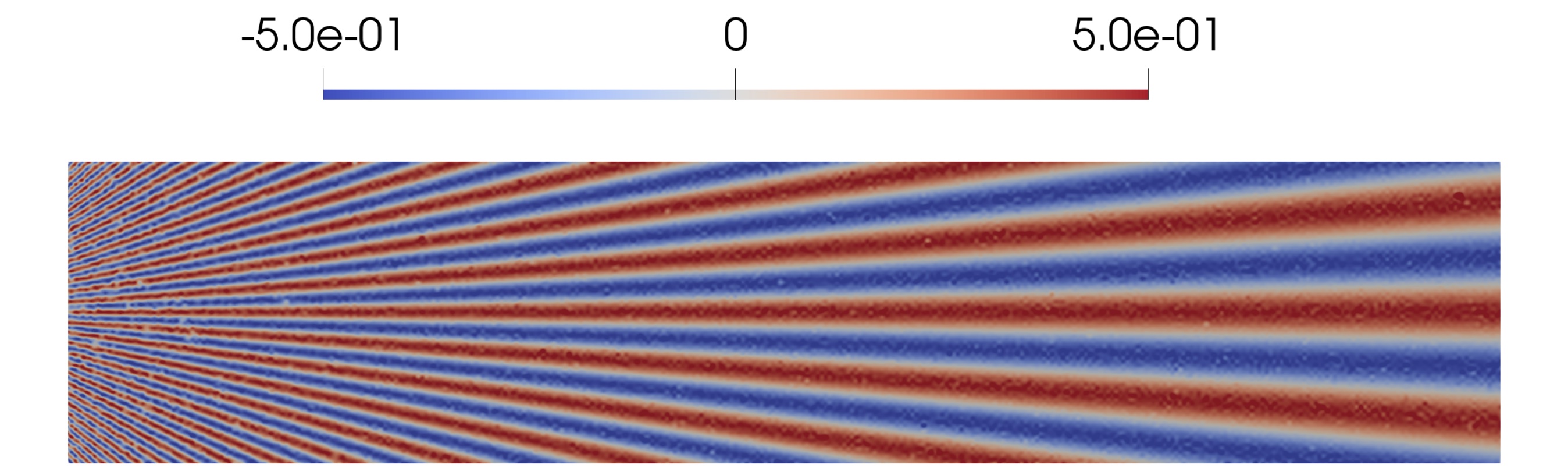}}\quad\quad
\subfigure[FE-Q4: $h=5$ (pixels)  ]{\includegraphics[scale=0.075]{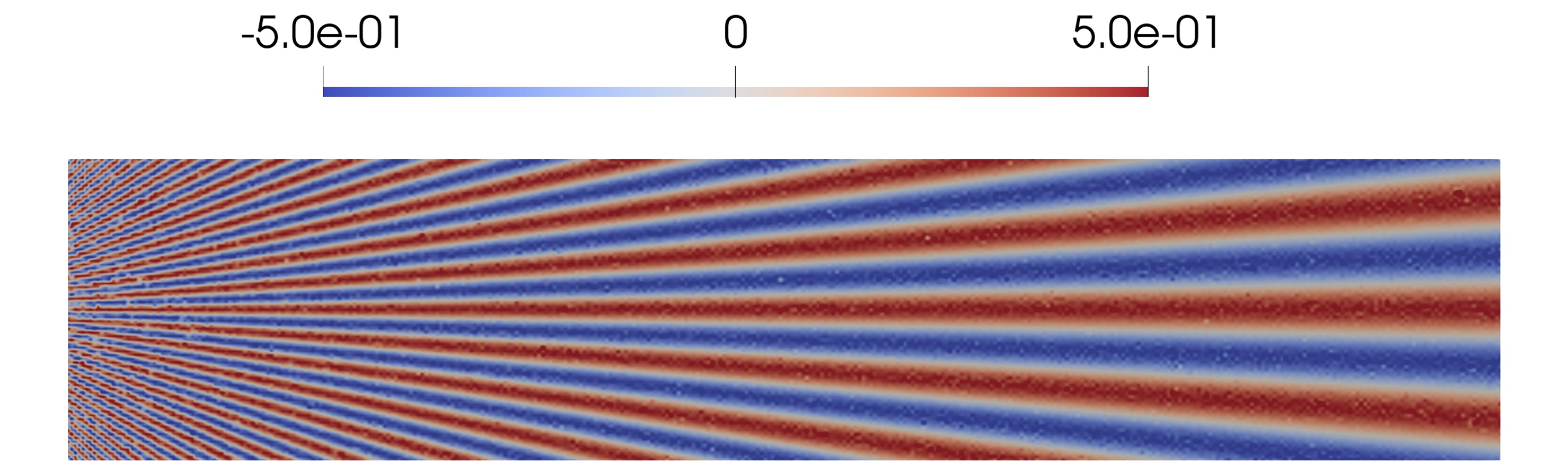}}\quad\quad\\
\subfigure[C-FE: $h=10$ (pixels) ]{\includegraphics[scale=0.075]{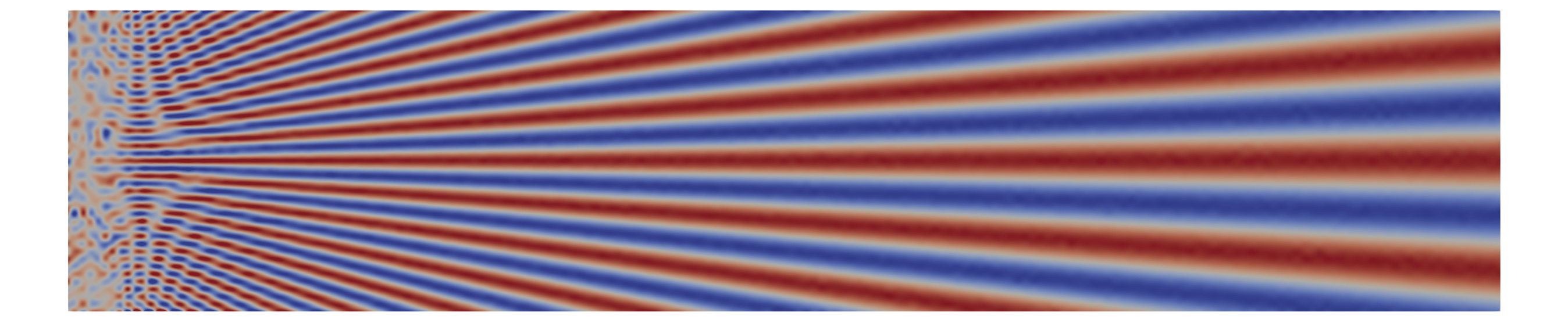}}\quad\quad
\subfigure[FE-Q4: $h=10$ (pixels)]{\includegraphics[scale=0.075]{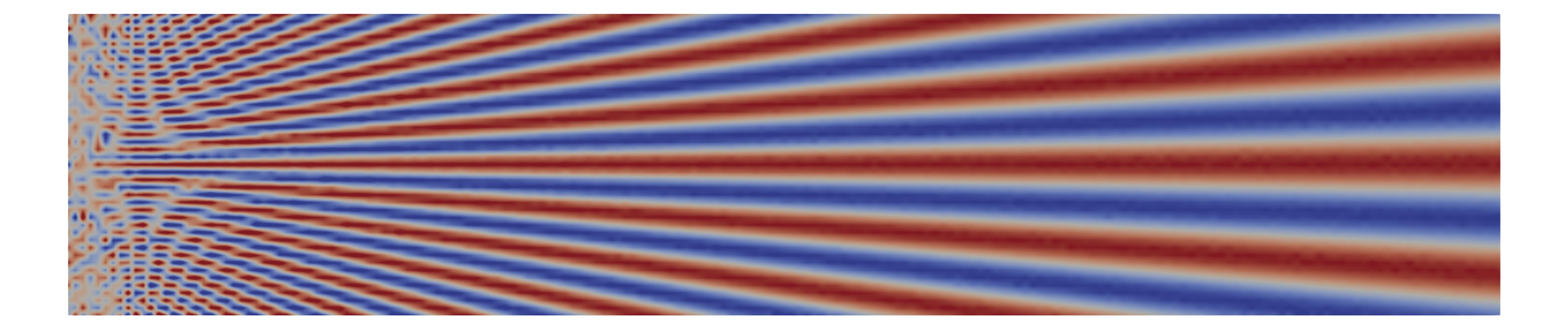}}\\
\subfigure[C-FE: $h=20$ (pixels)]{\includegraphics[scale=0.075]{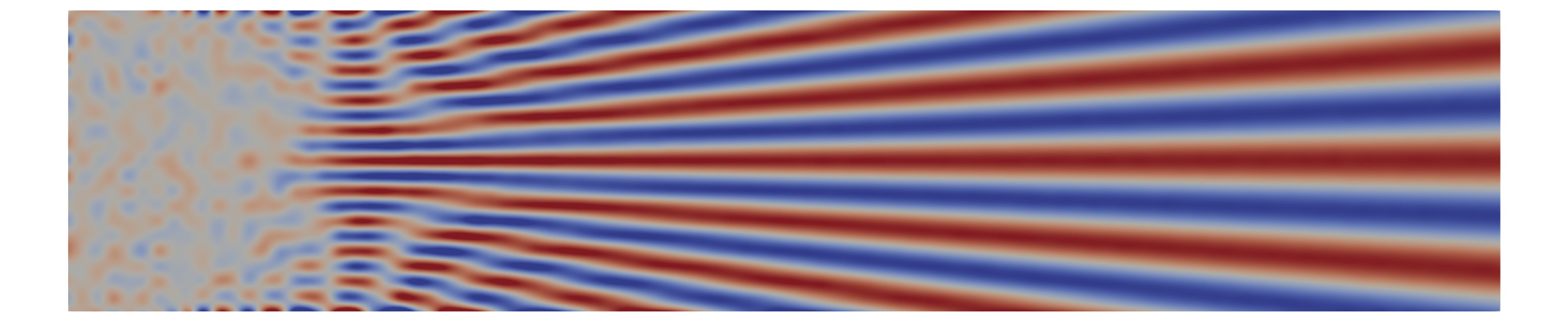}}\quad\quad
\subfigure[FE-Q4: $h=20$ (pixels) ]{\includegraphics[scale=0.075]{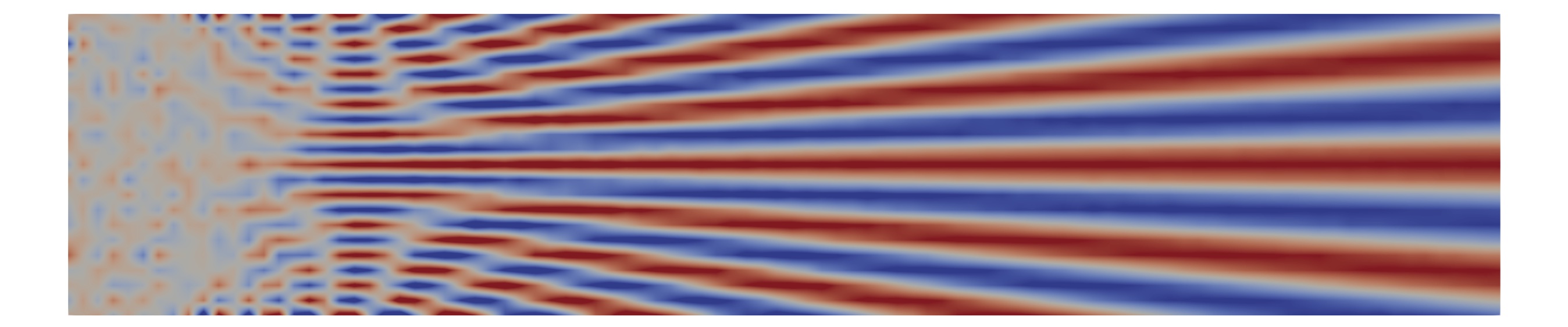}}\\
\caption{DIC results of the star1 example for the vertical displacement $v$. Three element sizes $h$ (pixels) are used in the solutions. FE-Q4 corresponds to the linear 4-node element.}
\label{fig:exam3uy}
\end{figure}

The proposed DIC method is applied to a set of benchmark problems of the DIC challenge 2.0 \cite{reu2022dic} that aims at providing a common framework for evaluating the performance of different DIC codes. The image data-set, called star1, is studied in this example. The star1 image is a "star" pattern with $\pm 5$ pixel sinusoidal vertical displacement amplitude and varying spatial frequency, details about the creation of the image can be found in \cite{reu2022dic} and the images can be downloaded at \cite{DICchallenge}. The initial reference and deformed images are illustrated in \figurename~\ref{fig:exam3uy}(a)(b). 

For the DIC analysis, we used the same graycale interpolation method as previous examples. The DIC results for the vertical displacement field using three different element sizes  are illustrated in \figurename~\ref{fig:exam3uy}(c-h). The overall trends of the results are consistent with those reported in \cite{reu2022dic}. As expected, the decreased element size can capture small scale variations of the displacement field in both C-FE and FE based DIC methods. However, by taking a closer look, C-FE still provides better results than FE under the same condition, especially in terms of smoothness and stability with respect to the decreasing element size. This is due to the higher order approximation and the non-local effect given by the convolution patch.

\begin{figure}[htbp]
\centering
\subfigure[Center line of the ZoI]{\includegraphics[scale=0.35]{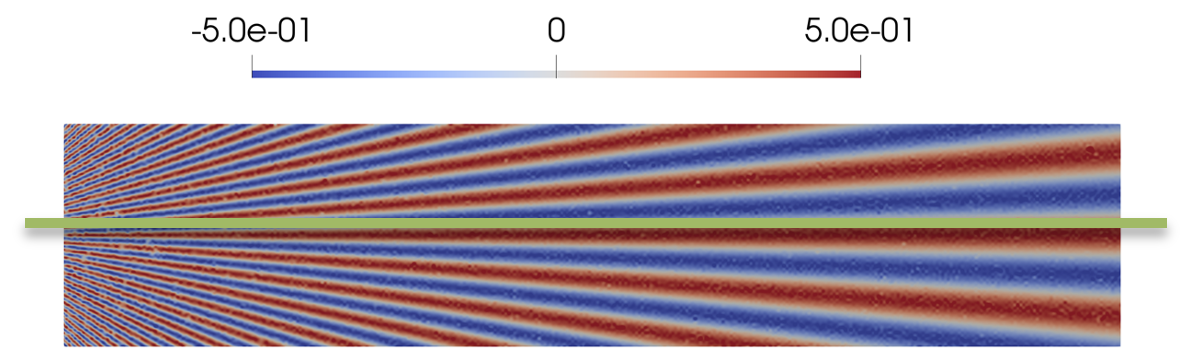}}\\
\subfigure[{Displacement, $h=5$ (px)}  ]{\includegraphics[scale=0.2]{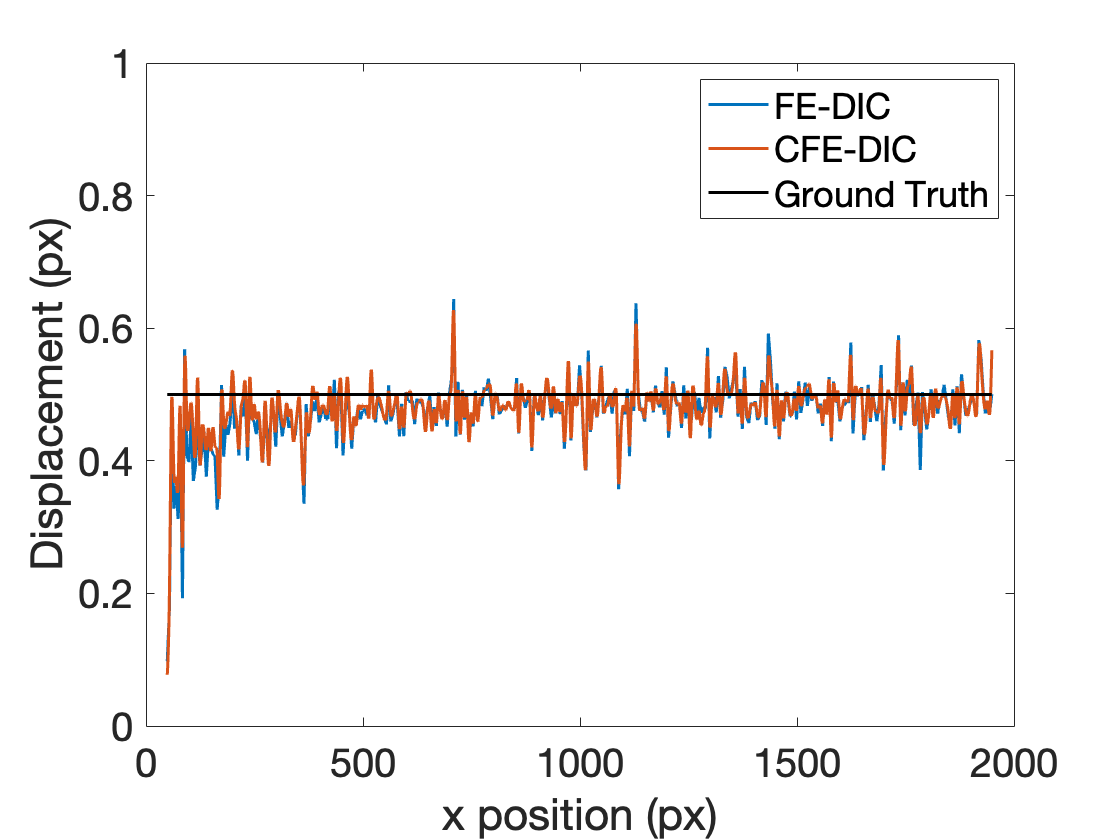}}
\subfigure[{Displacement, $h=10$ (px)}  ]{\includegraphics[scale=0.2]{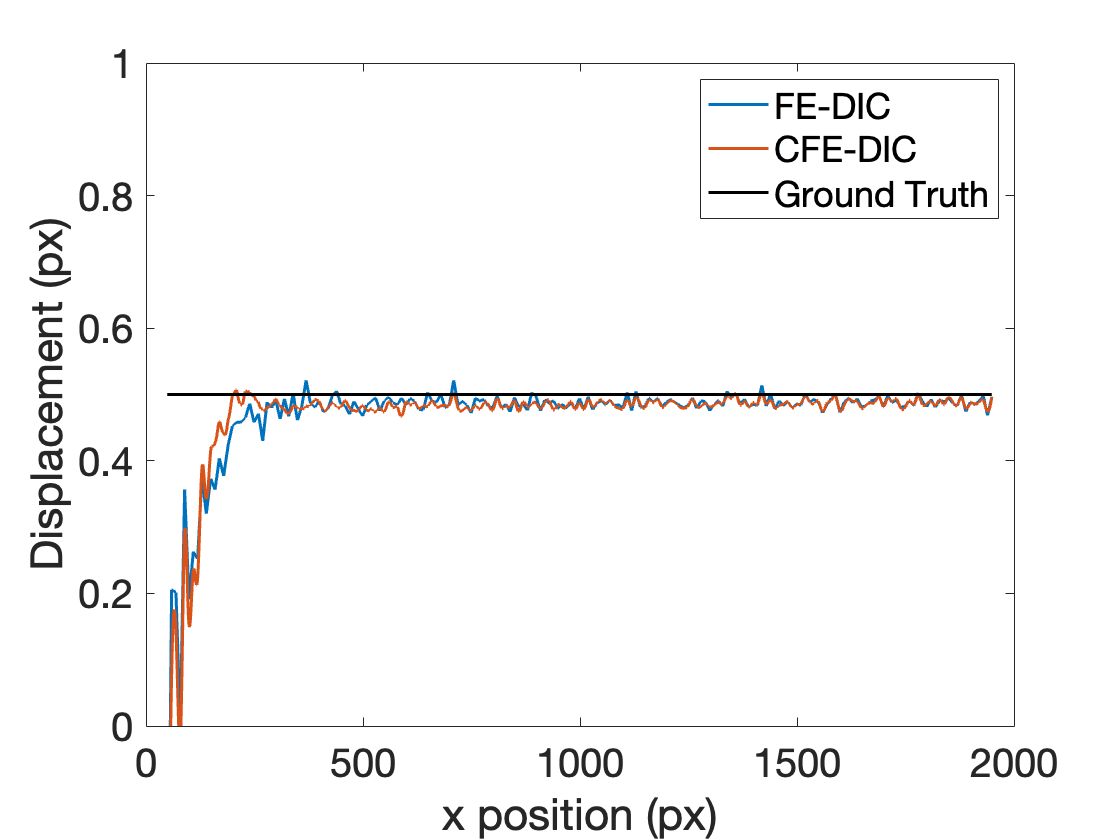}}\\
\caption{{Vertical displacement $v$ on the center line in the star1 example. px stands for pixel.} }
\label{fig:exam3line}
\end{figure}

\begin{figure}[htbp]
\centering
\subfigure[{Fitted curves for $h=10$ (px)}  ]{\includegraphics[scale=0.2]{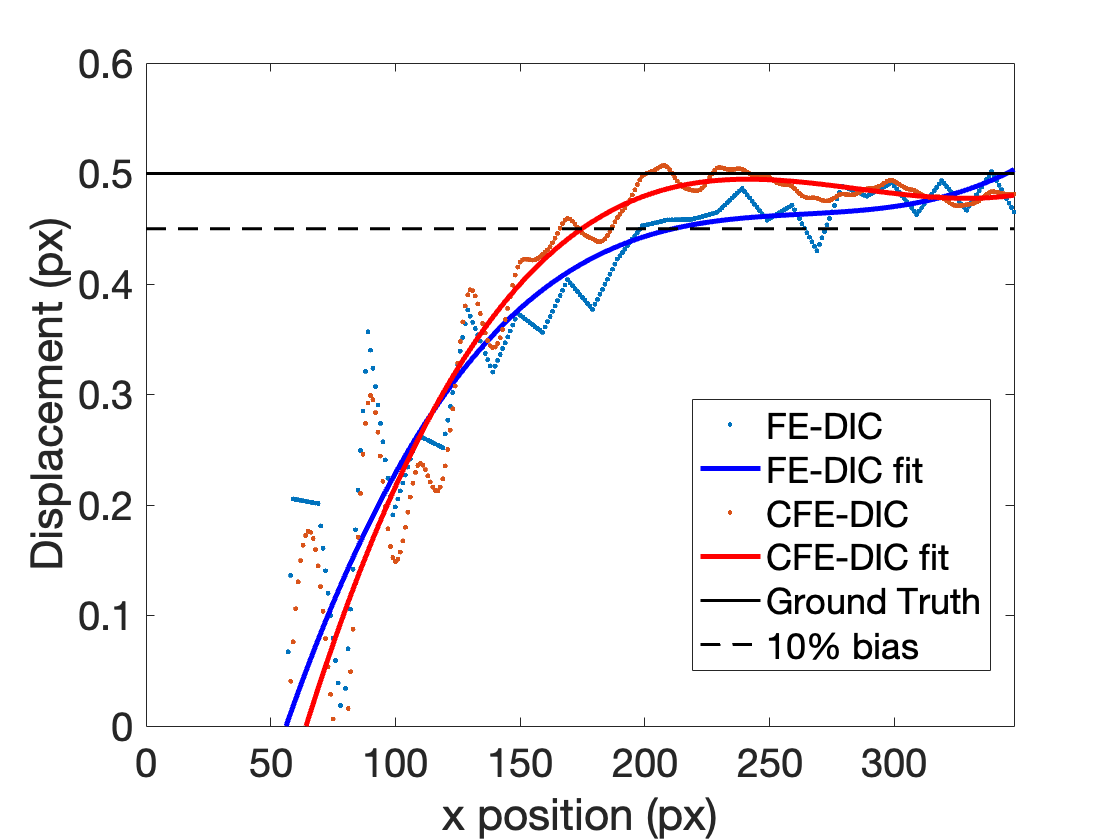}}
\subfigure[{Spatial resolution against element size }  ]{\includegraphics[scale=0.2]{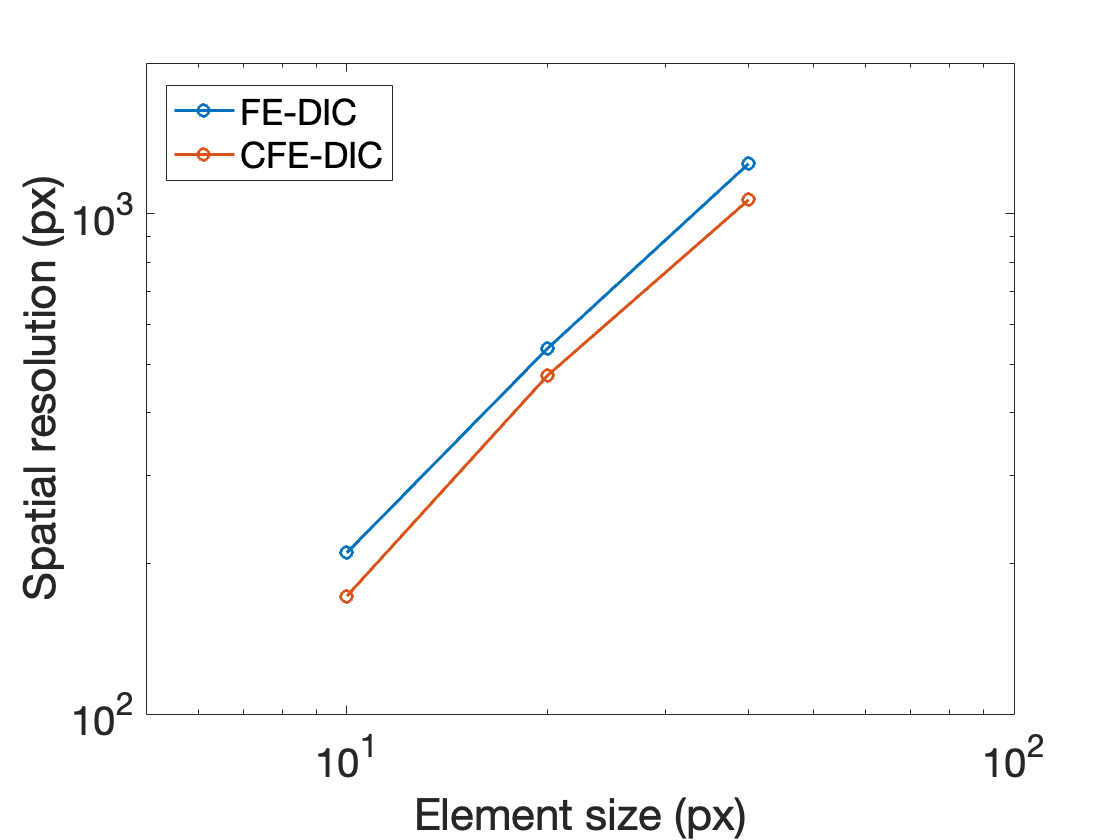}}\\
\caption{{Spatial resolution for the star1 example. px stands for pixel. Spatial resolution is defined as the position  where the fitted curve and the $10\%$-bias cutoff line intersect.}}
\label{fig:exam3SR}
\end{figure}

Now, let us look at the center line of the image, as shown in \figurename~\ref{fig:exam3line}. {Globally, the two methods, i.e., FE and C-FE, provide similar trends. When the element size is small (e.g., $h=5$), a stronger fluctuation presents in the measured displacements. When the element size is relatively large (e.g., $h=10$), the displacement gradually increases to the ground truth on the left. We can notice that the result given by C-FE has a faster increasing rate than that of FE. This feature can be quantified by the so-called spatial resolution \cite{reu2022dic}. Following the suggestions in \cite{reu2022dic}, we can first fit the row  data with polynomials, and then define the spatial resolution as the first point on the fitted curve that crossed the $10\%$ fractional attenuation line, as shown in \figurename~\ref{fig:exam3SR}(a). We can see that the C-FE based DIC has a better (smaller) spatial resolution. \figurename~\ref{fig:exam3SR}(b) shows the spatial resolutions with three different element sizes: $h=10, 20,$ and $40$. The improvement in spatial resolution demonstrates again the advantages of C-FE that provides a built-in convolution filter to regularize  DIC solutions. This improvement is highly desired, as spatial resolution is a crucial characteristic for DIC codes.}

\begin{figure}[htbp]
\centering
\subfigure[{Offset line 1, $h=10$ (px)}  ]{\includegraphics[scale=0.2]{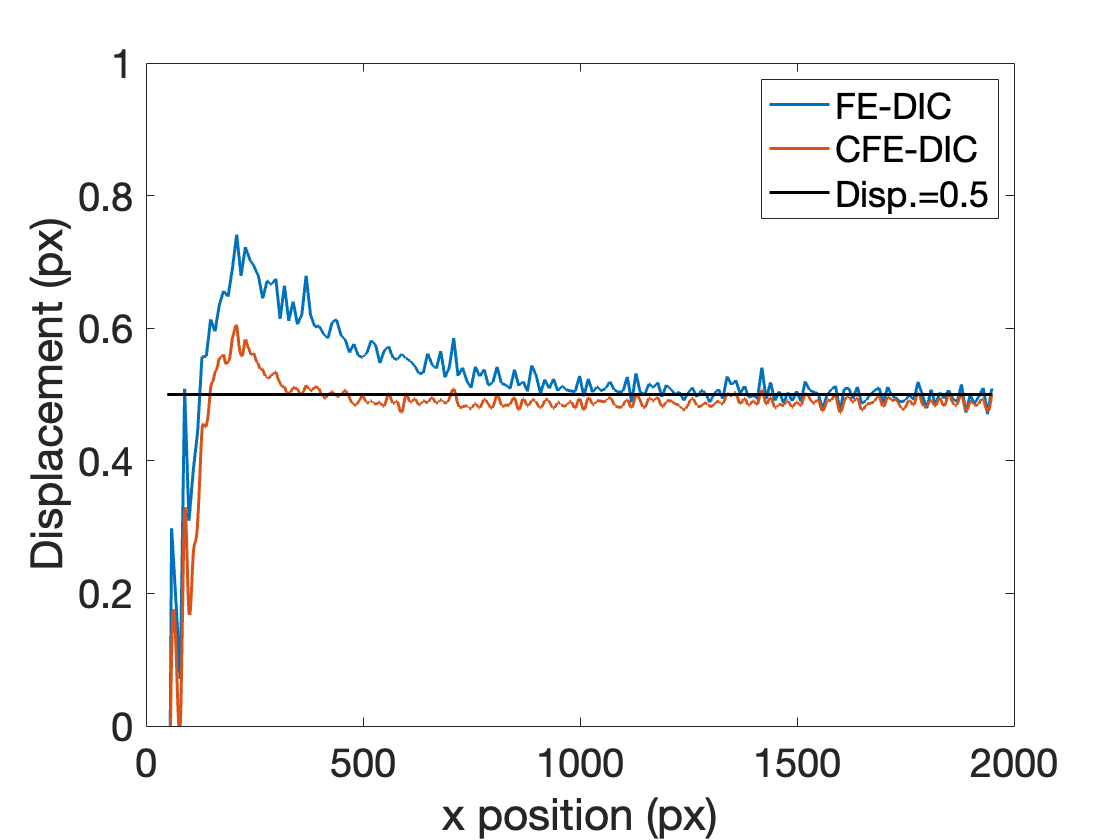}}
\subfigure[{Offset line 2, $h=10$ (px)}  ]{\includegraphics[scale=0.2]{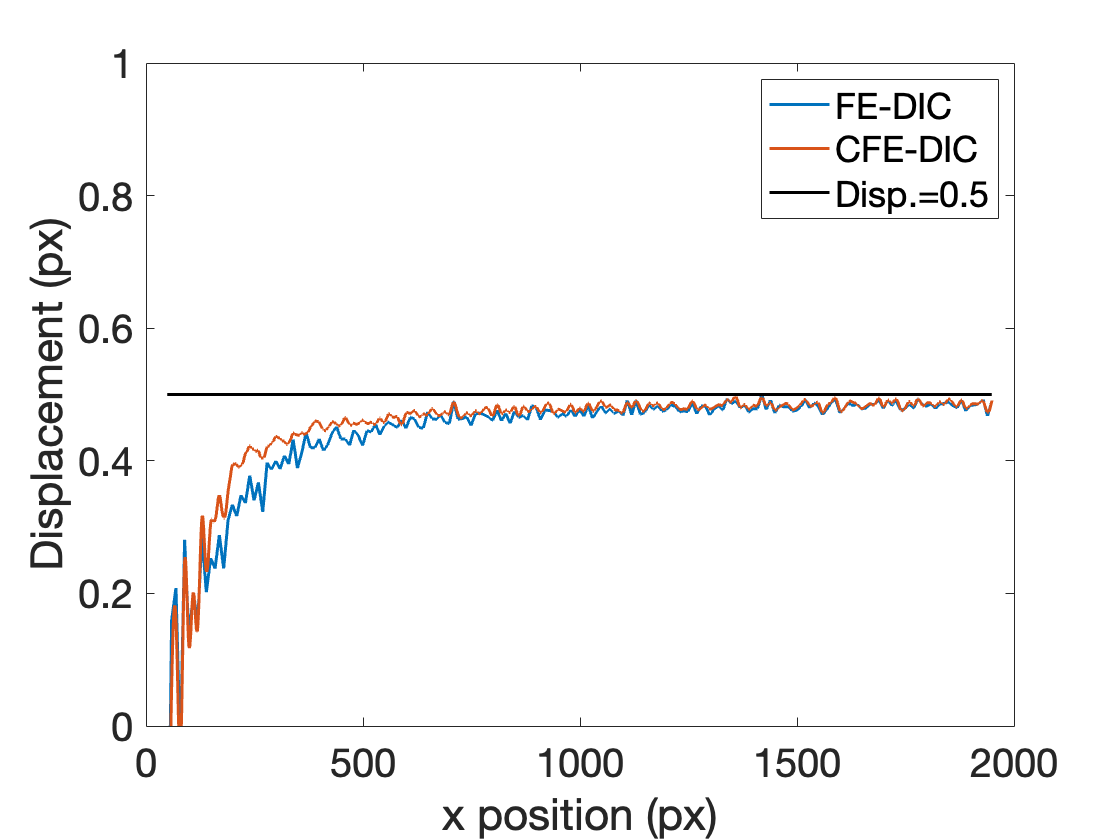}}\\
\caption{{Vertical displacement $v$ along two lines with small offsets to the center. px stands for pixel.}}
\label{fig:exam3offcenter}
\end{figure}

{In addition to the above line cut, we plotted the displacements along two different lines with small offsets to the center (see \figurename~\ref{fig:exam3offcenter}). We can see that the displacements still tend to approach $v=0.5$ but with different behaviors. Again, the C-FE results appear to behave better due to the built-in convolution filter.  }

\begin{figure}[htbp]
\centering
\subfigure[C-FE: $h=5$ (pixels)  ]{\includegraphics[scale=0.075]{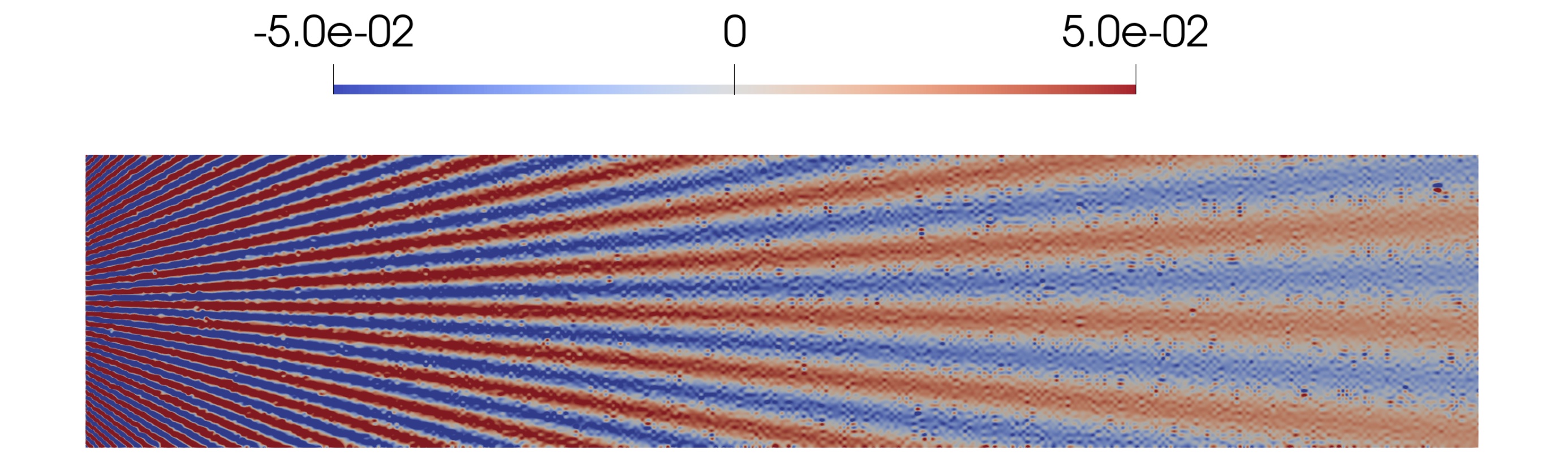}}\quad\quad
\subfigure[FE-Q4: $h=5$ (pixels)  ]{\includegraphics[scale=0.075]{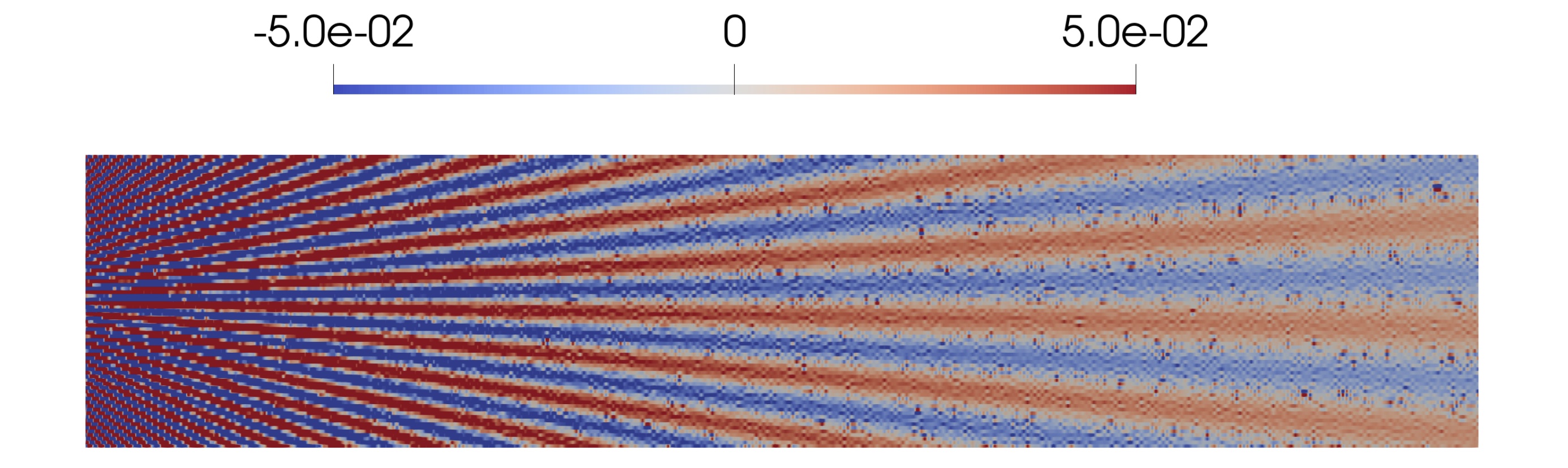}}\quad\quad\\
\subfigure[C-FE: $h=10$ (pixels) ]{\includegraphics[scale=0.075]{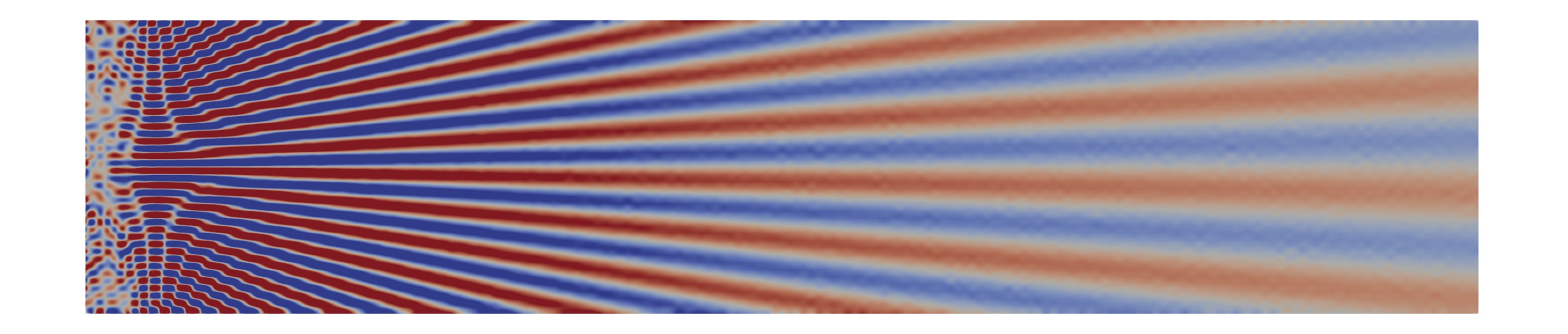}}\quad\quad
\subfigure[FE-Q4: $h=10$ (pixels)]{\includegraphics[scale=0.15]{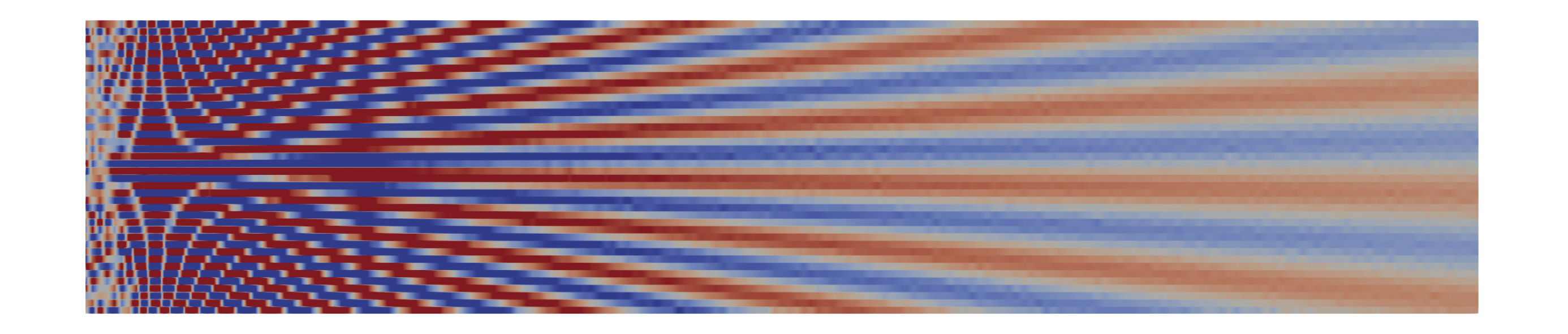}}\\
\subfigure[C-FE: $h=20$ (pixels)]{\includegraphics[scale=0.075]{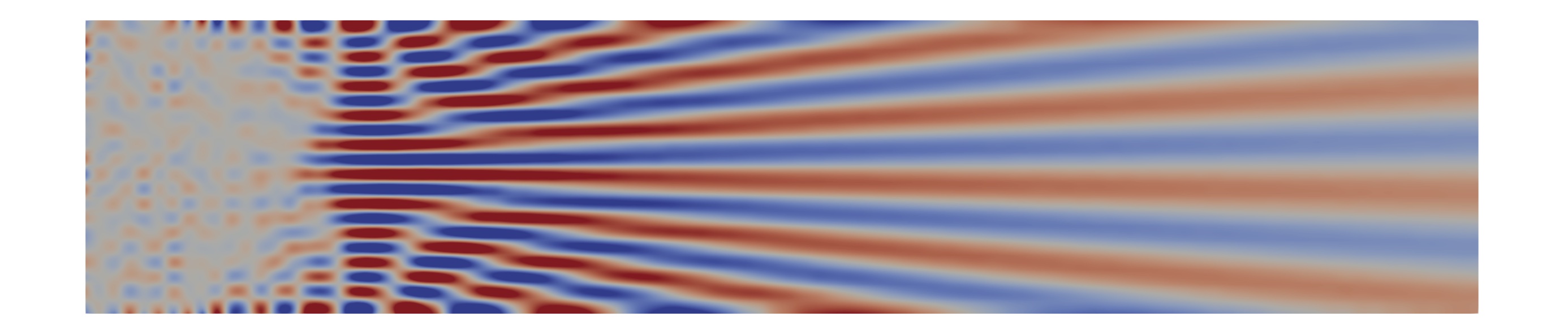}}\quad\quad
\subfigure[FE-Q4: $h=20$ (pixels) ]{\includegraphics[scale=0.15]{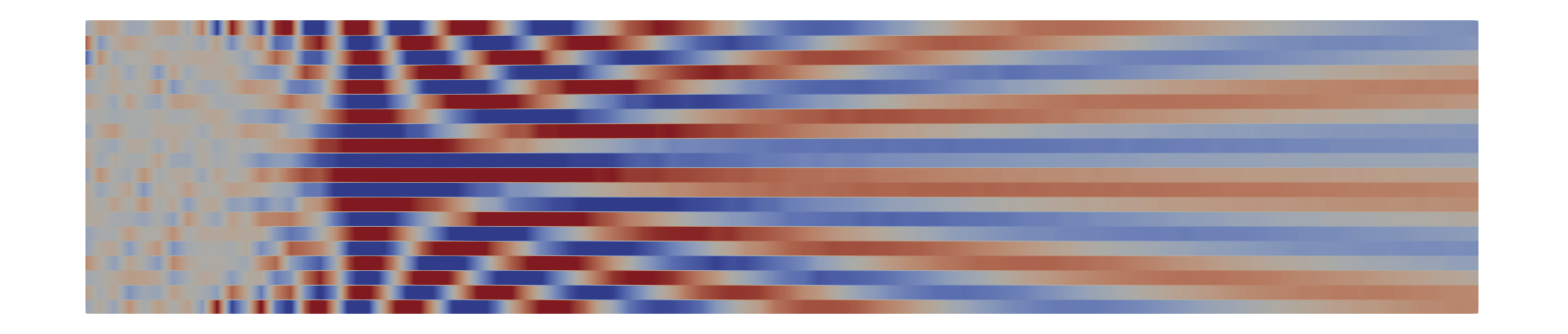}}\\
\caption{DIC results of the star1 example  for the strain $\varepsilon_{yy}$. FE-Q4 corresponds to the linear 4-node element.}
\label{fig:exam3duy}
\end{figure}

The strain results are depicted in \figurename~\ref{fig:exam3duy}. As expected, the C-FE results have much better smoothness and accuracy. This confirms again the advantages of high order approximations. Since the analytical expression of the displacement or strain field is not known, we did not compute the RMSE or $L^2$-error for this example.

\subsection{Example 4: DIC Challenge 2.0 star2}
The second data-set of images, namely star2, is also a "star" pattern with the same deformation field as star1, but includes a heteroscedastic image acquisition noise \cite{reu2022dic} modeled specifically on a FLIR (formerly PointGrey) 5 megapixel camera (GRAS-50S5C-C) \cite{sur2018rendering}. {The data-set includes 3 images: a noisy reference, an undeformed noise floor image, and a deformed noisy image. The reference and deformed images will be used to measure the diplacement field and calculate the spatial resolution. The undeformed noise floor image can be used to quantify the noise and so-called measurement resolution \cite{reu2022dic}, by comparing with the reference image. }

\begin{figure}[htbp]
\centering
\subfigure[Reference image]{\includegraphics[scale=0.075]{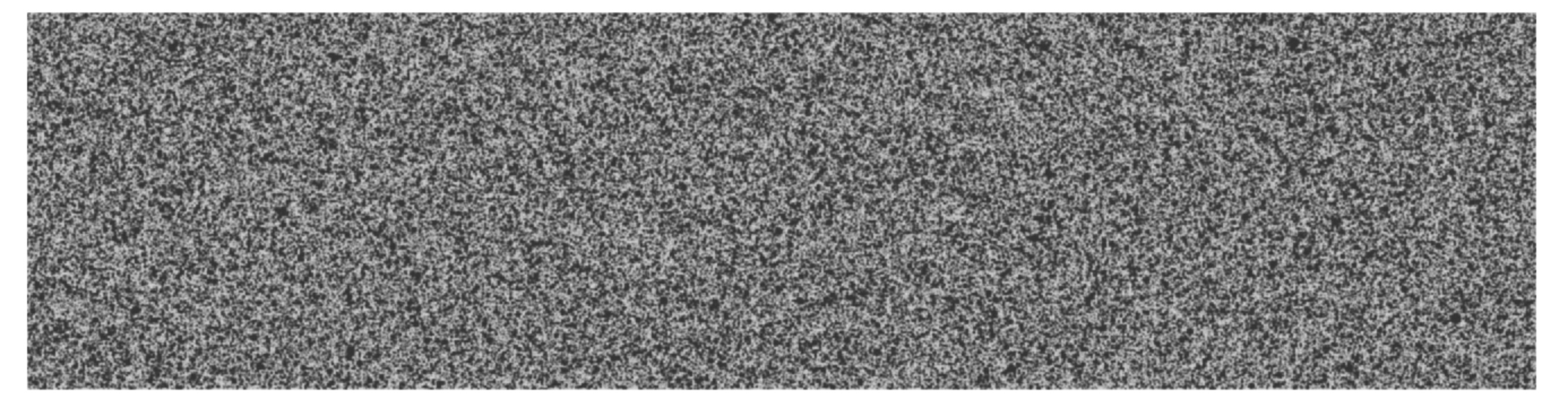}}\quad\quad
\subfigure[Deformed image ]{\includegraphics[scale=0.075]{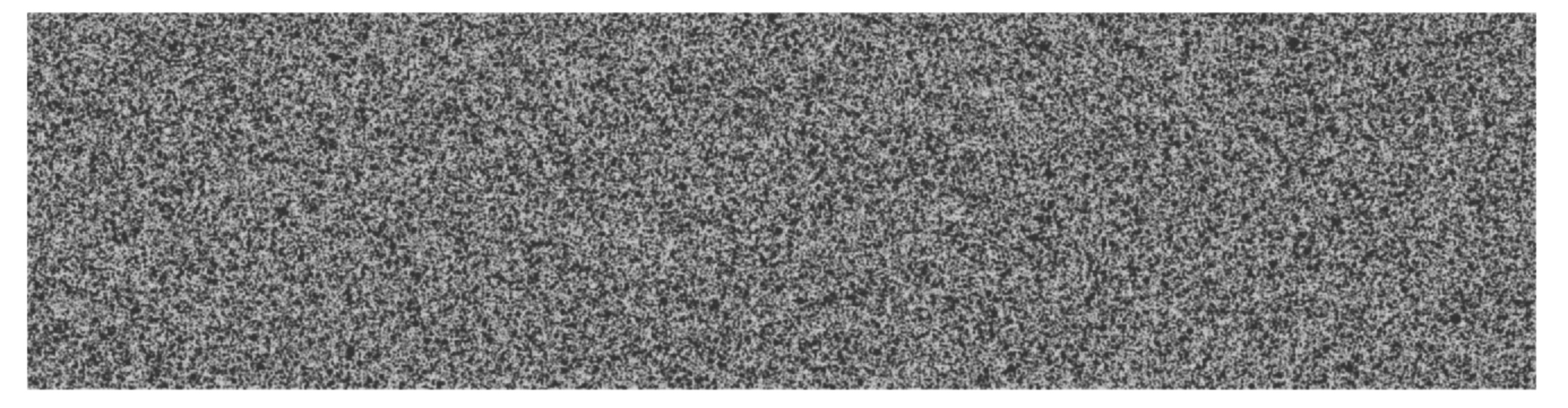}}\quad\\
\subfigure[C-FE: $h=5$ (pixels)  ]{\includegraphics[scale=0.075]{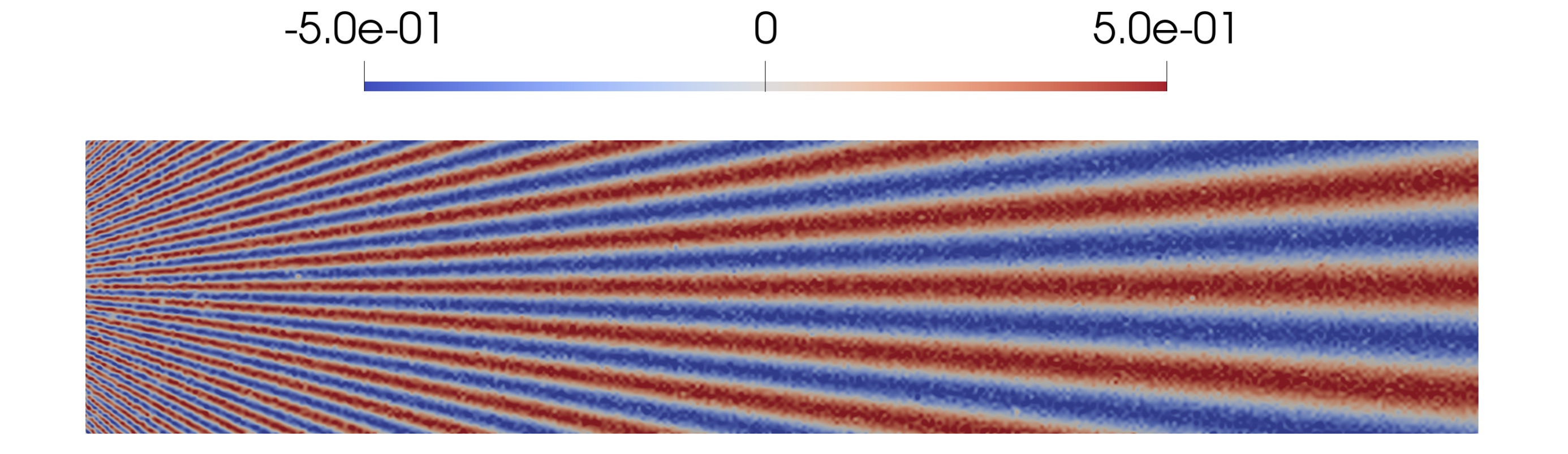}}\quad\quad
\subfigure[FE-Q4: $h=5$ (pixels)  ]{\includegraphics[scale=0.075]{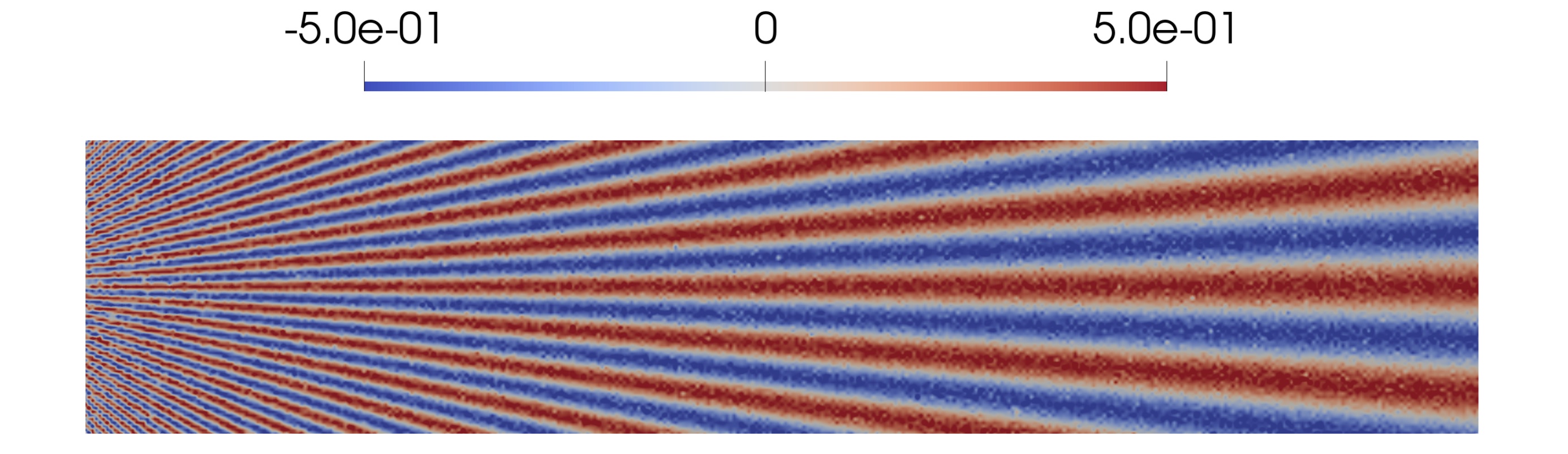}}\quad\quad\\
\subfigure[C-FE: $h=10$ (pixels) ]{\includegraphics[scale=0.075]{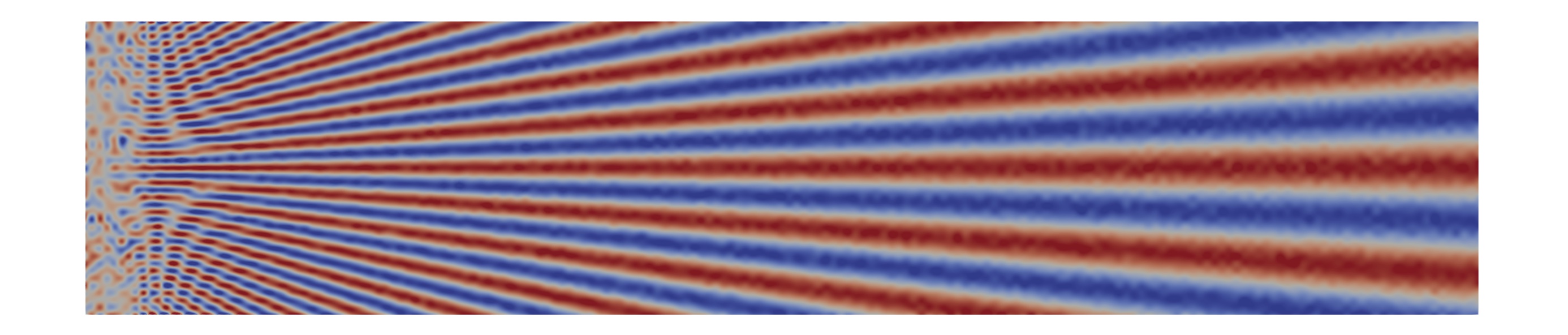}}\quad\quad
\subfigure[FE-Q4: $h=10$ (pixels)]{\includegraphics[scale=0.075]{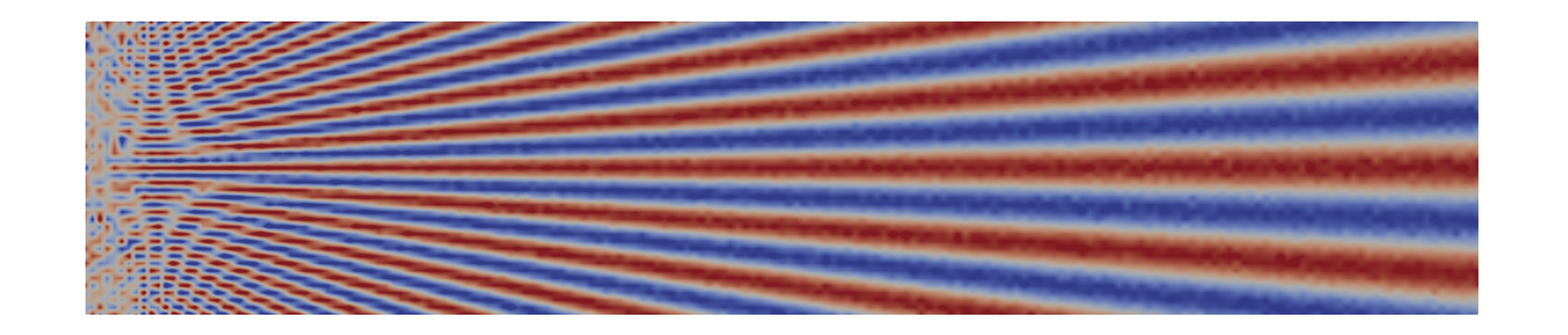}}\\
\subfigure[C-FE: $h=20$ (pixels)]{\includegraphics[scale=0.075]{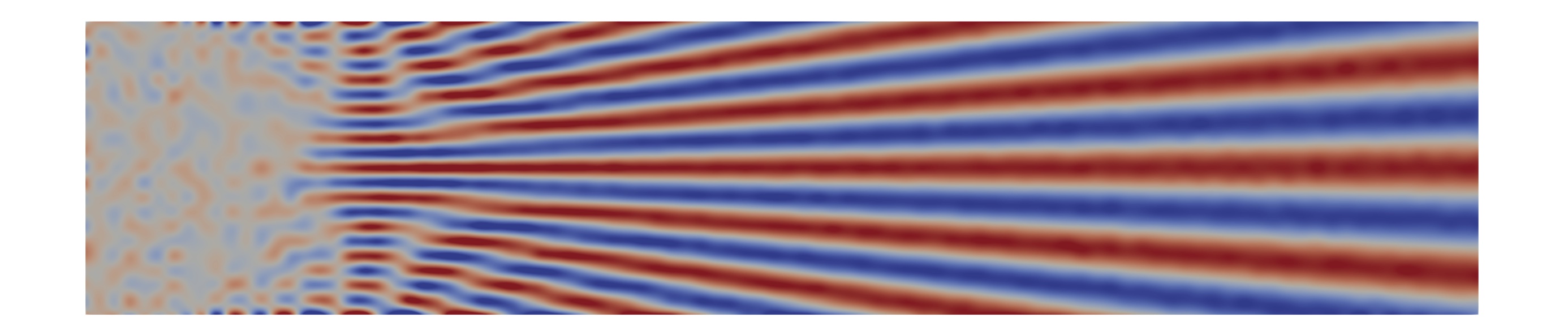}}\quad\quad
\subfigure[FE-Q4: $h=20$ (pixels) ]{\includegraphics[scale=0.075]{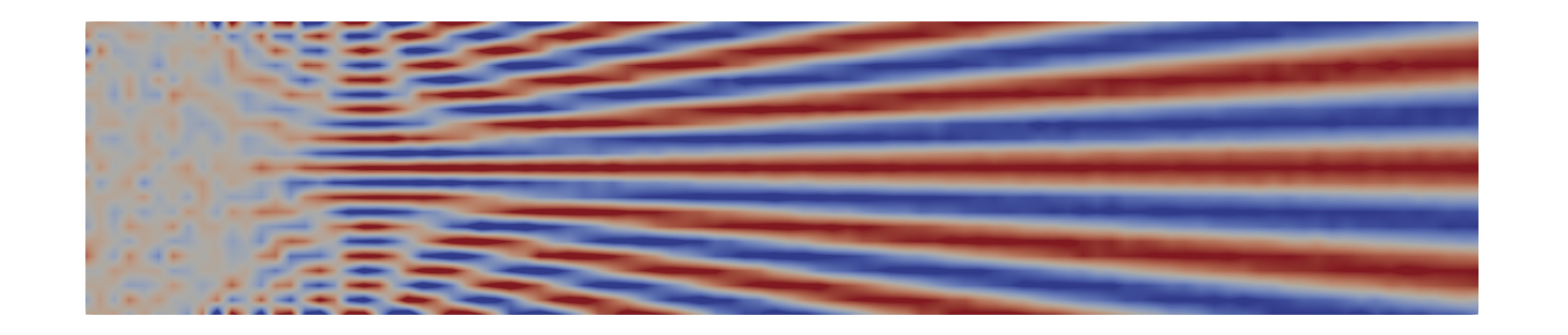}}\\
\caption{DIC results of the star2 example for the vertical displacement $v$. Three element sizes $h$ (pixels) are used in the solutions. FE-Q4 corresponds to the linear 4-node element.}
\label{fig:exam4uy}
\end{figure}

\begin{figure}[htbp]
\centering
\subfigure[Center line of the ZoI]{\includegraphics[scale=0.37]{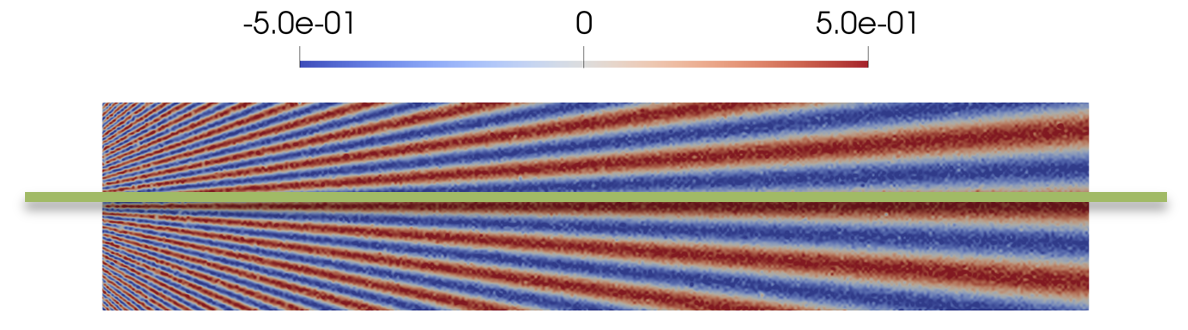}}\\
\subfigure[{Displacement, $h=5$ (px)} ]{\includegraphics[scale=0.2]{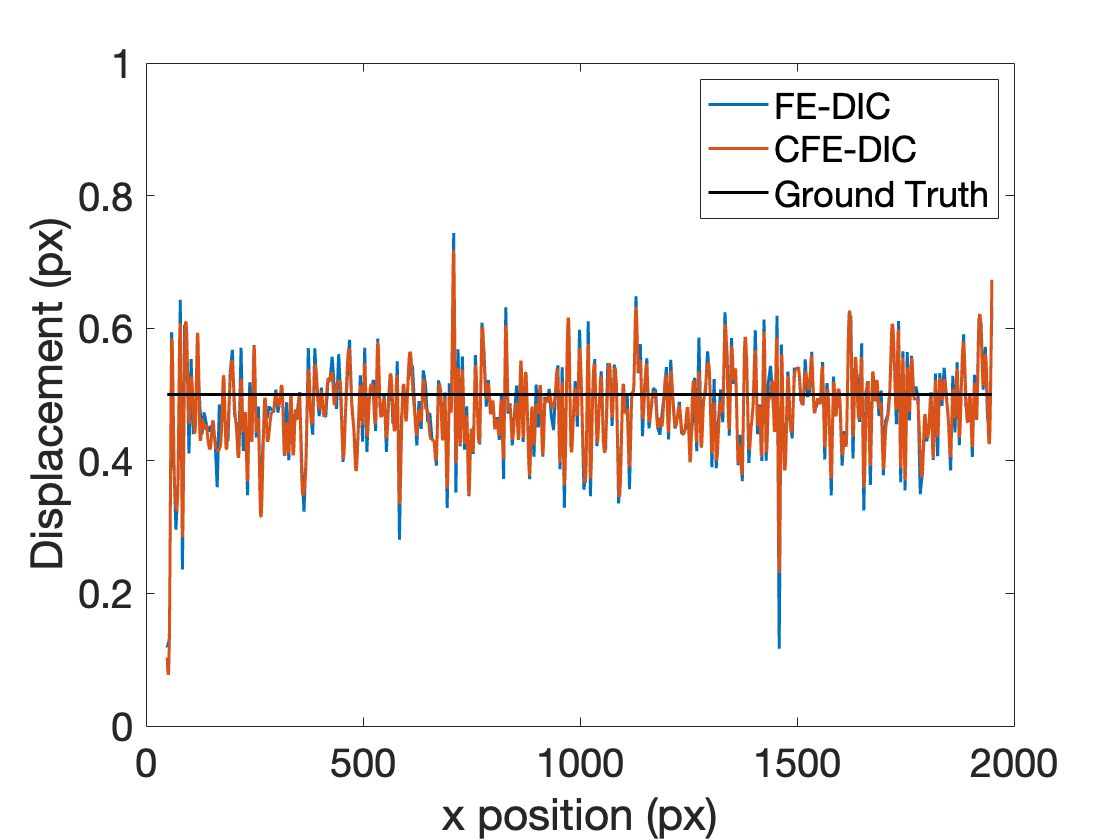}}
\subfigure[{Displacement, $h=10$ (px)}]{\includegraphics[scale=0.2]{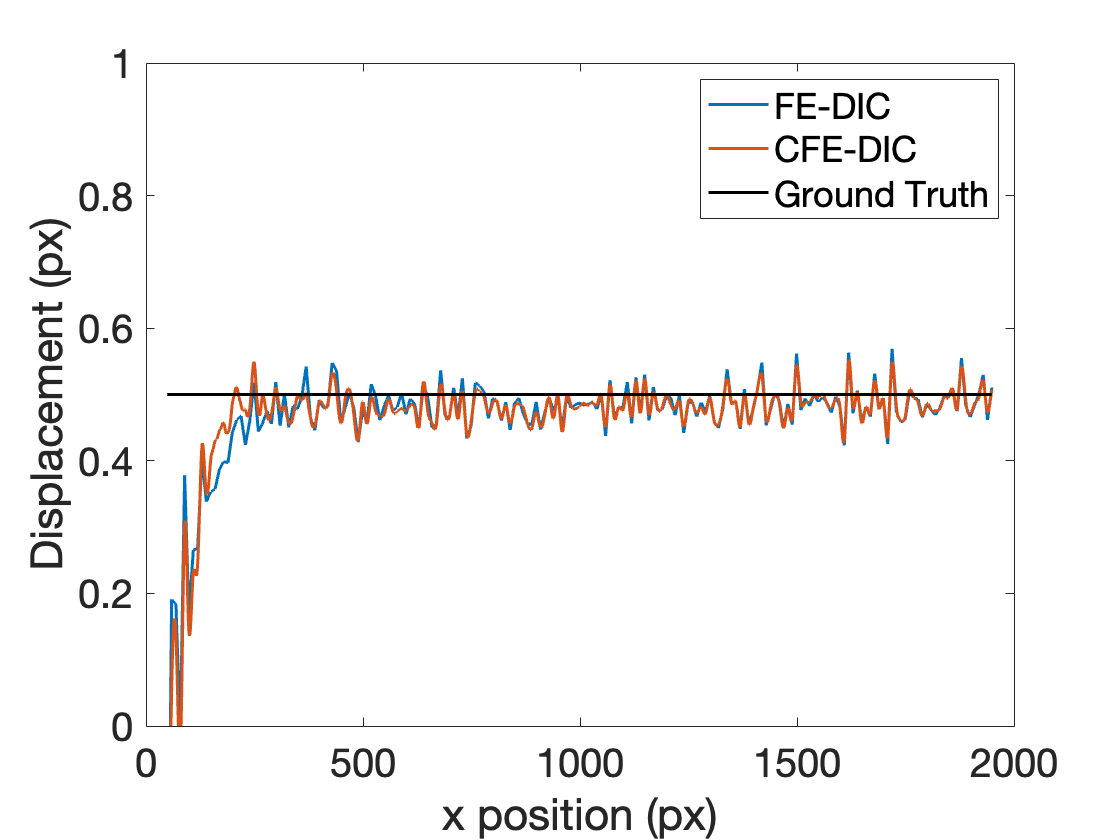}}
\caption{{Vertical displacement $v$ on the center line in the star2 example. px stands for pixel. }}
\label{fig:exam4line}
\end{figure}

The same DIC technique as previously is applied to this example and leads to the results in \figurename~\ref{fig:exam4uy}. {Here, the reference and deformed images are used to measure the displacement field.} As shown in the figure, both C-FE and FE results are affected by the noise, compared to the star1 results in previous section. Nevertheless, the C-FE  seems more robust with the noise, as illustrated in \figurename~\ref{fig:exam4line}.  By plotting the displacement samples over the center line, we can see the variation of the DIC results is amplified by the noise, however, the deviation of C-FE results seems smaller and the displacement samples remain closer to the ground truth. {The spatial resolution can be extracted from this line cut, as presented in the star1 example. We can expect similar spatial resolutions to star1, as the star2 example only differs with additional noise.}

\begin{figure}[htbp]
\centering
\subfigure[C-FE: $h=5$ (pixels)  ]{\includegraphics[scale=0.075]{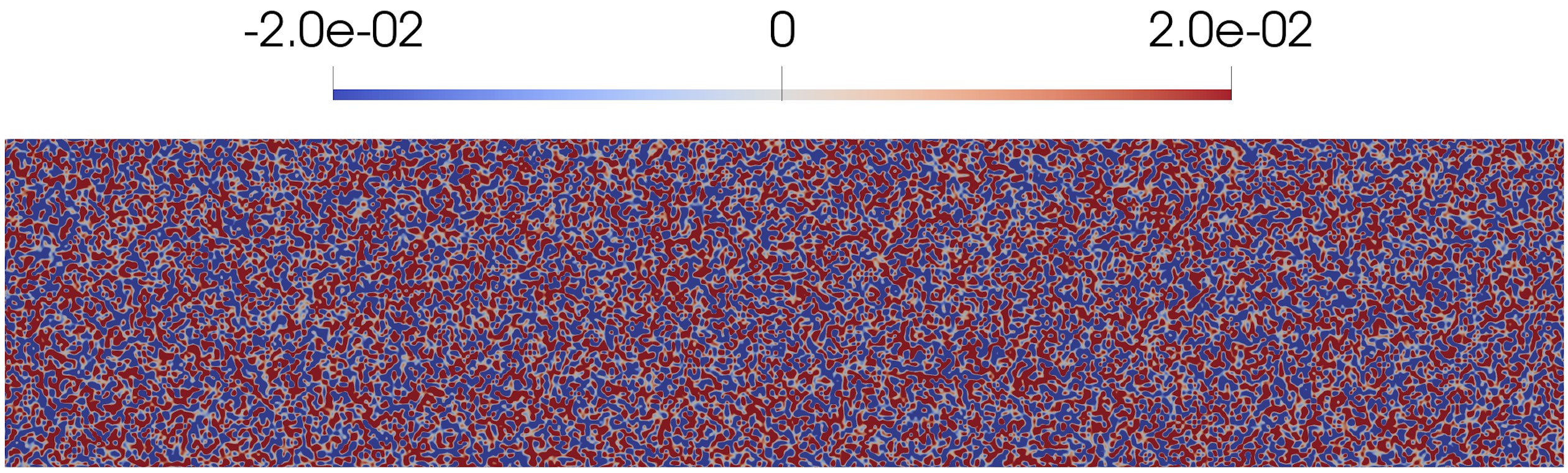}}\quad\quad\quad\quad
\subfigure[FE-Q4: $h=5$ (pixels)  ]{\includegraphics[scale=0.075]{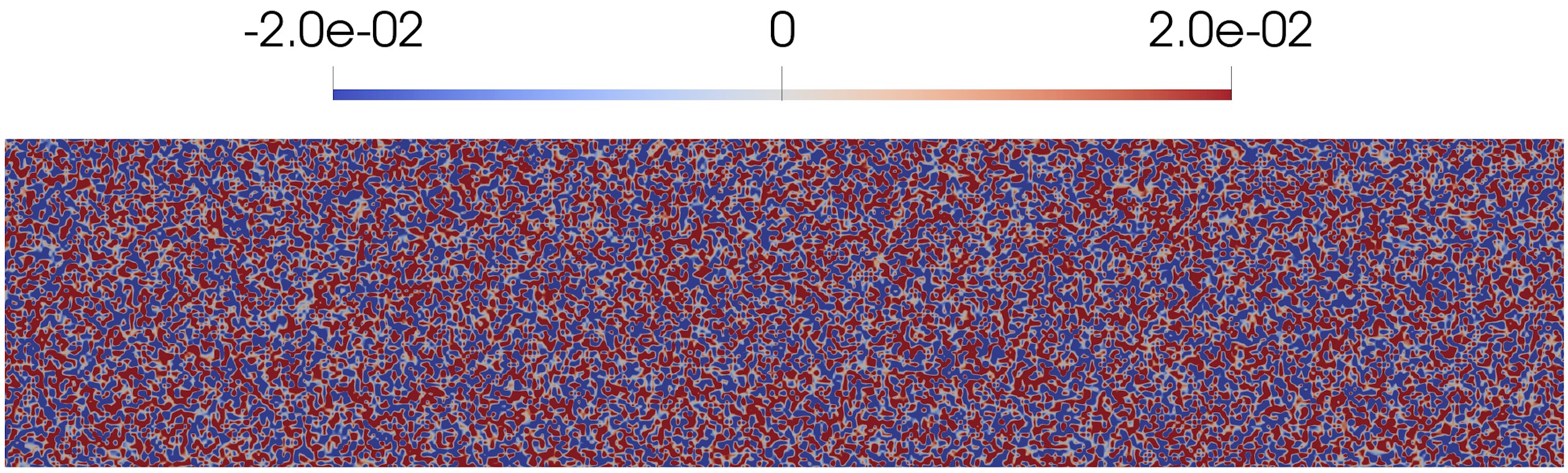}}\quad\quad\\
\subfigure[C-FE: $h=10$ (pixels) ]{\includegraphics[scale=0.075]{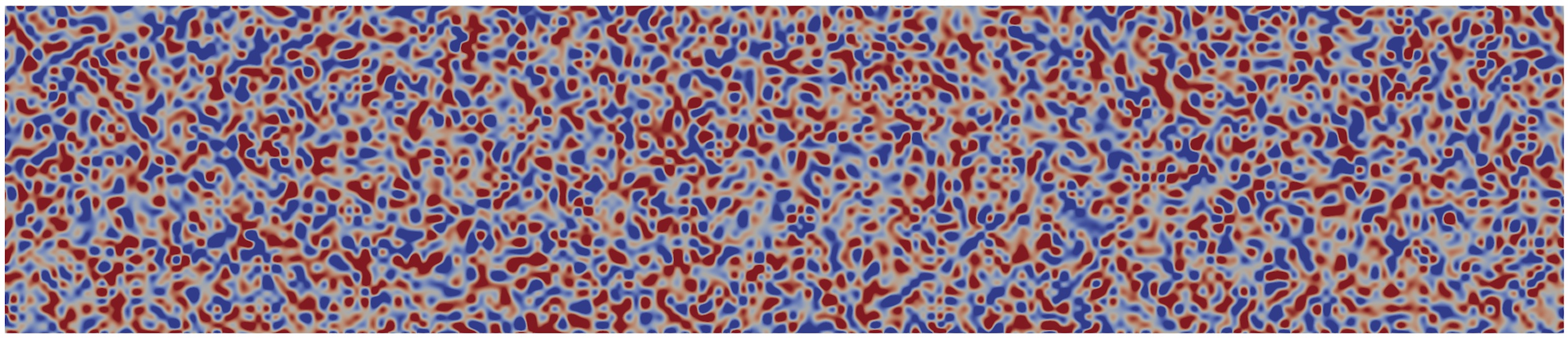}}\quad\quad\quad\quad
\subfigure[FE-Q4: $h=10$ (pixels)]{\includegraphics[scale=0.075]{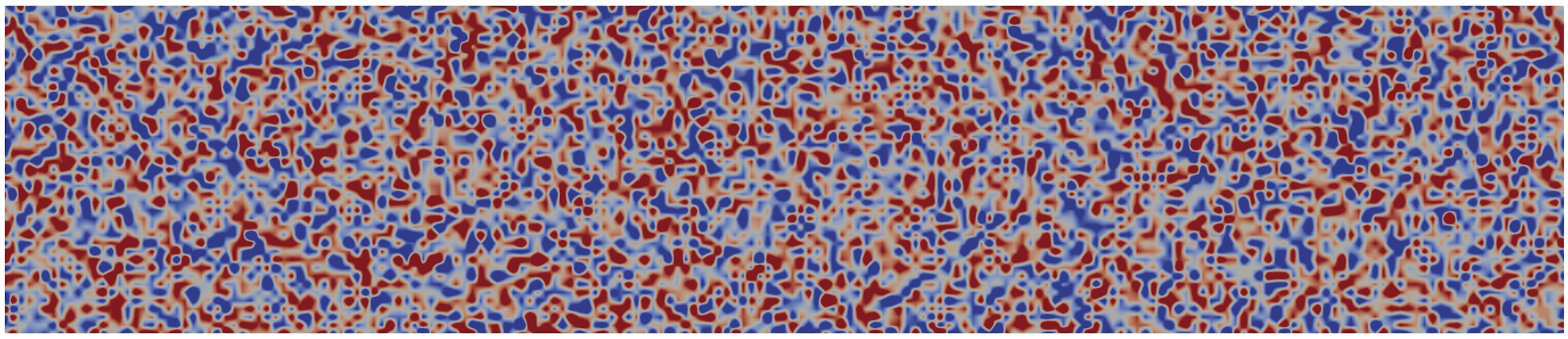}}\\
\subfigure[C-FE: $h=20$ (pixels)]{\includegraphics[scale=0.075]{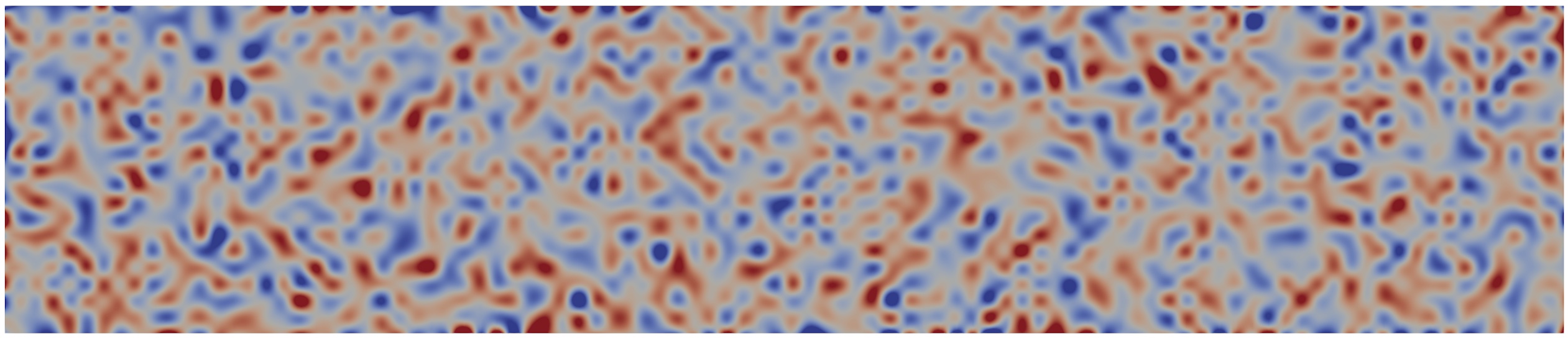}}\quad\quad\quad\quad
\subfigure[FE-Q4: $h=20$ (pixels) ]{\includegraphics[scale=0.075]{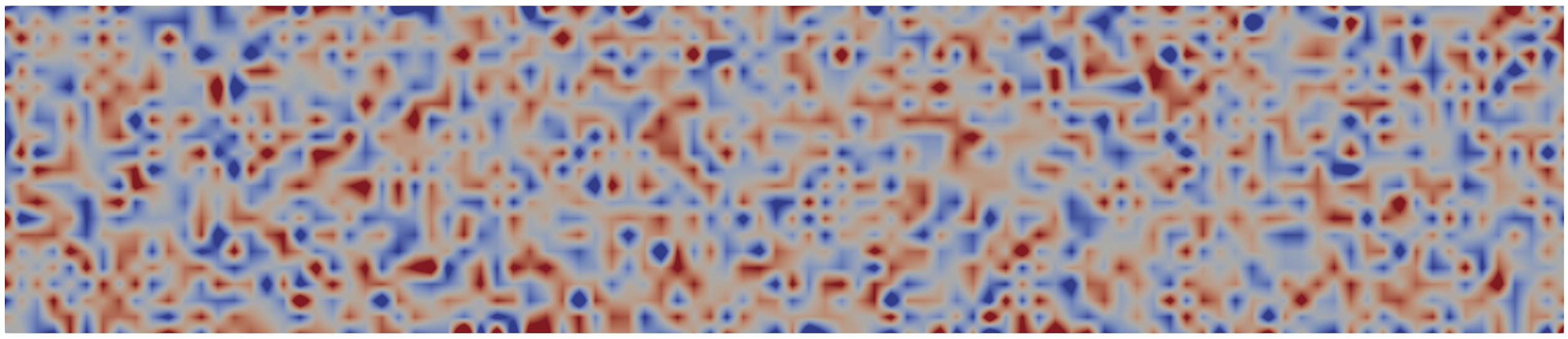}}\\
\caption{{DIC results of the displacement noise from the undeformed noise floor image. FE-Q4 corresponds to the linear 4-node element.}}
\label{fig:exam4noise}
\end{figure}

{\figurename~\ref{fig:exam4noise} illustrates the DIC results obtained with the reference and the undeformed noise floor image. As expected, the element size  has also a filtering effect on the noise. A large element size leads to a low frequency noise. Comparing the C-FE and FE results, we can see that the noise is somehow regularized by C-FE due to the built-in convolution filter. With this noise measurement, we can take sample noise data along the same line cut as displacement. This allows us to define a measurement resolution by taking the standard deviation of the row sample data.}

\begin{figure}[htbp]
\centering
\subfigure[{Spatial and measurement resolutions} ]{\includegraphics[scale=0.2]{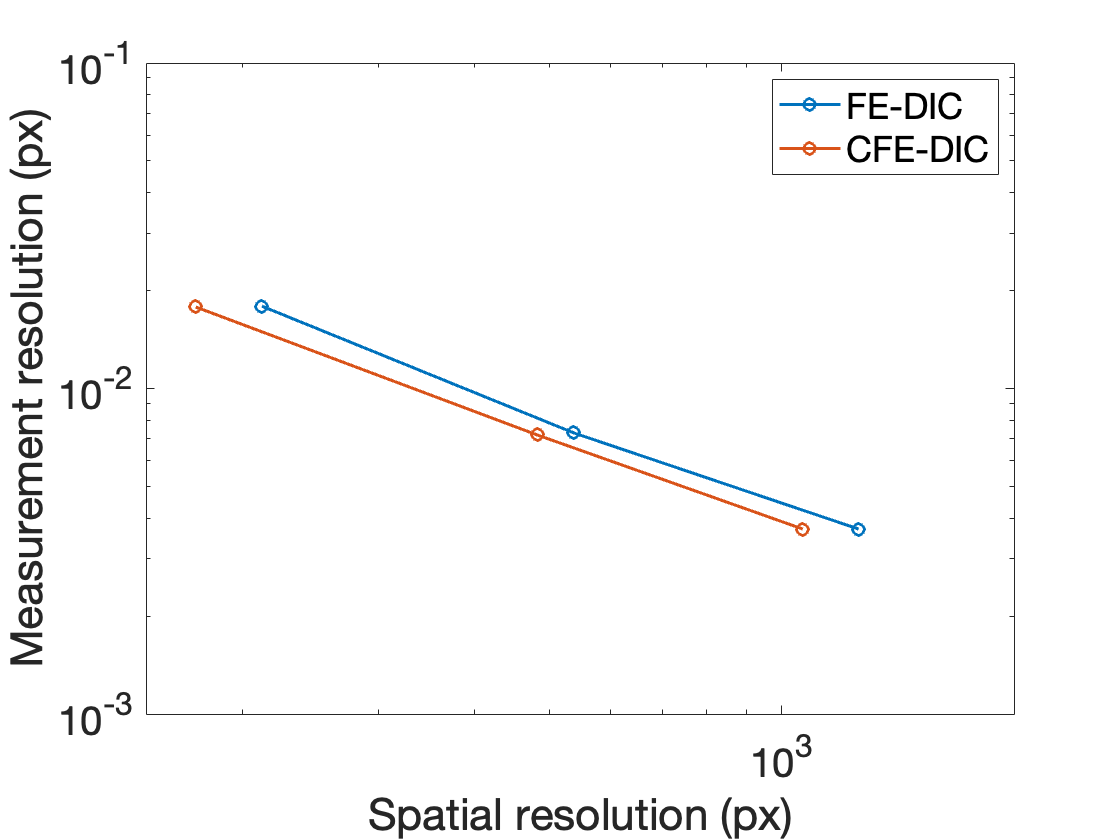}}
\subfigure[{Metrological Efficiency Indicator}  ]{\includegraphics[scale=0.2]{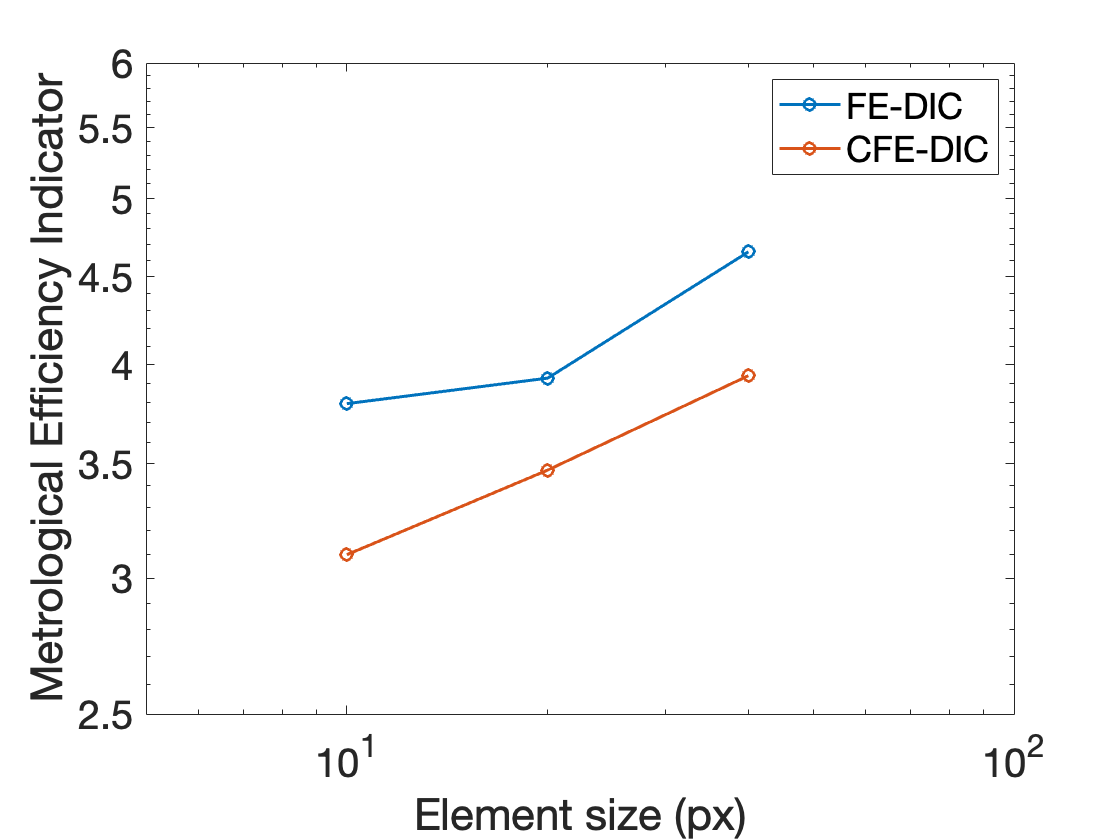}}\\
\caption{{Metrological Efficiency Indicator for the star2 example with 3 different element sizes: $h=10, 20$ and $40$ (px). px stands for pixel.}}
\label{fig:exam4MEI}
\end{figure}

{The extracted spatial and measurement resolutions for 3 element sizes $h=10, 20$ and $40$ are depicted in \figurename~\ref{fig:exam4MEI}(a). C-FE does provide smaller (better) resolutions for both metrics. As suggested in \cite{reu2022dic}, we can further calculate the Metrological Efficiency Indicator (MEI) as }
{\begin{equation}
\displaystyle
    \text{MEI}=\text{measurement resolution}\times \text{spatial resolution}
\end{equation}
}

{As shown in \figurename~\ref{fig:exam4MEI}(b), the C-FE based DIC has a smaller MEI than FE-DIC for the star2 example. This confirms again the superior performance of C-FE in the presence of noise. Note that the MEIs reported in \cite{reu2022dic} might not be comparable to ours as they might be calculated based on a different data-set or ZOI.}

\begin{figure}[htbp]
\centering
\subfigure[{Offset line, $h=5$ (px)}  ]{\includegraphics[scale=0.2]{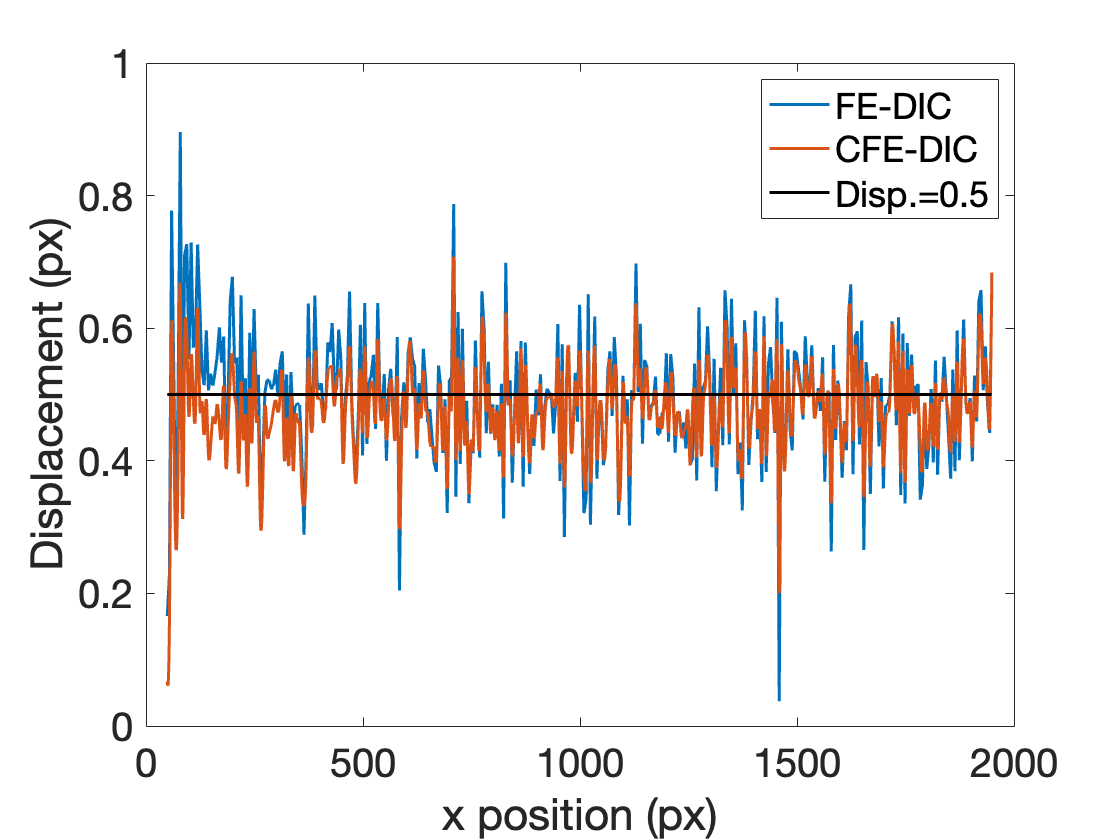}}
\subfigure[{Offset line, $h=10$ (px)}  ]{\includegraphics[scale=0.2]{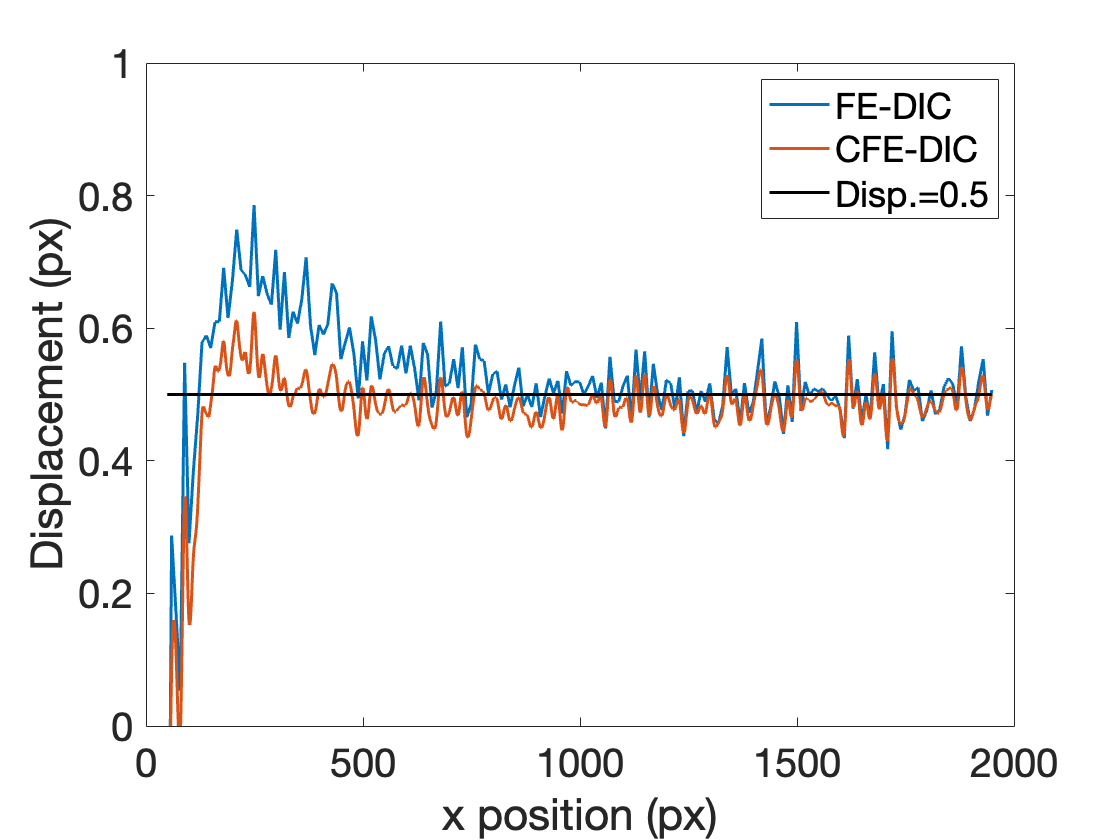}}\\
\subfigure[{Histogram of displacement data, $h=5$ (px)}  ]{\includegraphics[scale=0.2]{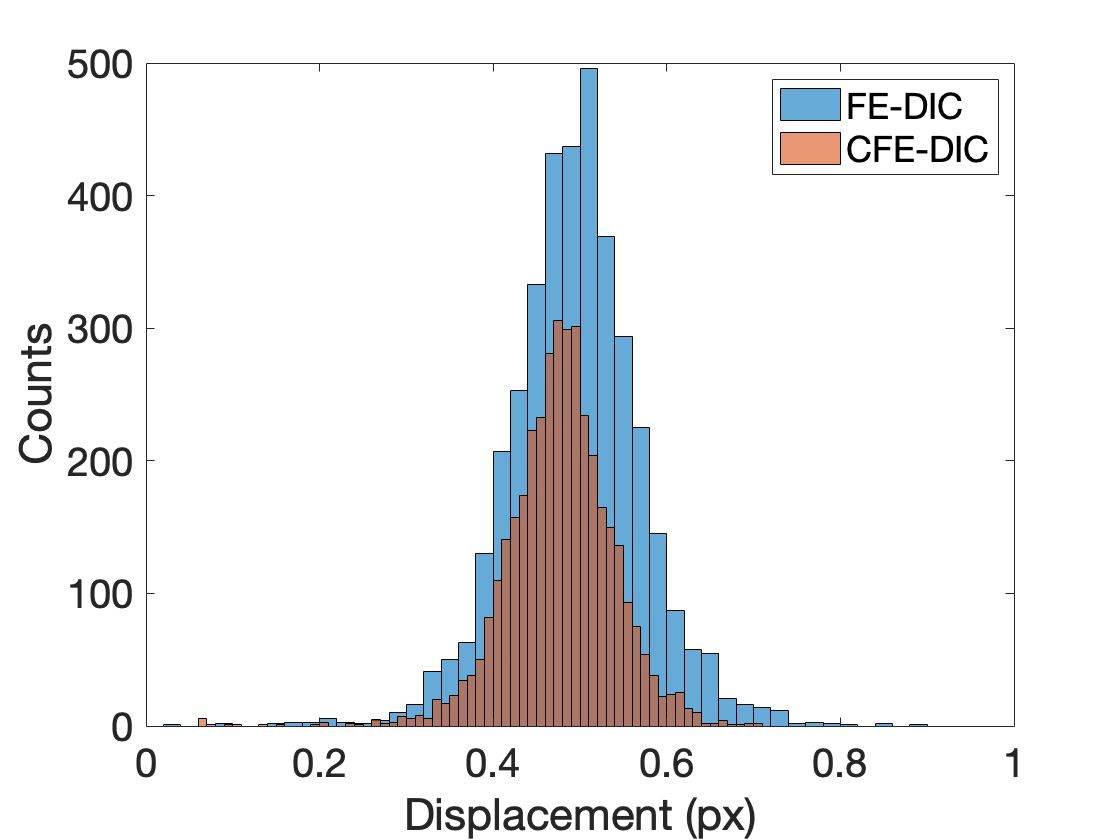}}
\subfigure[{Histogram of displacement data, $h=10$ (px)}  ]{\includegraphics[scale=0.2]{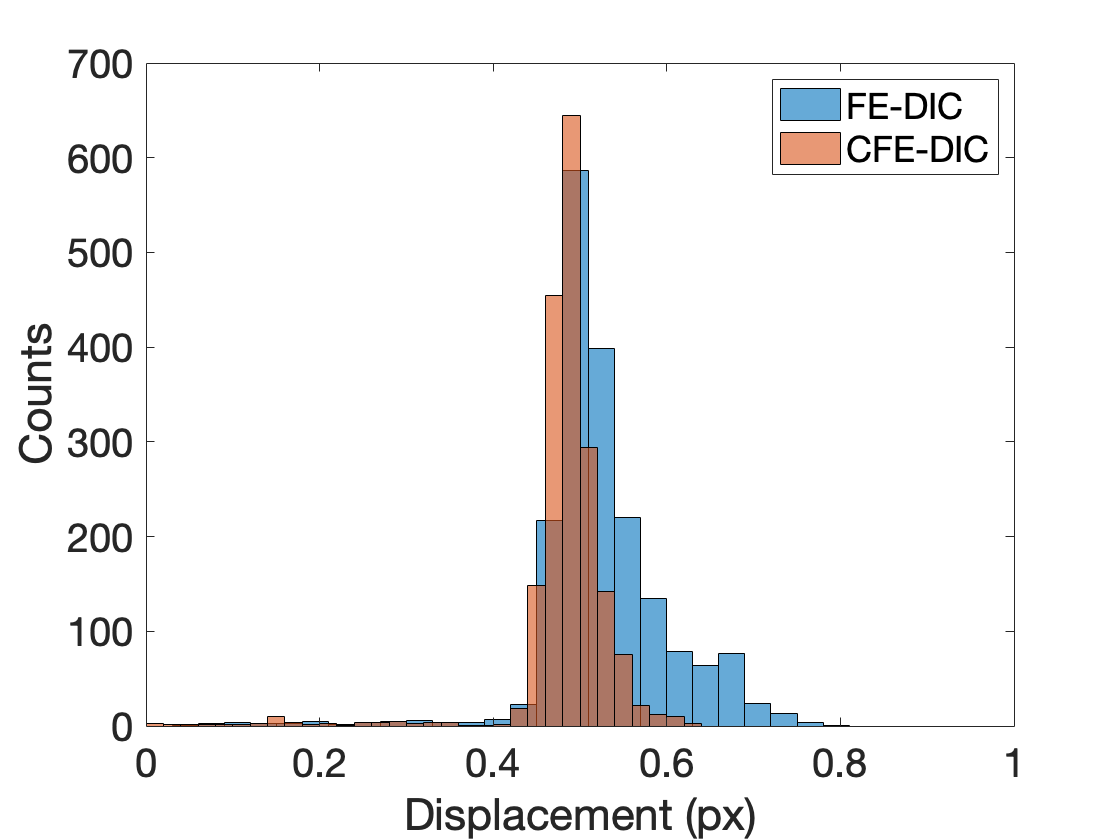}}\\
\caption{{Vertical displacement $v$ along an offset line. px stands for pixel.}}
\label{fig:exam4offcenter}
\end{figure}

To further investigate the influence of noise on the displacement field, we plotted the displacement data along another line cut with a small offset to the center. \figurename~\ref{fig:exam4offcenter} shows the displacement along this line and the histogram of the displacement data for two different meshes. We can observe the fluctuation induced by the noise for both C-FE and FE DIC results. Nevertheless, the C-FE results show a clearly smaller fluctuation, compared to the FE results.

\begin{figure}[htbp]
\centering
\subfigure[C-FE: $h=5$ (pixels)  ]{\includegraphics[scale=0.075]{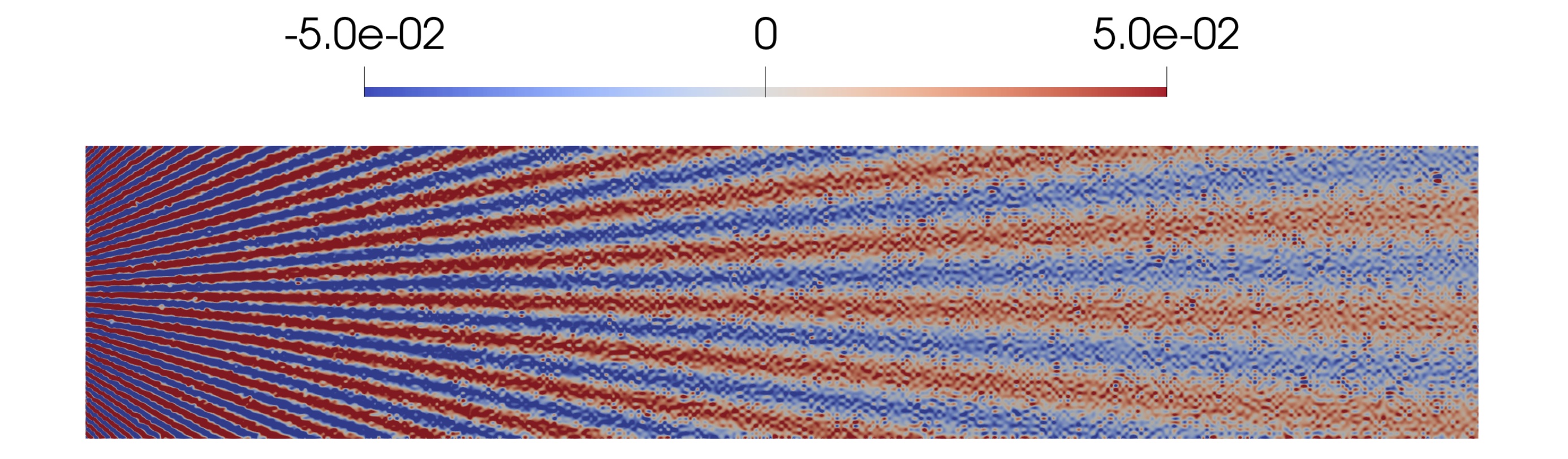}}\quad\quad
\subfigure[FE-Q4: $h=5$ (pixels)  ]{\includegraphics[scale=0.075]{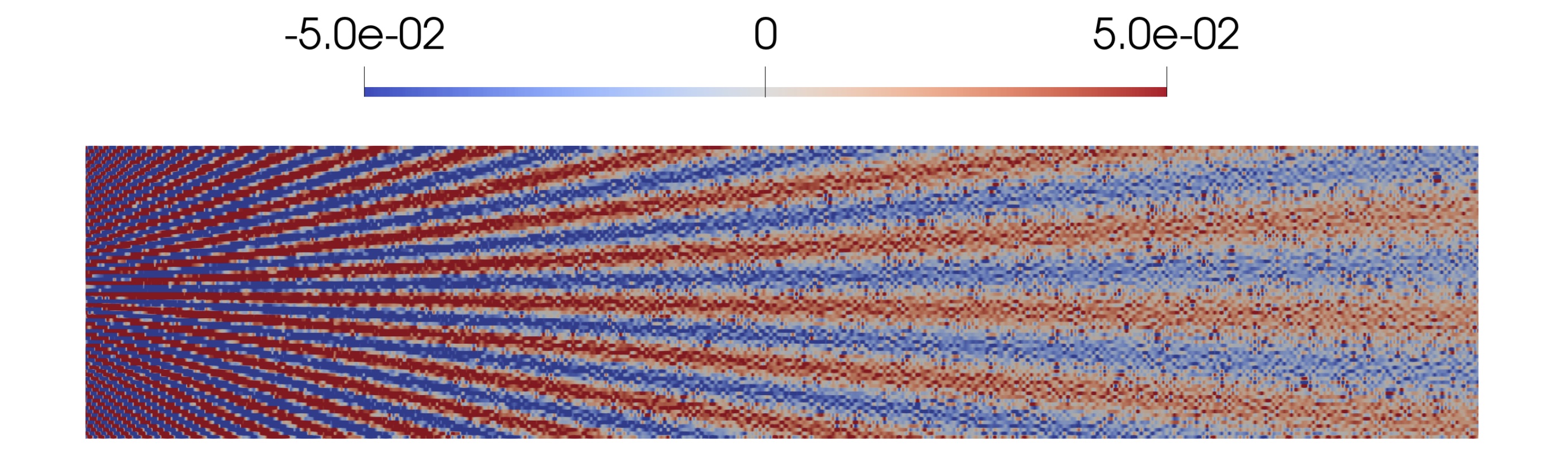}}\quad\quad\\
\subfigure[C-FE: $h=10$ (pixels) ]{\includegraphics[scale=0.075]{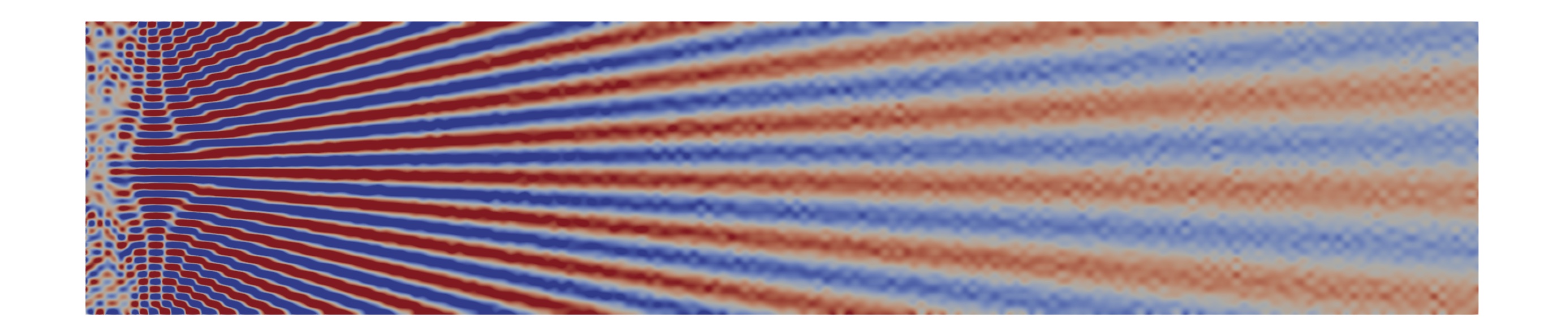}}\quad\quad
\subfigure[FE-Q4: $h=10$ (pixels)]{\includegraphics[scale=0.15]{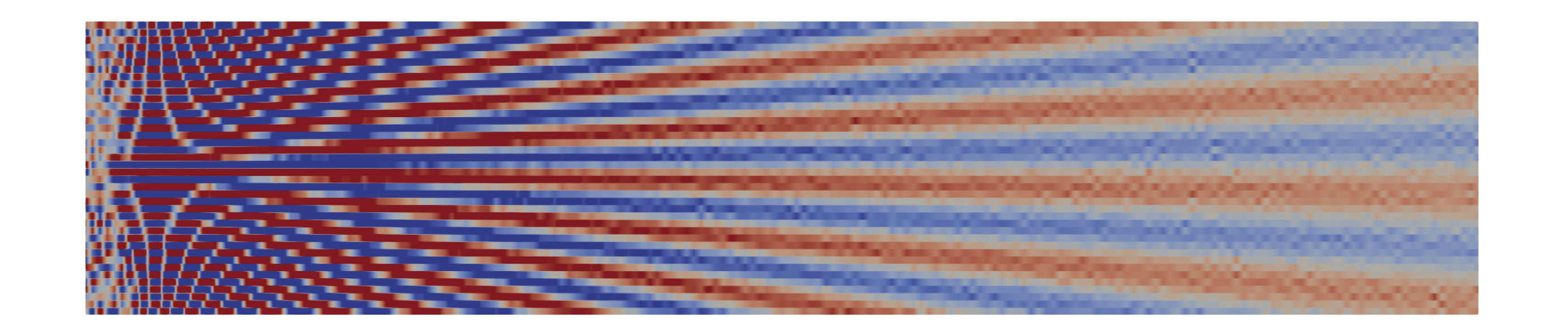}}\\
\subfigure[C-FE: $h=20$ (pixels)]{\includegraphics[scale=0.075]{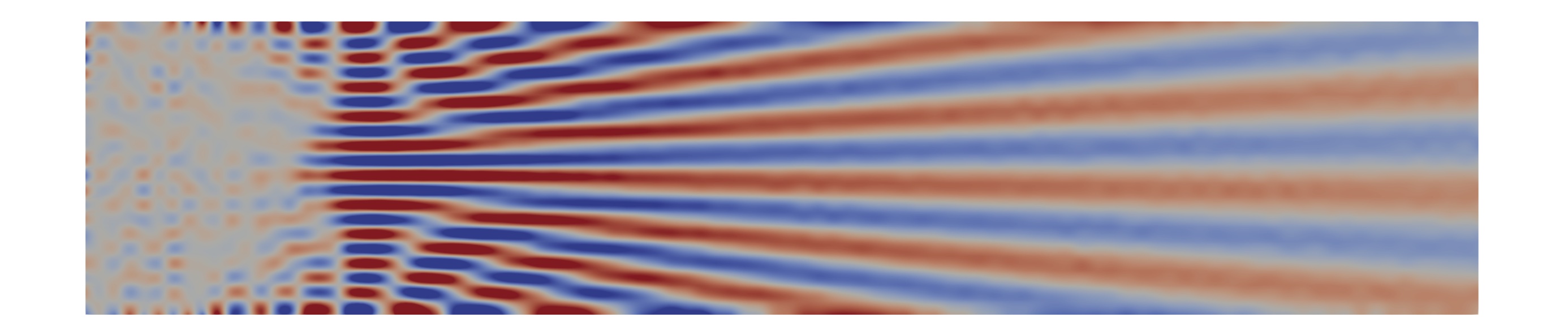}}\quad\quad
\subfigure[FE-Q4: $h=20$ (pixels) ]{\includegraphics[scale=0.15]{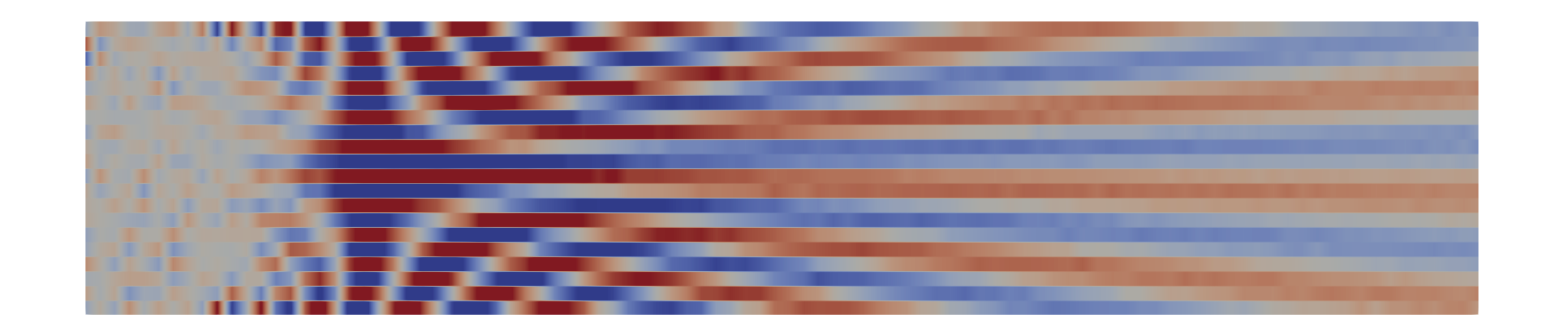}}\\
\caption{DIC results of the star2 example  for the strain $\varepsilon_{yy}$. FE-Q4 corresponds to the linear 4-node element.}
\label{fig:exam4duy}
\end{figure}

The {noise-induced} perturbations in displacement affect significantly the strain results. As shown in \figurename~\ref{fig:exam4duy}, the strain field becomes fuzzy with the noisy displacement. {Due to the smaller fluctuation}, the C-FE strain results still look better than that of FE. {This result confirms the performance and robustness of the C-FE based DIC for strain measurements}.

{We remark that the scope of this work is to demonstrate the benefits of using a more regular approximation provided by C-FE with respect to the FE approximation. A quantitative comparison to existing commercial DIC codes should be done at a later stage. This is because many of the existing DIC codes include an image processing stage or other additional numerical treatments prior to computing the DIC solutions, which is unclear to us at this time. The comparison between C-FE based DIC and an open-source DIC code will be performed in the near future.}

{\subsection{Discussion on the computational cost}}
{
The computational cost of C-FE mainly consists of two parts: one for computing the coefficients of the matrices $\boldsymbol{K}$ and $\boldsymbol{Q}$,  another for the solution of Eq. \eqref{eq:DIC-discrete}.  The first part is relatively easier to parallelize. If a massively parallel computing is used, such as graphics processing unit (GPU) computing, the cost associated with the first part is relatively low compared to the second part. This is confirmed by solving  PDEs with a GPU based C-FE implementation \cite{park2023convolution}. Therefore, we focus on the cost of the second part for solving Eq. \eqref{eq:DIC-discrete}.}

\begin{figure}[htbp]
\centering
\subfigure[{Computational cost versus the element size} ]{\includegraphics[scale=0.2]{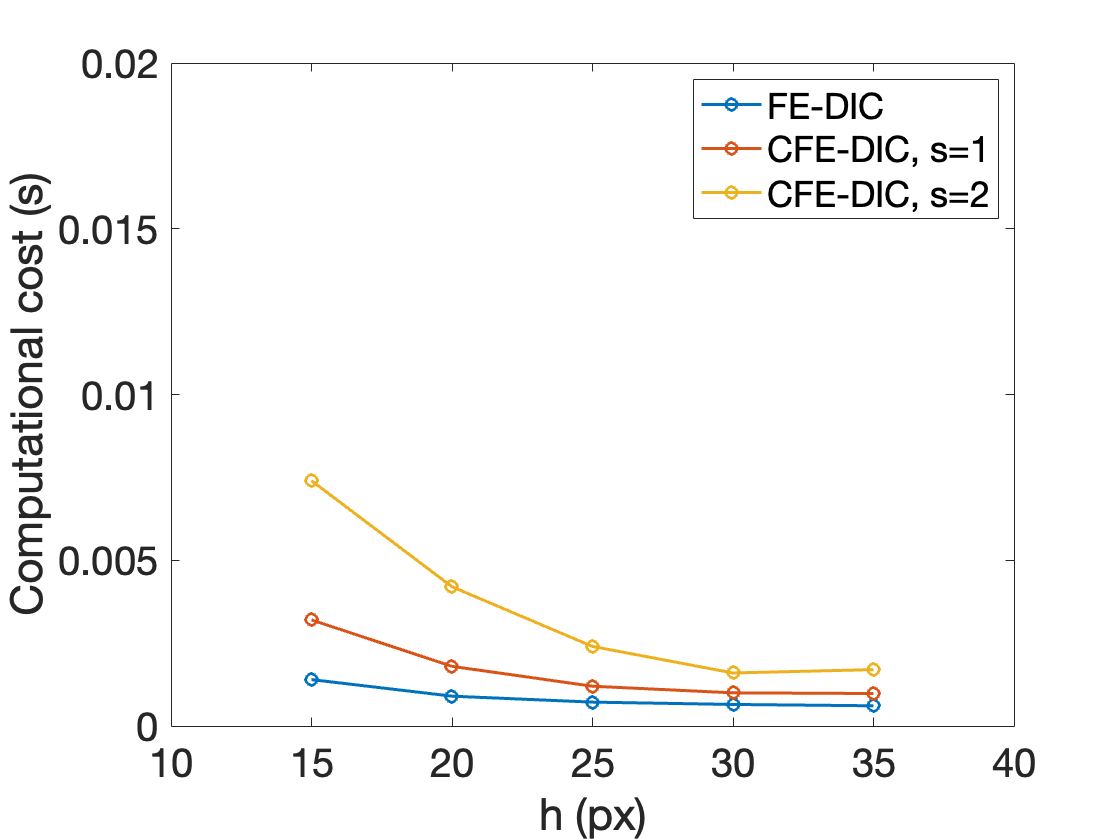}}
\subfigure[{Error versus the element size} ]{\includegraphics[scale=0.2]{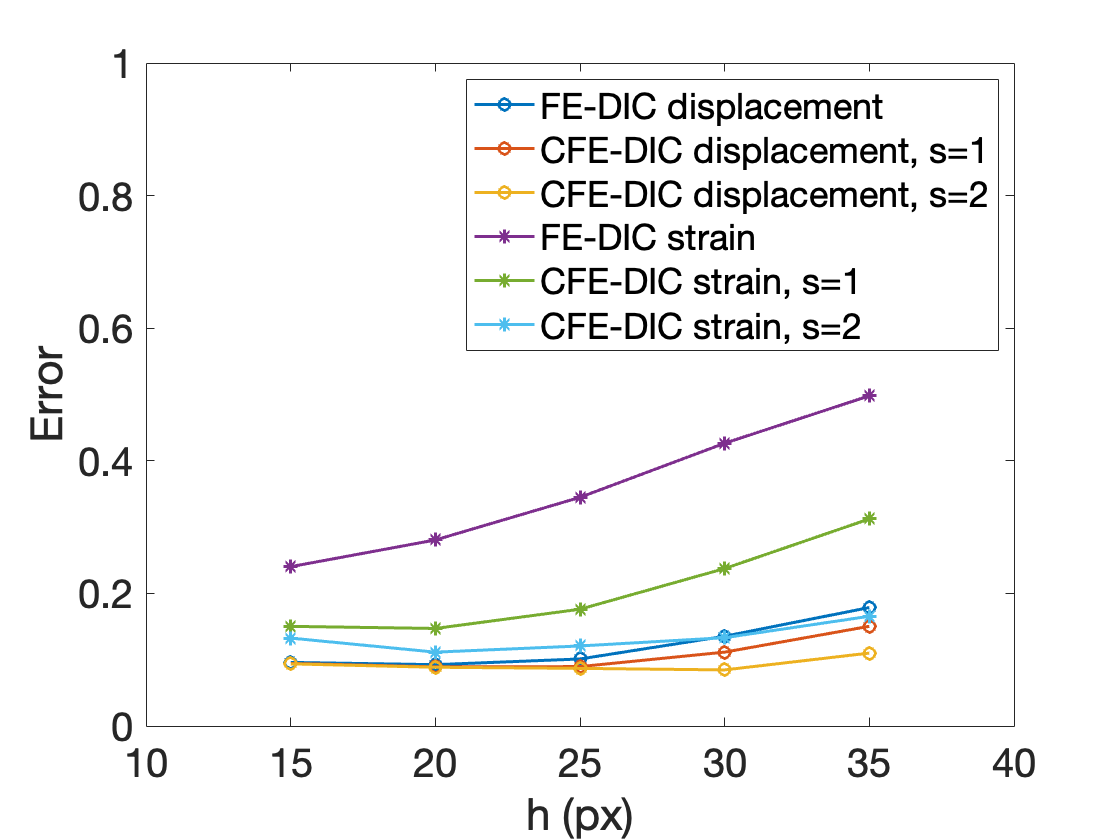}}
\caption{{Computational cost and accuracy of the proposed DIC method with different meshes. px stands for pixel.} }
\label{fig:cpucost}
\end{figure}

{
As mentioned earlier, C-FE might reduce the sparsity of the matrix $\boldsymbol{K}$ and can lead to higher computational costs for solving the inversion of the matrix, compared to standard FE. In order to make quantitative comparisons, the first example with analytical solutions presented in section \ref{sec:example1} is used for the test. The linear FE and C-FE DIC methods are tested for 5 different meshes: $h=15, 20, 25, 30, 35$ (pixels). In addition, we tested the C-FE DIC  with two sets of parameters: ($s=1, p=1, a=8$) and ($s=2, p=2, a=8$).  \figurename~\ref{fig:cpucost} reports the computational cost and $L^2$-norm error of the methods. Here, an iterative scheme, the preconditioned conjugate gradient (PCG) method pre-implemented in Matlab, is used for solving the inversion of the matrix with the tolerance $10^{-5}$ for both FE and C-FE. The computations are performed using Apple M2 chip with 24 GB memory.}

{
As shown in \figurename~\ref{fig:cpucost},  C-FE indeed increases the cost for solving the final discrete equation due to the reduced sparsity of matrix. With the current parameters, the cost for C-FE is about 2-4 times that of FE for the same mesh. However, if we look at a coarser mesh for C-FE, e.g., $h=25$ for $s=1$ or $h=30$ for $s=2$ {(as shown in \tablename~\ref{table:CPUvsErr})}, the computational time is approximately the same as that of FE at $h=15$ or $20$, while the accuracy of C-FE remains better than that of FE, especially for the strain measurement. Hence, C-FE can use a relatively coarser mesh to maintain the efficiency while keeping a higher accuracy than FE. {This seems hard to achieve with traditional quadratic Q8 FE, as its computational cost is still relatively high even with the slightly coarser mesh.}
}

\begin{figure}[htbp]
\centering
\subfigure[{FE} ]{\includegraphics[scale=0.13]{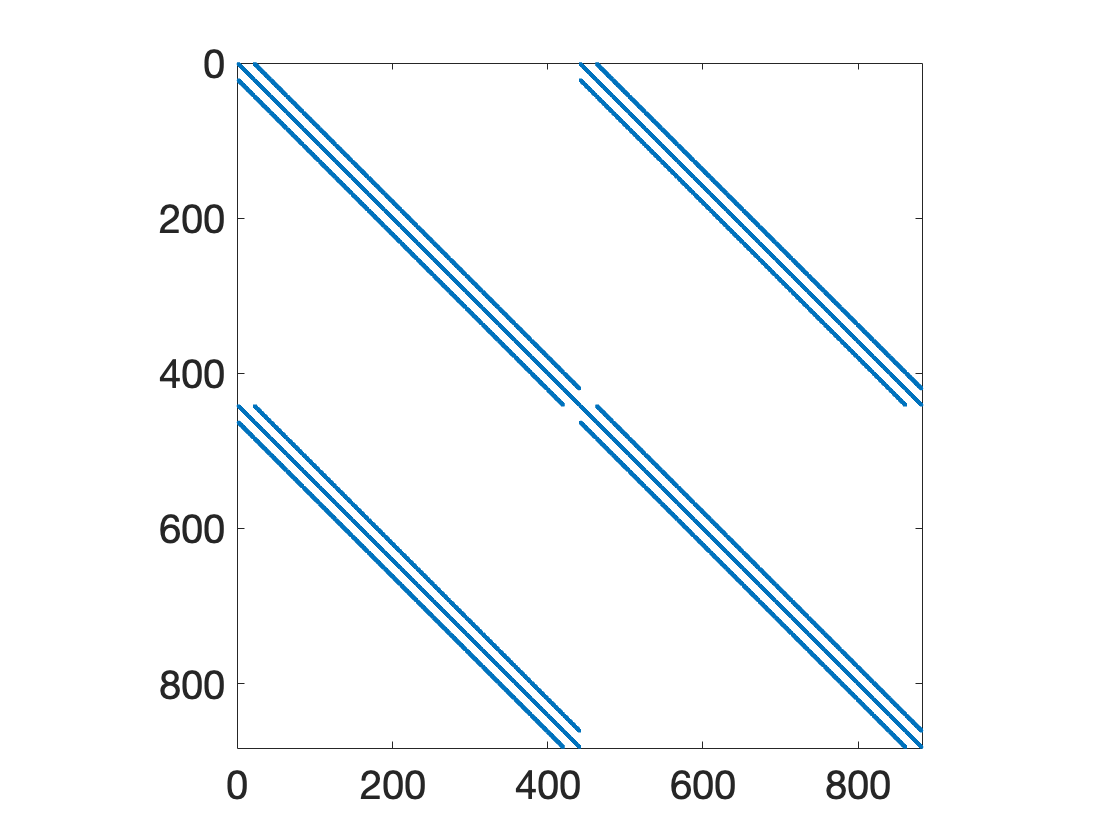}}
\subfigure[{C-FE, $s=1$} ]{\includegraphics[scale=0.13]{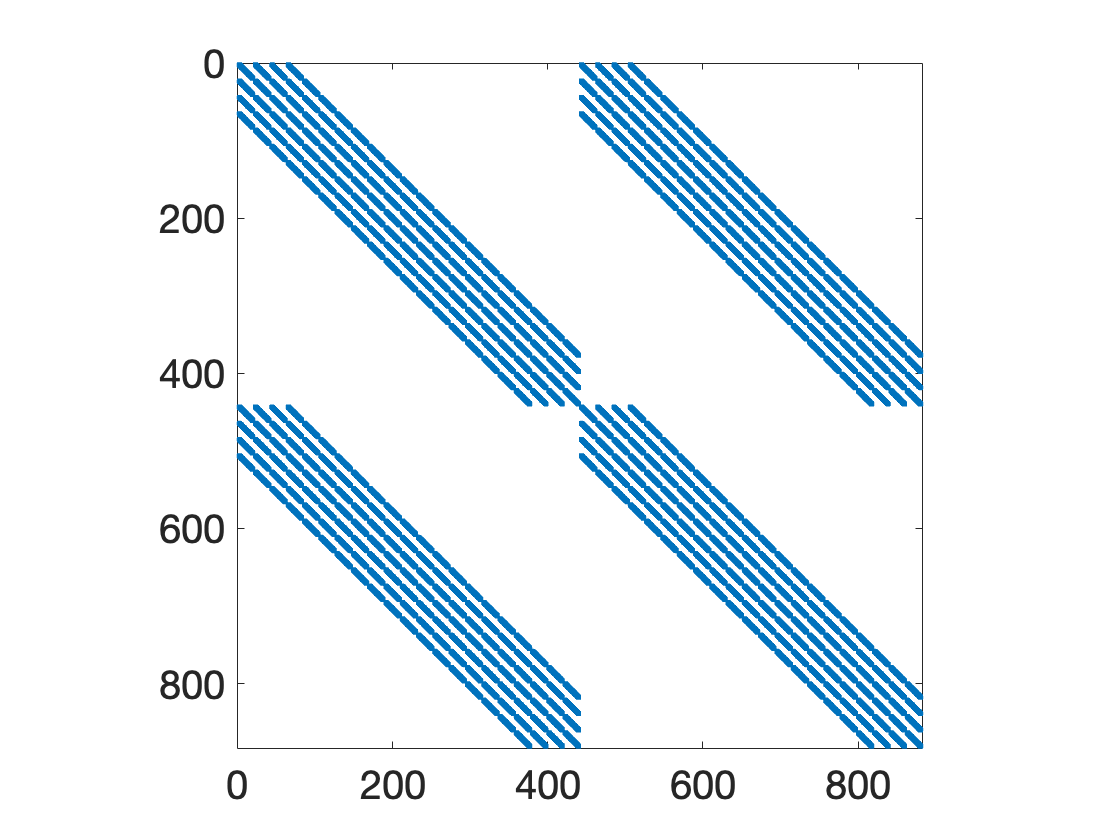}}
\subfigure[{C-FE, $s=2$} ]{\includegraphics[scale=0.13]{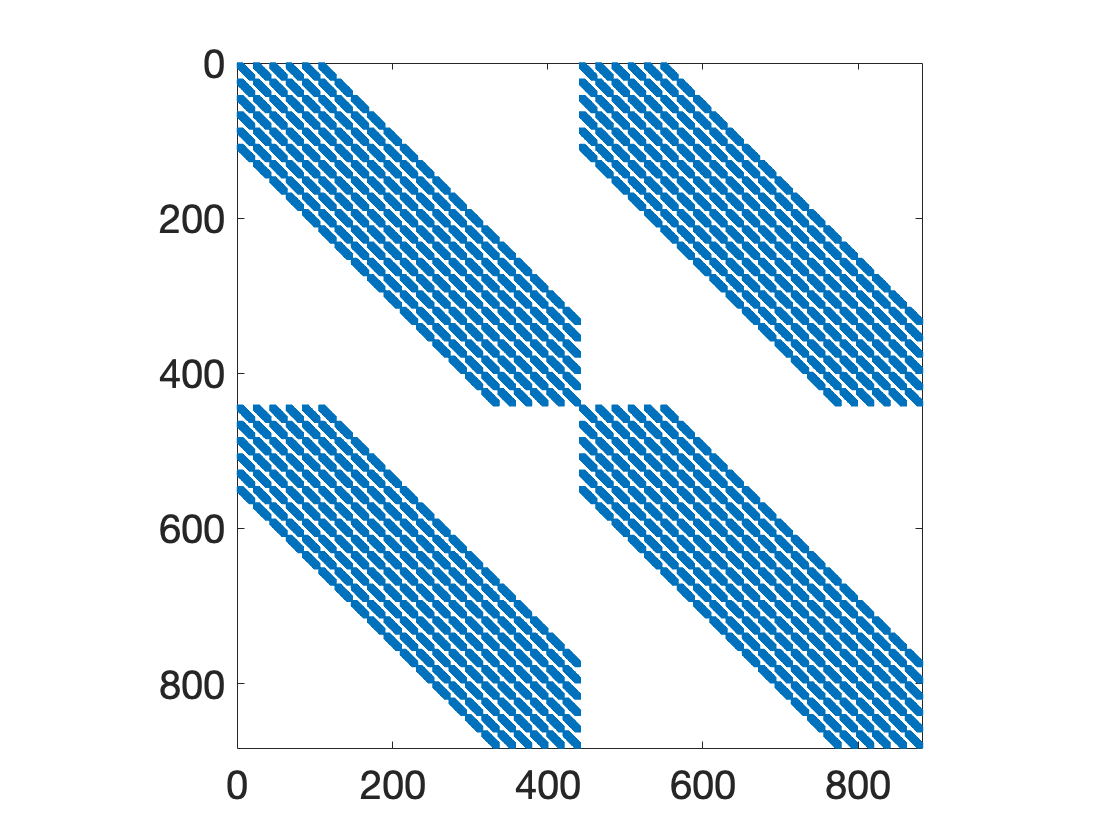}}
\caption{{Illustration of the sparsity of $\boldsymbol{K}$ for $h=20$. Blue dots indicate the non-zero values. The sparsity is 0.0191, 0.0937, and 0.2077 for FE, C-FE ($s=1$), and C-FE ($s=2$),  respectively.} }
\label{fig:sparsity}
\end{figure}

{Note that the computational cost of C-FE  can be further reduced if an adaptive patch size is used. For a better illustration of the influence of the patch size on the sparsity of the final discrete matrix $\boldsymbol{K}$, \figurename~\ref{fig:sparsity} shows a comparison between {the linear Q4} FE and C-FE. The sparsity should be reduced for an adaptive patch. In the future, we will explore more on this point and implement the C-FE DIC into GPU and compare different approaches within the context of parallel computing. }

\begin{table}[htbp]
\caption{{Computational cost and accuracy with selected parameters}}
\centering
\begin{tabular}{|c|c|c|c|c|c|c|}
\hline
 Method & Element size& Patch size   & $\epsilon_{L}(u)$ & $\epsilon_{L}(\varepsilon_{xx})$ & Time cost\\ \hline
FE & 15 & -   & {0.0959} & {0.2403} & 0.0014s \\ \hline
FE & 20 & -  & {0.0925} &{0.2806} & 0.0009s \\ \hline
C-FE & 25 & 1 & {0.0896}  & {0.1761}& 0.0012s \\ \hline
C-FE & 30 & 2 & {0.0846}  &{0.1332} & 0.0016s \\ \hline
FE-Q8 & 30 & -  & {0.0925} &{0.1577} & 0.0024s \\ \hline
\end{tabular}\\
\label{table:CPUvsErr}
\end{table}

\section{Conclusion}
A novel C-FE based DIC technique has been developed in this work,  based on a newly developed convolution approximation. The unique features of the proposed DIC method include: 1) arbitrarily high order approximations on a given FE mesh without adding more nodes; 2) highly smooth and accurate measurements for both displacement and strain; 3) flexible global/local adaptivity. The detailed implementation of the method has been discussed, including two sub-element grayscale smoothing and interpolation techniques. The performance of C-FE based DIC is expected to rely on the choice of the controlling parameters, including the polynomial order $p$, patch size $s$, and dilation $a$. The combination of parameters used in our work seems a good option for general cases, though their optimal values might need further investigations.

The proposed DIC method has been tested against various examples, including the DIC Challenge 2.0 benchmark problems, with comparison to FE based DIC.  The C-FE based DIC outperformed the usual FE-DIC in all the tested scenarios. The numerical results have confirmed that  C-FE can give highly smooth and accurate displacement and strain measurements, whereas FE only gives relatively lower order displacement approximations and usually fails to provide satisfactory strain results. In addition, the proposed DIC has shown excellent robustness with noisy images due to the convolution approximation. {These results demonstrate the great potential of C-FE. Further comparisons with existing DIC codes will be performed in the future in terms of both efficiency and accuracy.}

{Other} future studies of the method include the investigation of the optimal controlling parameters and their adaptivity as discussed before. The proposed DIC method falls into the family of global DIC techniques. A comparison of this method to local DIC approaches can also be interesting in future studies. The proposed method is general and can be extended to different types of elements and digital volume correlation analysis.

\section*{Acknowledgement}
YL would like to acknowledge the support of University of Maryland Baltimore County through the Summer Research Faculty Fellowship {and Strategic Awards for Research Transitions}. The authors also want to acknowledge the helpful discussions with Yujia Hu from {the} University of Shanghai for Science and Technology.

\begin{figure}[!htbp]
\centering
\includegraphics[scale=0.5]{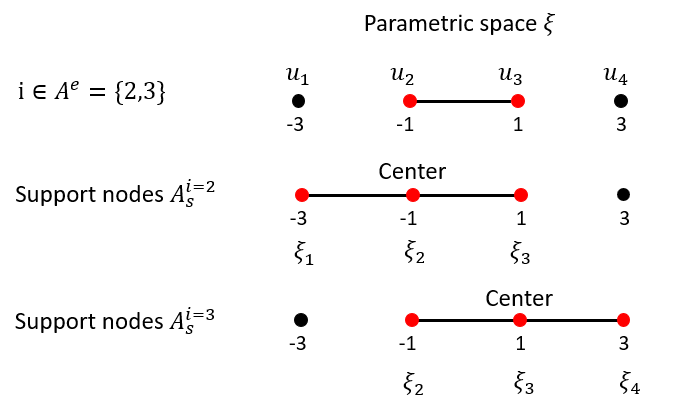}
\caption{Supporting nodes for a 1D convolution element with {the} patch size $s=1$ }
\label{fig:1DC-FEMnodes}
\end{figure}

\appendix
{\section{Radial basis interpolation}\label{apdx:radialbasis}}
Let us consider the following 1D case for illustration purposes
\begin{equation}
    \begin{aligned}           
    u^{{c}}({\xi})=&\sum_{i\in A^e}{N}_{{i}}({\xi})\sum_{j\in A^i_s} {W}^{{\xi}_i}_{a,j} ({\xi}) u_j\\
    \end{aligned}
\end{equation}
Then we can define the following interpolation as the part of the approximation centering around the $i$-th node in the element domain
\begin{equation}
\label{eq:radialbasis}     
    u^{i}({\xi})=\sum_{j\in A^i_s} {W}^{{\xi}_i}_{a,j} ({\xi}) u_j,
\end{equation}
where the supporting node set of $W$ is $A^i_s$ with a given patch size $s$.  \figurename~\ref{fig:1DC-FEMnodes} illustrates a 1D convolution element with {the} patch size $s=1$ and the two-node shape functions for $N_i$. Now the question is how to compute  $W$ based on the given supporting nodes.

Assuming the nodal solution value for the 4 nodes in \figurename~\ref{fig:1DC-FEMnodes} is $[u_1,u_2,u_3,u_4]$, we illustrate the radial basis interpolation procedure for the part centering around $i=2$. In this case, the parametric coordinates for the support nodes are $\{-3,-1,1\}$. Then we can consider the radial basis interpolation $u^{i=2}({\xi})$ has the following form
\begin{equation}      
    u^{i}({\xi})=\mathbf{\Psi}_a (\xi)\mathbf{k}+\mathbf{p}(\xi)\mathbf{l},
\end{equation}
where $\mathbf{\Psi}_a (\xi) $ is  a defined kernel function, which can be the reproducing kernel  or cubic spline kernel \cite{liu1995reproducing,chen2017reproducing} with the dilation parameter $a$, $\mathbf{p}(\xi)$ is the polynomial basis vector of {order $p$}, $\mathbf{k}=[k_1, k_2, k_3]^T$ and {$\mathbf{l}$ is the coefficient vector that helps to enforce the reproducing condition and the Kronecker delta property. Therefore, the size of $\mathbf{l}$ depends on the order of $\mathbf{p}$. In the case of a second order polynomial shown below,  $\mathbf{l}=[l_1, l_2, l_3]^T$.}  We give here {the} specific example for $\mathbf{\Psi}_a $  and $\mathbf{p}(\xi)$ using a cubic spline {kernel} and {the} second-order polynomial, {in which we can see that $a$ is  the dilation parameter that controls the window size (non-zero domain) of the cubic spline}
\begin{equation}      
\begin{aligned} 
   \mathbf{\Psi}_a (\xi)=[&{\Psi}_a (\xi-\xi_1),\  {\Psi}_a (\xi-\xi_2) , \ {\Psi}_a (\xi-\xi_3)]\\
   \text{where}\quad {\Psi}_a (\xi-\xi_I): &= {\Psi}_a (z) \quad \text{with} \quad z=\frac{|\xi-\xi_I|}{a}\\
   &=\begin{cases}
    \frac{2}{3} -4z^2+4z^3 \quad \forall z \in [0, \frac{1}{2}]\\
    \frac{4}{3} -4z+4z^2- \frac{4}{3}z^3 \quad \forall z \in [\frac{1}{2}, 1]\\
    0 \quad \forall z \in (1,+\infty)
   \end{cases},
\end{aligned}
\end{equation}
and 
\begin{equation}      
\begin{aligned} 
   \mathbf{p}=[1,\ \xi,\ \xi^2]
\end{aligned}
\end{equation}
Now we can compute $\mathbf{k}$ and $\mathbf{l}$ by enforcing the {conditions below} 
\begin{equation}   
\label{eq:reproducecondition}
\begin{aligned} 
\begin{cases}
    u^{i}({\xi_1})= u_1 \\
     u^{i}({\xi_2})= u_2 \\
    u^{i}({\xi_3})= u_3 \\
     {\sum_i {k}_i= 0}\\
     [\xi_1, \xi_2, \xi_3]\ \mathbf{k}=0\\
      [\xi_1^2, \xi_2^2, \xi_3^2]\ \mathbf{k}=0\\
   \end{cases},
\end{aligned}
\end{equation}
Solving the above equations gives the solution to $\mathbf{k}$ and $\mathbf{l}$, which reads
\begin{equation}      
\begin{aligned} 
\begin{cases}
    \mathbf{k}=\mathbf{K}\mathbf{u}\\
    \mathbf{l}=\mathbf{L}\mathbf{u}\\
   \end{cases},
\end{aligned}
\end{equation}
with 
\begin{equation}      
\begin{aligned} 
\begin{cases}
    \mathbf{u}=[u_1,\ u_2,\ u_3]^T\\
    \mathbf{L}=(\mathbf{P}^T\mathbf{R}_0^{-1}\mathbf{P})^{-1}\mathbf{P}^T\mathbf{R}_0^{-1}\\
    \mathbf{K}=\mathbf{R}_0^{-1}(\mathbf{I}-\mathbf{P}\mathbf{L})
   \end{cases},
\end{aligned}
\end{equation}
and 
\begin{equation}      
\label{eq:moment}
\begin{cases}
    \mathbf{R}_0=\begin{pmatrix}
     \mathbf{\Psi}_a (\xi_1)\\
      \mathbf{\Psi}_a (\xi_2)\\
       \mathbf{\Psi}_a (\xi_3)
    \end{pmatrix}=\begin{pmatrix}
     {\Psi}_a (\xi_1-\xi_1) &\  {\Psi}_a (\xi_1-\xi_2) &\  {\Psi}_a (\xi_1-\xi_3)\\
      {\Psi}_a (\xi_2-\xi_1) &\  {\Psi}_a (\xi_2-\xi_2) &\  {\Psi}_a (\xi_2-\xi_3)\\
       {\Psi}_a (\xi_3-\xi_1) &\  {\Psi}_a (\xi_3-\xi_2) &\  {\Psi}_a (\xi_3-\xi_3)
    \end{pmatrix}\\
    \\
    \mathbf{P}=\begin{pmatrix}
     \mathbf{p} (\xi_1)\\
      \mathbf{p} (\xi_2)\\
       \mathbf{p} (\xi_3)
    \end{pmatrix}=\begin{pmatrix}
     1 &\  \xi_1 &\  \xi_1^2\\
       1 &\  \xi_2 &\  \xi_2^2\\
        1 &\  \xi_3 &\  \xi_3^2\\
    \end{pmatrix}\\
   \end{cases},
\end{equation}
Finally, the radial basis interpolation with the computed coefficients reads
\begin{equation}      
  \begin{aligned}
      u^{i}({\xi}) = \mathbf{\Psi}_a (\xi)\mathbf{k}+\mathbf{p}(\xi)\mathbf{l} &= \mathbf{\Psi}_a (\xi)\mathbf{K}\mathbf{u}+\mathbf{p}(\xi)\mathbf{L}\mathbf{u}\\
      &= (\mathbf{\Psi}_a (\xi)\mathbf{K}+\mathbf{p}(\xi)\mathbf{L})\mathbf{u}\\
      &= {W}^{{\xi}_i}_{a,1}(\xi) u_1+{W}^{{\xi}_i}_{a,2} (\xi)u_2+{W}^{{\xi}_i}_{a,3}(\xi) u_3\\
      &= \sum_{j\in A^i_s} {W}^{{\xi}_i}_{a,j} ({\xi}) u_j ,  
  \end{aligned}
\end{equation}
where ${W}^{{\xi}_i}_{a,j}$ is obtained by identifying the corresponding coefficient of $u_j$. By analogy, we can compute the other convolution patch functions $W$ with the support $A^{i=3}_s$. {The} detailed mathematical derivation and analysis of the radial basis interpolation can be found in \cite{schaback2001characterization}.

The generalization of {the} above procedure to 2D cases is straightforward. {By} assuming $K$ is the number of nodes in the support $A^{i}_s$, the 2D cubic spline kernel $\mathbf{\Psi}_a (\boldsymbol{\xi})$  and second-order polynomial $\mathbf{p}(\boldsymbol{\xi})$ can be defined as
\begin{equation}      
\begin{aligned} 
   &\mathbf{\Psi}_a (\boldsymbol{\xi}) =[{\Psi}_a (\boldsymbol{\xi}-\boldsymbol{\xi}_1),\  {\Psi}_a (\boldsymbol{\xi}-\boldsymbol{\xi}_2),\dots , \ {\Psi}_a (\boldsymbol{\xi}-\boldsymbol{\xi}_K)]\\
   \text{with}\quad &{\Psi}_a (\boldsymbol{\xi}-\boldsymbol{\xi}_I): ={\Psi}_a (\xi-\xi_I){\Psi}_a (\eta-\eta_I)\\
\end{aligned}
\end{equation}
and 
\begin{equation}      
\begin{aligned} 
   \mathbf{p}=[1,\ \xi,\ \xi^2,\ \eta,\ \xi\eta,\ \eta^2]
\end{aligned}
\end{equation}
The remaining equations from \eqref{eq:reproducecondition} to \eqref{eq:moment} can be adapted accordingly. 

{
\section{{Illustration of the 1D convolution shape functions}}
\label{apdx:1DCFEshapefunction}
For a better understanding of the convolution shape function $\tilde{{N}}_k$, we illustrate here a 1D convolution approximation with the patch size $s=1$. Specifically, we use the example of \figurename~\ref{fig:1DC-FEMnodes} for the supporting nodes. The convolution patch function ${W}^{{\xi}_i}_{a,j}$ can be precomputed using the procedure described in \ref{apdx:radialbasis}. Recall that the general 1D convolution approximation is written as
\begin{equation}
\label{eq:C-FEM-1D}       
    u^{{c}}({\xi})=\sum_{i\in A^e}{N}_{{i}}({\xi})\sum_{j\in A^i_s} {W}^{{\xi}_i}_{a,j} ({\xi}) u_j
\end{equation}
In this specific example, the FE shape function nodal support set is $A^e=\{2,3\}$, and the nodal patch is $A^{i=2}_s=\{1,2,3\}$, $A^{i=3}_s=\{2,3,4\}$. Eq. \eqref{eq:C-FEM-1D} then becomes 
\begin{equation}
    \begin{aligned}           
    u^{c}({\xi}) & = \sum_{i\in A^e}{N}_{{i}}({\xi})\sum_{j\in A^i_s} {W}^{{\xi}_i}_{a,j} ({\xi}) u_j\\
    & = {N}_{{2}}(\xi){W}^{{\xi}_2}_{a,1}(\xi)u_1+({N}_{{2}}(\xi){W}^{{\xi}_2}_{a,2}(\xi)+{N}_{{3}}(\xi){W}^{{\xi}_3}_{a,2}(\xi))u_2\\
    &+({N}_{{2}}(\xi){W}^{{\xi}_2}_{a,3}(\xi)+{N}_{{3}}(\xi){W}^{{\xi}_3}_{a,3}(\xi))u_3+{N}_{{3}}(\xi){W}^{{\xi}_3}_{a,4}(\xi)u_4 \\
    & = \sum_{k\in A^e_s} \tilde{{N}}_k({\xi}) u_k,
    \end{aligned}
\end{equation}
where $A^e_s=\underset{i\in A^e}{\bigcup} A^{i}_s =\{1,2,3,4\}$. Therefore, there are in total 4 convolution shape functions for $s=1$. If $s=2$, we can expect $6$ shape functions, as shown in \figurename~\ref{fig:convshapefunctions}.}


\bibliographystyle{model1-num-names}
\bibliography{cDIC.bib}







\end{document}